\documentclass[aps,pra,reprint,showpacs,superscriptaddress,nofootinbib,longbibliography]{revtex4-1}

\setcitestyle{authoryear,round}

 \usepackage{subfigure}
 \usepackage{hyperref}
\usepackage{graphicx}
\usepackage{amssymb}
\usepackage{amsfonts}
\usepackage{dsfont}
\usepackage{amsmath}
\usepackage{multirow}
\usepackage{bm}

\newcommand{\be}{\begin{equation}}
\newcommand{\ee}{\end{equation}}
\newcommand{\bq}{\begin{eqnarray}}
\newcommand{\eq}{\end{eqnarray}}
\newcommand{\p}{\partial}
\newcommand{\epsgras}{\ensuremath{\bm{\epsilon}}}
\newcommand{\nablagras}{\ensuremath{\bm{\nabla}}}
\newcommand{\bc}{\begin{center}}
\newcommand{\ec}{\end{center}}
\newcommand{\beq}{\begin{equation}}
\newcommand{\eeq}{\end{equation}}
\newcommand{\bea}{\begin{eqnarray}}
\newcommand{\eea}{\end{eqnarray}}
\newcommand{\ind}[1]{_{\mbox{\scriptsize #1}}}
\newcommand{\GG}[1]{\ensuremath{\mathbf{#1}}}

\newcommand{\ds}{\displaystyle}
\newcommand{\ket}[1]{\ensuremath{| #1 \rangle}} 
\newcommand{\bra}[1]{\ensuremath{\langle #1 |}} 
\newcommand{\braket}[2]{\ensuremath{\langle #1 | #2 \rangle}} 
\newcommand{\np}{\ensuremath{n}}

\newcommand{\signgf}{s}
\newcommand{\BB}{\ensuremath{\boldsymbol{\mathcal{B}}}}

\newcommand{\secref}[1]{Sec.~\ref{#1}}
\newcommand{\secrefs}[1]{Sec.~\ref{#1}}
\newcommand{\Secref}[1]{Section~\ref{#1}}
\newcommand{\figref}[1]{Fig.~\ref{#1}}
\newcommand{\Figref}[1]{Figure~\ref{#1}}

\begin{document}

\title{Trapping atoms with radio-frequency adiabatic potentials}

 \author{H{\'e}l{\`e}ne Perrin}
 \affiliation{CNRS, UMR 7538, Universit\'e Paris 13, Sorbonne Paris Cit\'e, Laboratoire de physique des lasers, 99 avenue J.-B. Cl\'ement, F-93430 Villetaneuse}
 \author{Barry M.~Garraway}
 \affiliation{Department of Physics and Astronomy,
   University of Sussex, Falmer, Brighton, BN1 9QH,
   United Kingdom}

\begin{abstract}
In this chapter we review the field of radio-frequency dressed atom trapping. We emphasise the role of adiabatic potentials and give simple, but generic models of electromagnetic fields that currently produce traps for atoms at microkelvin temperatures and below. The paper aims to be didactic and starts with general descriptions of the essential ingredients of adiabaticity and magnetic resonance. As examples of adiabatic potentials we pay attention to radio-frequency dressing in both the quadrupole trap and the Ioffe-Pritchard trap. We include a description of the effect of different choices of radio-frequency polarisation and orientations or alignment. We describe how the adiabatic potentials, formed from radio-frequency fields, can themselves be probed and manipulated with additional radio-frequency fields including multi-photon-effects. We include a description of time-averaged adiabatic potentials. Practical issues for the construction of radio-frequency adiabatic potentials are addressed including noise, harmonics, and beyond rotating wave approximation effects.
\end{abstract}


\maketitle

\tableofcontents

\section{Introduction}
\label{sec:intro}
The manipulation of the external motion of atoms and molecules with light or magnetic fields represents one of the great successes of atomic physics in the past decades \citep{CohenDGOVO}, and has led to major discoveries acknowledged by several Nobel prizes. Laser cooling and trapping \citep{ChuNobel,CohenNobel,PhillipsNobel} gives access to a sample of billions of neutral particles, isolated from the environment, at temperatures in the microkelvin range. This results in a suppression of the Doppler effect, and cold atom samples have found their first applications in high precision measurements, in particular atomic clocks \citep{Bize2005} and atom interferometry \citep{Cronin2009}, including acceleration and rotation sensors \citep{Tino2014}. Thus the subject also now plays its role in the emergence of quantum technologies.

At these low temperatures the laser cooled atoms may then be confined in trapping potentials resulting from optical or magnetic fields. Whilst they would be immediately lost should they touch a normal physical container, we can avoid this problem by using, for example, magnetic fields, which allows us to keep the atoms confined for up to several minutes. Starting from this point, and applying the technique of evaporative cooling, which consists in the elimination of the most energetic particles confined in such conservative traps, has led to the first observation of Bose--Einstein condensation in dilute atomic gases by \cite{CornellNobel} and \cite{KetterleNobel}. This achievement has opened the now mature research field of ultracold quantum gases.

The study of ultracold atomic gases has given rich insights into many physical phenomena including superfluidity \citep{Leggett2006}, low-dimensional physics such as the Berezinskii--Kosterlitz--Thouless transition, many-body physics \citep{Bloch2008RMP}, and much more. This is made possible by the control over most of the relevant physical parameters. These include the temperature, the interaction strength, the dimension of the system and the confinement geometry. A famous example is the confinement of ultracold atoms in optical lattices \citep{Bloch2005}, which opens up the way to the quantum simulation of the electronic properties in solid state systems with ultracold atoms in periodic lattices as a model system \citep{Bloch2012}.

Both in the contexts of quantum sensors and quantum simulators, it becomes crucial to control precisely the confinement geometry of the sample to something far from the simple case of the harmonic trap. Optical lattices provide a well established way to realize periodic potentials, but there is also a need for other geometries such as double-wells, flat traps for two-dimensional gases or ring traps. Moreover, the dynamical control of the trapping potential is essential, for example, in the study of out-of-equilibrium dynamics in these systems. Adiabatic potentials arising from the dressing of atoms with a radio-frequency field provide widely tunable potentials for these purposes.

Radio-frequency (rf) fields are already routinely used in quantum gas experiments for the evaporative cooling stage in magnetic traps \citep{Ketterle1996a}. A rf `knife' is applied over a time of order a few seconds to induce a change in the atomic internal state and remove the most energetic atoms so that, on average, the remaining atoms have a lower average energy, leading to lower temperatures. The transfer of an energetic atom to an untrapped state is efficient if the rf amplitude is large enough, and this can be seen as a direct modification of the trapping potential due to the presence of the rf field. This rf field can then have its frequency continuously changed during the process, to adapt to the situation of fewer and colder trapped atoms. In effect then, we use our control of radio-frequency radiation to in turn control a microscopic system at ultra-cold temperatures. These basic ideas are also used for a next-level control of atomic systems through adiabatic potentials.

In this chapter we give an overview of an approach that broadens the scenario of rf evaporation to a much wider range of situations with radio-frequency adiabatic potentials that allow more complex trapping potentials needed for quantum simulation, atomtronics \citep{AmicoAtomtronics}, and sensing systems. We will recall that we can control frequency, amplitude, polarisation and the timing of rf radiation. We can have multiple rf fields.
All this control will allow us to fine-tune atomic potentials, such as barriers, without negative effects. It will allow us to split Bose--Einstein condensates for interferometry, even when they are close (microns) to noisy surfaces such as those found on atom chips. It will allow us to create ring traps for rotation physics and for sensing, dumb-bell traps, bubble traps, 2D traps and other exotic atom traps for the study of cold atomic gas in new topologies.

In order to get to grips with the adiabatic potentials in this Chapter, we first introduce the basic ideas of adiabaticity in \secref{sec:basics}. This will present the ideas that lead to a robust potential that can be modified by the experimentalist, and which the atoms follow as they move through space. To illustrate the approach to adiabaticity, we will include a classic non-adiabatic scenario: the Landau-Zener crossing. 
Then, in \secref{sec:magres}, we will discuss magnetic resonance in order to give the fundamentals of the interaction of our atoms with the radio-frequency field. In \secref{sec:adiabatic_potential} we combine the concepts of \secrefs{sec:basics} and \ref{sec:magres} to introduce the adiabatic potentials that are the core of the chapter. This includes a discussion of polarisation and `loading' (i.e.\ the method used to transfer cold atoms into the trap).
At this point the reader will have the needed basic theory and \secref{sec:examples_of_potentials} discusses significant examples of adiabatic potentials arising from some standard field configurations: i.e.\ from the Ioffe-Pritchard trap (\secref{sec:examples_IP}) and from the 3D quadrupole trap (\secref{sec:dressed_quad}). We include here (\secref{sec:ring-trap-quad}) a discussion on ring traps as this is an area of intriguing variety and potential applications.

The basic ideas and some examples are covered before the next \secrefs{sec:taaps} and \ref{sec:multipleRF} which look at some more complex ways of creating adiabatic potentials with even more variety. \Secref{sec:taaps} introduces the idea of modifying adiabatic potentials by moving and oscillating them in space at a frequency faster than the mechanical response of the atom. The resulting effective potential is governed by a wider set of parameters and has a new form which can allow us to make potentials that would not be allowed by Maxwell's equations applied to a static magnetic field. We give a didactic example in the section. 
Our second expansion of the basic ideas examines possibilities with multiple rf fields in \secref{sec:multipleRF}. This can involve one rf field used to make an adiabatic potential, and another one probing it. This can be used for spectroscopy of the trap. Equally, a second rf field can be used to cool down a rf dressed trap (rf evaporative cooling again). Or, three different frequencies can be used to make a double-well potential (or a double shell), and this idea can be extended to more potential wells with additional fields at different frequencies. 
Finally, we look at some practical matters in \secref{sec:practical}, that is, issues of trap lifetime, strong rf and alignment of polarisation directions, and then we conclude in \secref{sec:conclusion}.

\section{Basic concepts}
\label{sec:basics}

 \subsection{Concept of adiabaticity}
 \label{sec:adiabaticity}
If you have ever heard a violin or a guitar make a sliding note just from the slide of the fingers on the neck of the instrument you'll have an idea of a classical version of adiabatic following. In the instrument case, an eigenmode (or modes) of the string are excited, and subsequently the properties of that mode are changed by moving the finger at the boundary. In particular the frequency of the mode changes, and provided the change in eigenmode pitch is not too rapid, the excitation, e.g.\ of a fundamental, \emph{follows} the mode. That is, if we started with a fundamental excitation at low frequency, we obtain a fundamental excitation at a higher frequency after shortening the string. This would not be true if the slide is too quick: in that case, starting with only a fundamental excitation, we would obtain excited harmonics after the slide. 
The situation is very similar for a quantum system as we see below in \secref{sec:an-adiabatic-trap}.
This is best exemplified with a two-state quantum system such as found for many simplified atomic systems, or for a spin-half particle undergoing an interaction. (We will examine systems with more than two levels in the next sections where atoms interacting with magnetic fields naturally provide multi-level systems with an angular momentum quantum number $F= 1/2, 1, 3/2, 2, 5/2, 3, \cdots$.) The two-level system then can be parameterised by a Hamiltonian of the form
\begin{equation}
  \label{eq:H-TLS}
\hat  H(t) = \begin{bmatrix}
        -\alpha(t) & \beta(t)\\
         \beta(t)  & \alpha(t)
       \end{bmatrix}.
\end{equation}
Here the parameter $\beta(t)$ represents the coupling between two energy levels which have energies $\mp\alpha(t)$. Everything is allowed to be time-dependent in the Hamiltonian (which immediately makes the problem hard to solve exactly, apart from some special cases). We don't include here a constant energy term; if we do, it affects both levels in the same way and it just adds a time-dependent phase factor to all the states. In the example of the magnetic spin problem, the parameter $\alpha(t)$ might arise from the interaction of the magnetic dipole with a (slowly) time-varying magnetic field, and the parameter $\beta(t)$ might arise from the additional interaction with a radio-frequency (rf) field, which causes a coupling of the Zeeman states with energies $\pm\alpha(t)$. The parameter $\beta(t)$ might also have a time-varying amplitude.

As in the case of the sliding guitar string, we want to start with a particular eigenenergy (frequency) and then see if we can `follow' it as the Hamiltonian parameters change. At an instant of time $t$ the eigenvalues of the Hamiltonian of Eq.~\eqref{eq:H-TLS} are found by diagonalisation with a unitary operator $\hat U$ such that
\begin{equation}
  \label{eq:H-diagonalised}
\hat  H_A(t) = \begin{bmatrix} -\mathcal{E}(t) & 0 \\
                            0          &\mathcal{E}(t) 
           \end{bmatrix}
 = 
      \hat U^\dagger(t) \hat H(t)  \hat U(t)
 . 
\end{equation}
Here, the instantaneous eigenenergies are given by $\pm\mathcal{E}$ where we easily find that 
\begin{equation}
\mathcal{E}(t) = \sqrt{ \alpha^2(t) + \beta^2(t)  }.
\label{eq:eigenenergy}
\end{equation}
Then the adiabatic principle states that the system state `follows' the eigenstate of Eq.~\eqref{eq:H-TLS}, with instantaneous energies $\pm\mathcal{E}$, provided the changes in time of $\alpha(t)$ and $\beta(t)$ are slow enough \citep{MessiahEnglish}. But how slow is `slow enough'? We can get a more precise idea of `slowness' by looking at the actual Schr\"odinger dynamics in the adiabatic basis. That is, if the original Hamiltonian of Eq.~\eqref{eq:H-TLS} satisfies the Schr\"odinger equation
\begin{equation}
  \label{eq:TLS-SchrodingerEQ}
  i \hbar \frac{\partial}{\partial t} \Psi(t) = \hat H(t) \Psi(t) 
\end{equation}
where $\Psi(t)$ is a two-component spinor, then to obtain the equivalent, exact, equation in the adiabatic basis we substitute for $\hat H(t)$ from Eq.~\eqref{eq:H-diagonalised}, such that 
$   \hat H(t) =     \hat U(t) \hat H_A(t)  \hat U^\dagger(t) $,
and transform the states to the adiabatic basis through
\begin{equation}
  \label{eq:Adiabaticstates}
    \Psi(t) = \hat U(t)  \Psi_A(t).
\end{equation}
Then we obtain 
\begin{equation}
  \label{eq:Schrodinger_rot}
  i \hbar \frac{\partial}{\partial t} \Psi_A(t) = \hat H_A(t) \Psi_A(t)
  -i\hbar \hat U^\dagger \frac{\partial}{\partial t} \hat U  \Psi_A(t)
\end{equation}
so that we see that the actual, effective Hamiltonian in the adiabatic basis, is
\begin{equation*}
  \hat H_A(t)  
  -i\hbar\hat U^\dagger \frac{\partial}{\partial t} \hat U ,
\end{equation*}
which consists of the adiabatic Hamiltonian $\hat H_A(t)$, as intended, as well as an additional term, to be regarded (here) as a \emph{correction}. In the case of the two-component Hamiltonian of Eq.~\eqref{eq:H-TLS} we can find that the correction term has the form
\begin{equation}
  \label{eq:non-adia-correction}
   \begin{bmatrix} 0& \gamma(t)\\ \gamma^*(t)&0
   \end{bmatrix}
 \end{equation}
which plays the role of the \emph{non-adiabatic coupling} between the eigenstates.  The non-adiabaticity parameter $\gamma(t)$ is given by
\begin{equation}
  \label{eq:expression-gamma-non-adia}
   \gamma(t) = -i\frac{\hbar}{2}\, \frac{\dot{\alpha}(t)\beta(t) - \alpha(t)\dot{\beta}(t)}{\mathcal{E}^2(t)}\, .
\end{equation}
Thus, given that the exact Hamiltonian in the adiabatic basis has the eigenenergy $\mathcal{E}(t)$ on the diagonal and the non-adiabatic coupling $\gamma(t)$ on the off-diagonals, it is common to use as a simple measure of adiabaticity the requirement that \citep{MessiahEnglish}
\begin{equation}
  \label{eq:adia-criterion}
  |\gamma(t)| \ll \mathcal{E}(t).
\end{equation}
This is then a simple measure of how slow is `slow enough'.

\subsection{The Landau-Zener paradigm}
\label{sec:land-zener-paradigm}

At present we imagine that the time dependence of the Hamiltonian of Eq.~\eqref{eq:H-TLS} is due to an external controlling parameter such as a magnetic field strength. However, we shall shortly see that the time dependence can also arise from the dynamics of an atom in a potential (\secref{sec:an-adiabatic-trap}, below), i.e.\ the motion of an atom can itself generate the time dependence in the parameters. 
In essence, if the position of an atom is given by some dynamics as $x_0(t)$, and a potential for the atom is given by $U_0(x)$ we can make a semi-classical approximation by letting (see Eq.~\eqref{eq:H-TLS}) $\alpha(t) \longrightarrow U_0(x_0(t))$, which implies $\dot\alpha=\dot x_0\partial_x U_0$. As a result, we note that Eq.~\eqref{eq:adia-criterion} now appears as an adiabaticity condition depending on the atomic velocity $\dot x_0$. This is a half-way house to a full quantum treatment to be considered in \secref{sec:LZlosses}.

\begin{figure}[t]
\centering
\subfigure[Population in \ket{2}]{\includegraphics[width=0.75\linewidth]{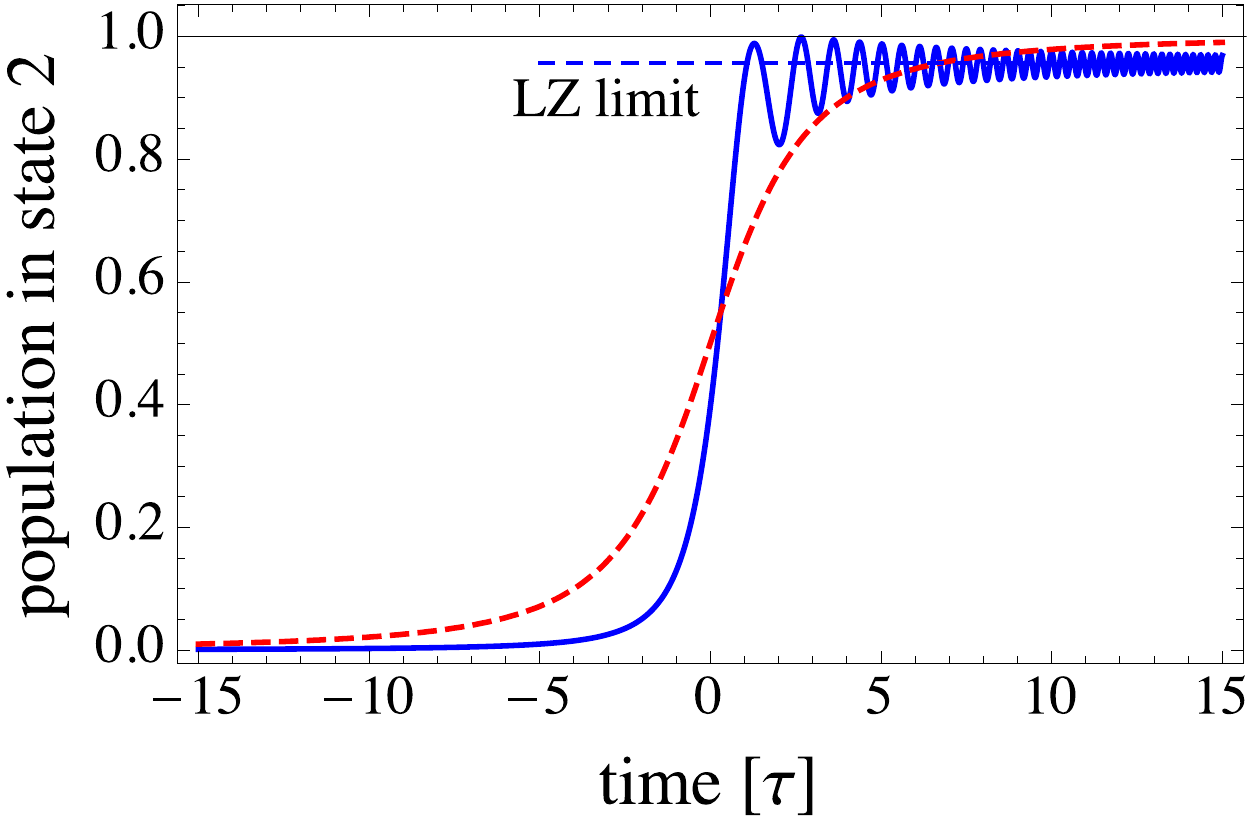}}\\[3mm]
\subfigure[Energy]{\includegraphics[width=0.75\linewidth]{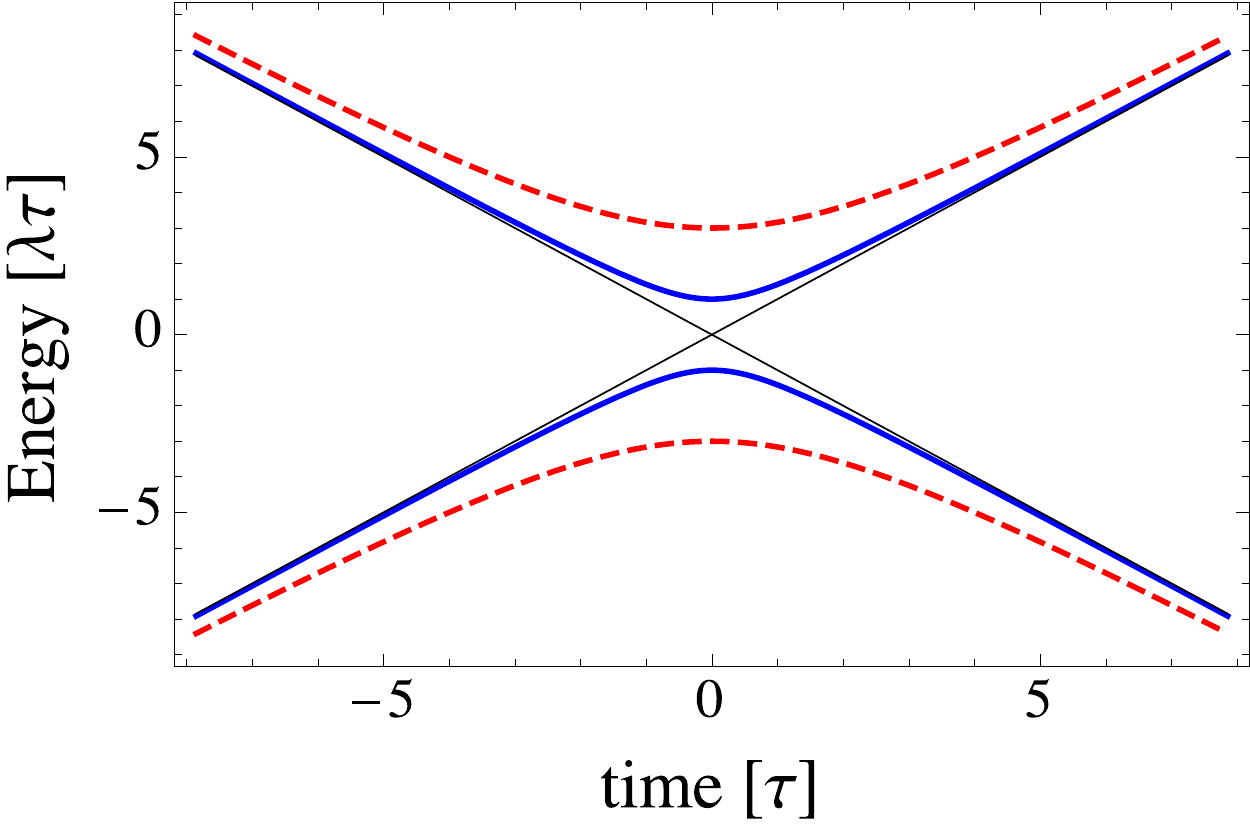}}
\caption{\label{fig:LZ}
Hamiltonian evolution during a linear sweep, starting from state \ket{1} at $t=-\infty$, for two different values of the coupling: $V=\lambda\tau$ corresponding to $\Lambda=1$ (blue full line) and $V=3\lambda\tau$ (red dashed line, $\Lambda=9$). The time unit is $\tau=\sqrt{\hbar/\lambda}$. Upper graph: Evolution of the population $P_2$ in state \ket{2}. The expected $t\to\infty$ Landau-Zener limit, $P_2=1-\exp(-\pi\Lambda)$, is indicated as a dashed blue line in the case $V=\lambda\tau$. Lower graph: energies of the bare states (thin black line) and of the adiabatic states $\pm\mathcal{E}(t)$.
}
\end{figure}

A simple example of a specific Hamiltonian in the form of Eq.~\eqref{eq:H-TLS} in the basis $\left\{\ket{1},\ket{2}\right\}$ of the bare states is the Landau-Zener model, where the Hamiltonian of Eq.~\eqref{eq:H-TLS} takes the form
\begin{equation}
  \label{eq:H-TLS-LZ}
  \hat H(t) = \begin{bmatrix}
        -\lambda t & V\\
         V  & \lambda t
       \end{bmatrix}
\end{equation}%
where the constant $\lambda$ describes the rate of change of potential with time and the coupling $V$ between states \ket{1} and \ket{2} is assumed to be a constant. 
This is an example of Eq.~\eqref{eq:H-TLS} with $\alpha(t)\rightarrow\lambda t$ and $\beta \rightarrow V$.
This model can be an approximation to the dynamics of a particle passing through a region of rf resonance at approximately constant speed. The adiabatic energies of Eq.~\eqref{eq:H-TLS-LZ} are $\mathcal{E}(t) = \sqrt{ (\lambda t)^2 + V^2 }$ as illustrated in \figref{fig:LZ}.
The related non-adiabaticity parameter $\gamma(t)$, Eq.~\eqref{eq:expression-gamma-non-adia}, is given by the expression
\begin{equation}
  \label{eq:LZ-gamme}
   \gamma(t) = -i\frac{\hbar}{2} \frac{\lambda V }{\left( \lambda^2 t^2 + V^2   \right)}.
\end{equation}
We saw in Eq.~\eqref{eq:adia-criterion} that the ratio $|\gamma(t)|/\mathcal{E}(t)$ should be small for an adiabatic process. In the case of the Landau-Zener model we find that $ |\gamma(t)|/\mathcal{E}(t) = \hbar \lambda V / [ 2 ( \lambda^2 t^2 + V^2 )^{3/2} ]$. We see that this ratio has its largest value at $t=0$, where $ |\gamma(0)|/\mathcal{E}(0) = \hbar \lambda  / ( 2 V^2 )$. This, in turn, implies that to be adiabatic we would like a strong coupling $V$ and/or a `low' rate of in the potential, $\lambda$ so that $\hbar \lambda  / ( 2 V^2 ) \ll 1 $. Conventionally, we define a parameter $\Lambda$ such that \citep{Suominen1991}
\begin{equation}
  \label{eq:BigLam:defn}
  \Lambda = \frac{V^2}{\hbar \lambda}
\end{equation}
and then we have $\Lambda \gg 1 $ for highly adiabatic behaviour.

The Landau-Zener model \citep{Landau1932,Zener1932} is a nice example of an exactly solvable and non-trivial time-dependent quantum mechanical problem. The time-dependent solution to Eq.~\eqref{eq:H-TLS-LZ} involves the parabolic cylinder, or Weber functions and is quite complex \citep{Zener1932,Vitanov1996}. 
A typical time-evolution is shown in \figref{fig:LZ}a for a medium value of adiabaticity parameter, $\Lambda=1$, where most, but not all, of the probability of the initial state is transferred to the other state of the system around $t=0$.
However, despite the complex time-evolution, the probability amplitudes in the long time limit (starting from $t\rightarrow-\infty$) have a very simple form which can be used as a model for practical applications. The  probability for the system to remain in its initial eigenstate is given by 
\begin{equation}
P = \exp(-\pi\Lambda)
\,.
\label{eq:LZ-Pstay}
\end{equation}
When the adiabaticity parameter $\Lambda$ is high, this probability becomes exponentially small, which happens because of adiabatic transfer out of the bare state at the point of the crossing of bare levels. 

The adiabatic transfer to the other state is understood by looking at the bare energies $\alpha(t)$ and adiabatic energies $\mathcal{E}(t)$ which are illustrated in \figref{fig:LZ}b. Although the bare state energies $\pm\alpha(t)$ cross over at $t=0$, the adiabatic energies do not, and a state following the adiabatic path ends up changing its bare state by avoiding the bare state `crossing' at $t=0$.

\subsection{Application to adiabatic traps}
\label{sec:an-adiabatic-trap}
This situation can be applied to magnetic spin problems in several ways. For example, an atomic or nuclear spin in a magnetic field can be subjected to a radio-frequency field far below resonance. If the frequency is `slowly' swept through the resonance to far above resonance the spin population is transferred from one spin state to the opposite state in a process known as adiabatic rapid passage (ARP). The situation is analogous to \figref{fig:LZ} with the change in bare state energy mapped to the frequency of the exciting field. This means that the measure of `slowly' is given by Eq.~\eqref{eq:adia-criterion}, which for a linear chirp requires a big value of $\Lambda$ as given by Eq.~\eqref{eq:BigLam:defn}.

Another example, which is highly relevant to this article, occurs if the radio-frequency field has constant frequency (and amplitude) in time, but the `static' magnetic field slowly varies in time so that a passage through resonance again takes place. Again, this can lead to the so-called adiabatic rapid passage, but a point of special interest is the case where the magnetic field is \emph{spatially} varying. This realises a potential, $U_0(x)$, see \secref{sec:magnetic-interaction}, down which the atom can accelerate, until it reaches a resonance region such as that found in \figref{fig:LZ}b. Indeed the time co-ordinate of the original Landau-Zener model can now be mapped to space through $t\longrightarrow x/v$. This means that an atom traveling from the left, following the upper adiabatic (red) curve of \figref{fig:LZ}b will find that provided its velocity $v$ is not too high in the critical Landau-Zener crossing region, it remains on the adiabatic potential after the crossing and starts decelerating on the upwards potential. Eventually, the kinetic energy of the atom is lost and it turns around and starts accelerating back to the left side. This cycle will repeat and thus the atom is trapped, adiabatically, by the magnetic resonance location, see \secref{sec:adiabatic_potential}. The condition for remaining trapped, over one cycle of oscillation, is that the probability given by Eq.~\eqref{eq:LZ-Pstay} should be very small. To evaluate this, the parameter $\lambda$ in Eq.~\eqref{eq:BigLam:defn} should be replaced by $v \frac{\p U_0}{\p x}$ where the gradient of potential $\frac{\p U_0}{\p x}$ is to be evaluated at the resonance location. This will be discussed more precisely in \secref{sec:LZlosses}.

\section[Introduction to magnetic resonance]{Introduction to magnetic resonance in classical and quantised descriptions}
\label{sec:magres}

\subsection{Angular momentum operators and rotation}
\label{sec:spin-operators-spin-rot}

In this section, for the sake of fixing the notations, we recall simple results on rotation of angular momentum operators, which we will use in this review, and which play a key role in finding and utilising the adiabatic states. The example we have in mind is an atom having a nuclear spin \GG{\hat I}, an orbital angular momentum \GG{\hat L} and an electronic spin \GG{\hat S}. The total angular momentum operator relevant to the interaction with a weak magnetic field is $\GG{\hat F} = \GG{\hat J}+\GG{\hat I} = \GG{\hat L}+\GG{\hat S}+\GG{\hat I}$. For example, for a rubidium 87 atom in its $5S_{1/2}$ ground state, we have $I=3/2$, $L=0$ and $S=1/2$, such that $J=1/2$ and $F \in \{|I-J|,...|I+J|\}$: $F=1$ or $F=2$, which are the two hyperfine states of the atomic ground state. For the purpose of this review, where spins will interact with static magnetic fields or radio-frequency fields, we consider a fixed value $F$ of the angular momentum quantum number. In the following, in order to lighten up the writing, we will extend the term `spin' to the total angular momentum \GG{\hat F}.

\subsubsection{Spin operators}
\label{sec:spin-operators-1}

A spin operator, which by convention here we will take as the total atomic angular momentum $\mathbf{\hat{F}}$, is a vector operator (dimension $\hbar$) associated to the quantum number $F$. $F\ge 0$ is an integer for bosonic particles, or a half integer for fermions. The projection of $\mathbf{\hat{F}}$ along any axis, represented by a unit vector $\mathbf{u}$, is denoted as $\hat{F}_\mathbf{u} = \mathbf{\hat{F}}\cdot\mathbf{u}$, and is an operator in the space of spin vectors.

Given a quantisation axis $\GG{e}_z$, we can find a basis where both $\mathbf{\hat{F}}^2$ and $\hat{F}_z$ are diagonal. The spin eigenstates are labelled \ket{F,m} where $m \in \{-F,-F+1,\dots,F-1,F\}$, with eigenvalues given by
\begin{eqnarray}
&&\mathbf{\hat{F}}^2\ket{F,m} = F(F+1) \hbar^2 \ket{F,m},\\
&&\hat{F}_z\ket{F,m} = m \hbar \ket{F,m}.\label{eq:Fzeigenstate}
\end{eqnarray}
We note that the quantum numbers have the labels $F$ and $m$ here. The label $m_F$ is often chosen instead of $m$, but here we choose the latter label in order to simplify some of our later expressions.

We also introduce the rising and lowering operators $\hat{F}_+$ and $\hat{F}_-$, defined as
\beq
\hat{F}_\pm = \hat{F}_x \pm i\hat{F}_y.
\eeq
It is clear from their definition that $[\hat{F}_\pm]^\dagger = \hat{F}_\mp$. Their commutation relations with $\hat{F}_z$ are:
\beq
[\hat{F}_z,\hat{F}_\pm] = \pm\hbar\hat{F}_\pm.
\eeq
From these relations, we can deduce their effect on \ket{F,m}, which is, in the case of $\hat F_+$, to increase $m$ by one unit, whereas in the case of $\hat F_-$ the quantum number $m$ is decreased by one unit. That is:
\beq
\hat{F}_\pm\ket{F,m} = \hbar \sqrt{F(F+1)-m(m\pm1)}\ket{F,m\pm1}.
\eeq

\subsubsection{Rotation operators}
\label{sec:spin-operators-2}

From now on, as we will concentrate on operators which do not change the value of $F$, we will further simplify the state notation and use \ket{m}, where the number $F$ is implicit, and we may also use $\ket{m}_z$ to emphasise that the quantisation axis is chosen along $z$. Conversely, an eigenstate of $\hat{F}_\GG{u}$ will be labeled $\ket{m}_\GG{u}$.

The operator which allows us to transform $\ket{m}_z$ into $\ket{m}_{z'}$ where $z'$ is a new quantisation axis is a rotation operator. The rotation around any unit vector \GG{u} by an angle $\alpha$ is described by the unitary operator
\beq
\hat{R}_\GG{u}(\alpha) = \exp\left[-\frac{i}{\hbar}\alpha\GG{\hat{F}}\cdot\GG{u}\right].
\eeq
This derives in fact from the fact that \GG{\hat{F}} is the generator of infinitesimal rotations in the Hilbert space, see for example \citep{CohenQuantiqueAnglais} or \citep{WalravenAtomicPhysics}. The inverse rotation, by an angle $-\alpha$, is described by its Hermitian conjugate: $\hat{R}_\GG{u}(-\alpha) = [\hat{R}_\GG{u}(\alpha)]^{-1} = [\hat{R}_\GG{u}(\alpha)]^\dagger$.
Starting from an eigenstate $\ket{m}_z$ of $\hat{F}_z$, the effect of $\hat{R} = \hat{R}_\GG{u}(\alpha)$ is to give the corresponding eigenstate $\ket{m}_{z'}=\hat{R}^\dagger\ket{m}_z$ of the rotated operator $\hat{F}_{z'}=\hat{R}^\dagger \hat{F}_z\hat{R}$:
\bea
&&\ket{m}_z = \hat{R}\ket{m}_{z'} \nonumber\\
&&\Rightarrow  \hat{F}_{z'}\ket{m}_{z'} = \hat{R}^\dagger \hat{F}_z \ket{m}_z = m\hbar \hat{R}^\dagger\ket{m}_z = m\hbar \ket{m}_{z'}.\nonumber
\eea

The rotation by the sum of two angles is simply the product of the two rotations around the same axis:
$$
\hat{R}_\GG{u}(\alpha+\beta) = \hat{R}_\GG{u}(\alpha)\hat{R}_\GG{u}(\beta).
$$
However, as the spin projections onto different axes do not commute, the composition of rotations around different axes do not commute. A useful formula is the decomposition of a rotation around any vector \GG{u} in terms of rotations around the basis axes $(x,y,z)$. If the spherical angles describing the direction of the unit vector \GG{u} are $(\theta, \phi)$ such that
\beq
\GG{u}=\sin\theta\left(\cos\phi\,\GG{e}_x+\sin\phi\,\GG{e}_y\right)+\cos\theta\,\GG{e}_z,
\eeq
we can write
\beq
\hat{R}_\GG{u}(\alpha) = \hat{R}_z(\phi)\hat{R}_y(\theta)\hat{R}_z(\alpha)\hat{R}_y(-\theta)\hat{R}_z(-\phi),
\label{eq:rot_decomp}
\eeq
where $\hat{R}_i$ stands for $\hat{R}_{\GG{e}_i}$. Starting from the right hand side, the two first rotations put \GG{u} on top of $z$, the central operator makes the rotation by $\alpha$ around $z$, and the two last operators bring back \GG{u} to its original position.

\subsubsection{Rotation of usual spin operators}
\label{sec:spin-operators-3}

In the following, we will need to transform hamiltonians $\hat{H}$ through rotations, where the Hamiltonian is a sum of spin operators, and calculate operators such as $\hat{R}^\dagger \hat{F}_\GG{u'}\hat{R}$, where $\hat{R} = \hat{R}_\GG{u}(\alpha)$ and the direction $\GG{u'}$ is in general different from $\GG{u}$. Here we give its effect in simple cases. Let us first consider rotations by $\alpha$ around the quantisation axis $z$, such that $\hat{R} = \hat{R}_z(\alpha)$.
\begin{eqnarray}
\left[ \hat{R}_z(\alpha) \right]^\dagger \hat{F}_z\hat{R}_z(\alpha) &=& \hat{F}_z,\label{eq:rotateFz}\\
\left[ \hat{R}_z(\alpha) \right]^\dagger \hat{F}_\pm\hat{R}_z(\alpha) &=& e^{\pm i\alpha}\hat{F}_\pm.\label{eq:rotateFpm}\\
\left[ \hat{R}_z(\alpha) \right]^\dagger \hat{F}_x\hat{R}_z(\alpha) &=& \cos\alpha\,\hat{F}_x - \sin\alpha\,\hat{F}_y,
\end{eqnarray}

The general expression for the rotated operator $\hat{R}_\GG{u}^\dagger \hat{F}_\GG{u'}\hat{R}_\GG{u}$ can be deduced from these equations.

\begin{figure}[t]
\centering\includegraphics[width=0.7\linewidth]{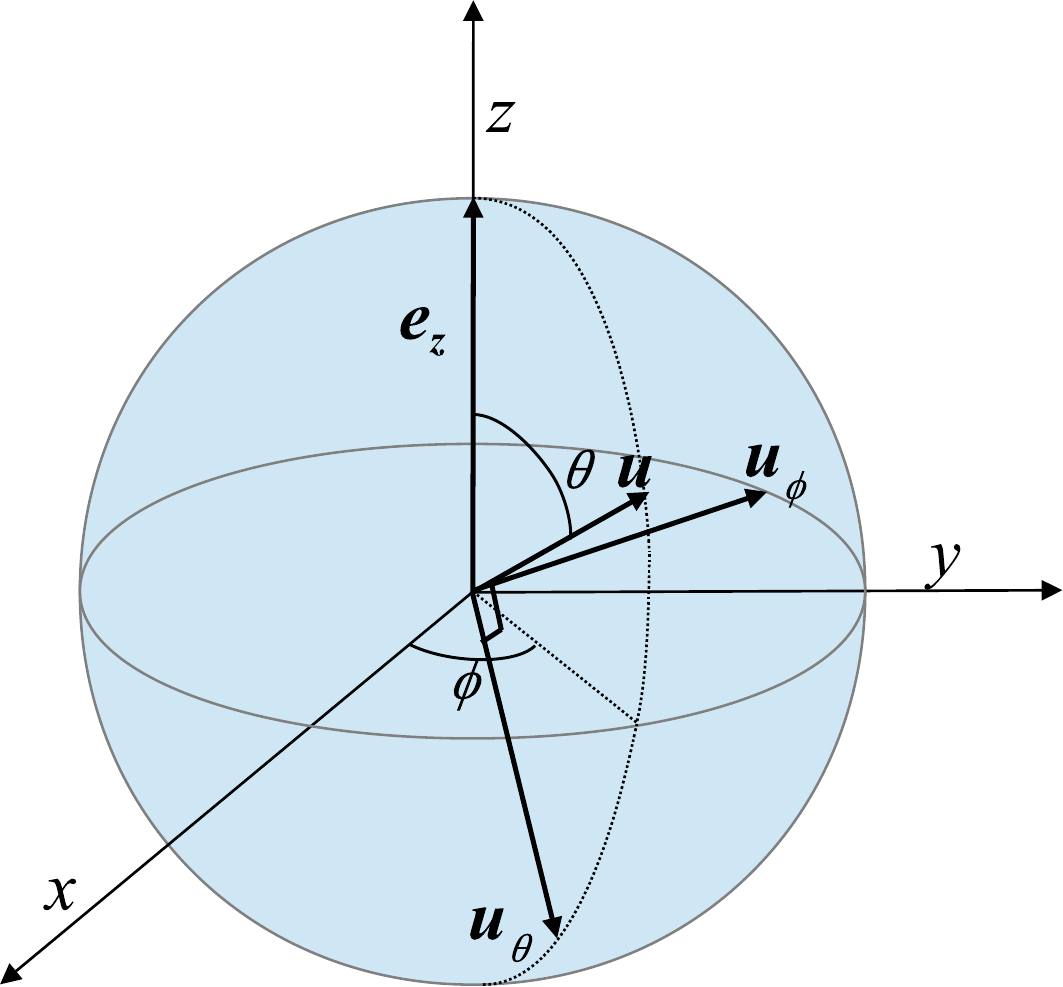}
\caption{Orientation of the basis $(\GG{u},\GG{u}_\theta,\GG{u}_\phi)$ relative to the basis $(\GG{e}_z,\GG{e}_x,\GG{e}_y)$.}
\label{fig:sphere}
\end{figure}

\subsubsection{Time-dependent rotations}
\label{sec:time-depend-rotatations}

For the description of an atom moving in an inhomogeneous magnetic field, we will need to deal with time-dependent rotation angles, and with the derivatives of the $\hat R$ operators. After using Eq.~\eqref{eq:rot_decomp} and some straightforward arithmetic, we come to the following expression involving $\partial_t \hat{R}_\GG{u}(\alpha)$:
\bea
i\hbar\hat{R}^\dagger_\GG{u}(\alpha)\partial_t \hat{R}_\GG{u}(\alpha) \,&&= \dot{\alpha}\hat{F}_\GG{u}\nonumber\\
&& +(1-\cos\alpha)\left[- \dot\theta\hat{F}_\GG{u_\phi} + \dot\phi\sin\theta\hat{F}_\GG{u_\theta}\right] \nonumber\\
&& + \sin\alpha\left[\dot\phi\sin\theta\hat{F}_\GG{u_\phi} + \dot\theta \hat{F}_\GG{u_\theta}\right],\label{eq:diff_rot_op}
\eea
where $(\GG{u},\GG{u_\theta},\GG{u_\phi})$ form an orthonormal basis, see \figref{fig:sphere} with
\beq
\GG{u_\theta}=\cos\theta\left(\cos\phi\,\GG{e}_x+\sin\phi\,\GG{e}_y\right)-\sin\theta\,\GG{e}_z
\eeq
and
\beq
\GG{u_\phi}=-\sin\phi\,\GG{e}_x+\cos\phi\,\GG{e}_y.
\eeq

The time variation of $\hat{R}$ can be recast using the rising and lowering operators $\hat{F}_{\GG{u},\pm} = \hat{F}_\GG{u_\theta} \pm i \hat{F}_\GG{u_\phi}$ with respect to the eigenstates of $\hat{F}_\GG{u}$, under the form
\bea
&&i\hbar\hat{R}^\dagger_\GG{u}(\alpha)\partial_t \hat{R}_\GG{u}(\alpha) = \nonumber\\
&&\dot{\alpha}\hat{F}_\GG{u} +\sin(\alpha/2)\left[(\dot\theta - i\dot\phi\sin\theta)\,e^{i\frac{\alpha}{2}}\hat{F}_{\GG{u},+} + h.c.\right].
\label{eq:diff_rot_op_pm}
\eea

This last expression makes clear that, if the direction \GG{u} around which the rotation is performed varies with time, the time derivative of the rotation operator now involves also spin projections along directions orthogonal to \GG{u}
\beq
(\dot\theta - i\dot\phi\sin\theta)\sin\frac{\alpha}{2}\,e^{i\frac{\alpha}{2}}\hat{F}_{\GG{u},+},
\label{eq:outcoupling}
\eeq
which couple different eigenstates of $\hat{F}_\GG{u}$ with an amplitude
\beq
\sqrt{\dot\theta^2+\dot\phi^2\sin^2\theta}\,\sin\frac{\alpha}{2}.
\label{eq:outcoupling_amplitude}
\eeq

 \subsection{Magnetic interaction}
\label{sec:magnetic-interaction}

Consider an atom with a total spin $F$ in its ground state. The Land\'e factor is labelled as $g_F$. In the presence of a static and homogeneous magnetic field $\mathbf{B}_0 = B_0\,\mathbf{e}_z$ which we use to set the quantisation axis $z$, i.e. $B_0$ is positive, the magnetic dipolar interaction between the atomic spin and the field reads
\begin{equation}
\hat{H}_0 = \frac{g_F \mu_B}{\hbar} \mathbf{B}_0 \cdot \hat{\mathbf{F}} = \frac{g_F \mu_B}{\hbar} B_0 \hat{F}_z.
\label{eq:hB0}
\end{equation}
Here, $\mu_B$ is the Bohr magneton. From Eq.~\eqref{eq:Fzeigenstate}, we get the first order Zeeman energy of each magnetic sublevel $\ket{F,m}_z$:
\beq
E_{m} = m g_F \mu_B B_0.
\eeq

We see that because $m$ can have different signs the Zeeman energies $E_{m}$ can also have different signs. This means that when we later introduce a field gradient, we find that some $m$ state, with a positive energy, lead to forces which tend to return the atom to regions of weaker field: these states are called low field seeking states as a result.
The low field seeking states, used for magnetic trapping in an inhomogeneous magnetic field, are the states such that $m>0$ for a positive $g_F$, or $m<0$ if $g_F<0$. We introduce
\beq
\signgf=g_F/|g_F|=\pm 1
\label{eq:signgf}
\eeq
as being the sign of $g_F$. The low field seeking states are those with $\signgf m > 0$, with a positive energy $E_{m} = |m| \hbar\omega_0$, where the frequency spacing between Zeeman levels, the Larmor frequency, is defined as 
\beq
\omega_0 = |g_F| \mu_B B_0/\hbar.
\label{eq:Larmor}
\eeq

 \subsection{Theory of magnetic resonance and coupling to a classical rf field}
\label{sec:theory-magnetic-resonance}

Let us now consider the coupling of these Zeeman states with a near resonant rf field, i.e. one of frequency $\omega$ close to $\omega_0$. In a first approach, we will concentrate on a linearly polarised rf field. Other polarisation configurations will be discussed in \secref{sec:polarization}. While the electric field coupling is negligible at rf frequencies, the magnetic field couples to the atomic spin through the magnetic dipolar interaction.

The rf field is usually produced by a coil or a wire in the vicinity of the atomic sample. Its amplitude is usually larger than a mG, such that the atomic spin interacts with many rf photons. For this reason, the rf field can be described by a classical coherent field. We will first present in \secref{sec:classical_field} a semi-classical analysis where a classical field couples to a quantum spin. A full quantum treatment for the field and the spin will be derived in \secref{sec:quantum_field}. We will see in \secref{sec:quantum_spectro} that a quantum description for the rf field provides a simple interpretation for the possible transitions between dressed states.
In this section, we will assume that both the static magnetic field and the rf field are spatially homogeneous. In this case, the atom-field coupling term does not depend on the external variables of the atom, whose evolution is thus decoupled from the spin variables. The case of spatially dependent fields is essential for the realisation of adiabatic potentials, though. It will be treated in \secref{sec:adiabatic_potential}.

\subsubsection{A spin coupled to a classical field}
\label{sec:classical_field}
In this section, we consider a linearly polarised, homogeneous, classical rf field, along some direction $\epsgras$, where $\epsgras$ is a real unit vector. The magnetic component of the classical rf field is $\mathbf{B}_1(t)=B_1\cos (\omega t+\phi)\,\epsgras$, where $\phi$ is the field phase at initial time. It couples to the atomic spin through the Hamiltonian
\beq
\hat{V}_1 = \frac{g_F \mu_B}{\hbar} \mathbf{B}_1 \cdot \hat{\mathbf{F}}\cos (\omega t+\phi)
\eeq
as in Eq.~\eqref{eq:hB0}.
The total Hamiltonian describing the atomic state is then
\bea
\hat{H} &=& \hat{H}_0 + \hat{V}_1 \\
&=& \frac{g_F \mu_B}{\hbar} \mathbf{B}_0 \cdot \hat{\mathbf{F}}+ \frac{g_F \mu_B}{\hbar} \mathbf{B}_1 \cdot \hat{\mathbf{F}}\cos (\omega t+\phi).\nonumber
\eea

We now use the direction of the static field $\mathbf{B}_0 = B_0\,\mathbf{e}_z$ as quantisation axis $z$ ($B_0>0$). The presence of a component of the rf field along $z$ only slightly modifies the result and will be discussed in \secref{sec:misalignment}. In this section, we will consider instead the special, and effective case where $\mathbf{B}_1$ is polarised linearly and in a direction orthogonal to $z$ (known as $\sigma$ polarization), such that we can set the axis $x$ parallel to the rf field: $\mathbf{B}_1(t) = B_1 \cos(\omega t+\phi) \, \mathbf{e}_x$ ($B_1>0$). In the end this turns out to be an optimum choice for a linearly polarised rf field, see \secref{sec:misalignment}. The time-dependent Hamiltonian now reads:
\beq
\hat{H} = \signgf\left[\omega_0 \hat{F}_z+ 2 \Omega_1 \cos (\omega t+\phi) \hat{F}_x\right].
\eeq
We have introduced the Rabi frequency
\beq
\Omega_1 = \frac{|g_F| \mu_B B_1}{2 \hbar}.
\label{eq:Rabi_linear}
\eeq
Using the $\hat{F}_\pm$ operators, the Hamiltonian can be written as
\bea
\hat{H} = \signgf\omega_0\hat{F}_z &+& \signgf \frac{\Omega_1}{2}\left[ e^{-i(\omega t+\phi)} \, \hat{F}_+ +e^{i(\omega t+\phi)} \, \hat{F}_- \right.\nonumber\\
&+&\left.e^{-i(\omega t+\phi)} \, \hat{F}_- + e^{i(\omega t+\phi)}\,\hat{F}_+\right]\nonumber
\eea
where we recall that $\signgf=\pm1$, see Eq.~\eqref{eq:signgf}.

The evolution of the atomic spin state $\ket{\psi}$ through this Hamiltonian is given by the Schr\"odinger equation $i\hbar \partial_t \ket{\psi} = \hat{H}\ket{\psi}$. The first term of the Hamiltonian is responsible for a spin precession around the $z$ axis at frequency $\omega_0$, a in direction determined by $\signgf$. The other terms couple different $\ket{m}_z$ states and induce transitions. These transitions will be resonant for $\omega = \omega_0$. To emphasise this point, it is useful to write the Hamiltonian in the basis rotating at the frequency $\signgf\omega$ around $z$. We introduce the rotated state
\beq
\ket{\psi\ind{rot}} =\hat{R}^\dagger\ket{\psi} \quad \mbox{or} \quad \ket{\psi} = \hat{R}\ket{\psi\ind{rot}},
\label{eq:psirot}
\eeq
where $\hat{R}=\hat{R}_z\left[\signgf(\omega t+\phi)\right]$.
This rotated state now obeys a Schr\"odinger equation $i\hbar \partial_t \ket{\psi\ind{rot}} = \hat{H}\ind{rot}\ket{\psi\ind{rot}}$, where
\beq
\hat{H}\ind{rot} = - i\hbar\hat{R}^\dagger\left[\partial_t \hat{R}\right] + \hat{R}^\dagger\hat{H}\hat{R}.
\label{eq:hrot}
\eeq
Using Eqs.~\eqref{eq:rotateFz}, \eqref{eq:rotateFpm} and \eqref{eq:diff_rot_op}, we obtain
\bea
\hat{H}\ind{rot} &=&  - \signgf\delta\hat{F}_z + \signgf\frac{\Omega_1}{2}\left[e^{i(\signgf-1)\omega t} \, \hat{F}_+ +e^{-i(\signgf-1)\omega t} \, \hat{F}_- \right]\nonumber\\
&+&\signgf\frac{\Omega_1}{2}\left[ e^{i(\signgf+1)\omega t}\,\hat{F}_+ + e^{-i(\signgf+1)\omega t} \, \hat{F}_-  \right].\label{eq:Hrot}
\eea
where we have introduced the detuning $\delta = \omega-\omega_0$.

Depending on the sign of $\signgf$, either the two first terms (for $\signgf=1$) or the two last terms (for $\signgf=-1$) in the brackets become static in this rotating frame. On the other hand, the two other terms evolve at high frequency $\pm2\omega$.

We are interested in the limit of near resonant rf coupling, so that the detuning $\delta$ is very small as compared to the rf frequency $\omega$. If we also assume that the rf coupling $\Omega_1$ is much smaller than $\omega$, we can apply the rotating wave approximation (RWA). In this approximation, the time dependent terms of Eq.~\eqref{eq:Hrot}, which evolve at a frequency much larger than the other frequencies in the problem, are effectively time-averaged to zero. 
The choice of which terms are dropped from Eq.~\eqref{eq:Hrot} depends on the sign $\signgf$.
We are then left with a time-independent Hamiltonian
\beq
\hat{H}\ind{eff} = - \signgf\delta\hat{F}_z + \signgf\frac{\Omega_1}{2}\left(\hat F_+ + \hat F_- \right) = \signgf\left(-\delta\hat{F}_z + \Omega_1\hat F_x\right),
\label{eq:Heff}
\eeq
which can be written in terms of a very simple spin Hamiltonian, in the spirit of Eq.~\eqref{eq:hB0}:
\beq
\hat{H}\ind{eff} = \Omega\hat{F}_\theta.
\eeq
Here, we have defined the frequency splitting through the \emph{generalised Rabi frequency}
\beq
\Omega = \sqrt{\delta^2 + \Omega_1^2},
\label{eq:gen_Rabi_freq}
\eeq
and the spin projection $\hat{F}_\theta = \hat{\mathbf{F}}\cdot\mathbf{e}_\theta$, where $\mathbf{e}_\theta = \cos\theta \, \mathbf{e}_z + \sin\theta\,\mathbf{e}_x$, is
\beq
\hat{F}_\theta = \cos\theta\,\hat{F}_z + \sin\theta\,\hat{F}_x,
\eeq
with
\beq
\theta = \arccos \left( -\frac{\delta}{\Omega} \right) + \frac{\signgf - 1}{2}\pi.
\label{eq:theta_dress}
\eeq
$\hat{H}\ind{eff}$ is the Hamiltonian of a spin interacting with a static, effective, magnetic field, pointing in the direction $\signgf\mathbf{e}_\theta$
\beq
\mathbf{B}\ind{eff} = \frac{\hbar\Omega}{g_F\mu_B}\mathbf{e}_\theta.
\eeq
$\hat F_\theta$ is linked to $\hat F_z$ by the rotation operator of angle $\theta$ around $y$:
$$
\left[ \hat{R}_y(\theta) \right]^\dagger \hat{F}_\theta\hat{R}_y(\theta) = \hat F_z, \quad \mbox{or} \quad \hat F_\theta = \hat R_y(\theta)F_z\hat R_y(-\theta).
$$
The eigenstates of $\hat{H}\ind{eff}$ are spin states $\ket{m}_\theta$ in the new basis of quantisation axis set by the angle $\theta$. They can be deduced from the initial states in the basis with quantisation axis $z$ through the same rotation:
\beq
\ket{m}_\theta = \hat R_y(\theta)\ket{m}_z = e^{-i\theta\hat{F}_y/\hbar}\ket{m}_z.
\label{eq:rotated_state}
\eeq
The two states $\ket{m}_\theta$ and $\ket{m}_z$ are mapped and have the same spin projection $m\hbar$ in their respective basis:
\beq
\hat{F}_\theta\ket{m}_\theta = m\hbar\ket{m}_\theta \quad \mbox{and} \quad \hat{F}_z \ket{m}_z = m\hbar\ket{m}_z.
\eeq
Their eigenenergies thus read:
\beq
E_{m}' = m \hbar\Omega.
\eeq
With our convention of the orientation of the new quantisation axis $\mathbf{e}_\theta$ with respect to the effective magnetic field $\mathbf{B}\ind{eff}$, the low field seeking states, with the highest energies, are always those with $m>0$, whatever the sign of $g_F$. As a result, for detunings $\delta$ very large as compared to $\Omega_1$ and negative, the extreme low field seeking state $\ket{m = \signgf F}_z$ is connected to (i.e. has the largest projection onto) the low field seeking state $\ket{m = F}_\theta$ in the new basis, while for $\delta\gg\Omega_1$ and $\delta>0$, $\ket{m = F}_\theta$ is connected to $\ket{m = -\signgf F}_z$.

The system can be flipped from $\ket{sF}_z$ to $\ket{-sF}_z$ with a linear sweep of the rf frequency by following adiabatically the state $\ket{F}_\theta$, in a way similar to the two-level system presented in the introduction, see \figref{fig:LZ}. The adiabaticity criterion given at Eq.~\eqref{eq:adia-criterion} can be recast under a very simple formula:
\beq
|\dot\theta|\ll\Omega.
\label{eq:adia_thetadot}
\eeq
This condition expresses that changes in the orientation $\theta$ of the adiabatic state must be slower than the effective Rabi frequency $\Omega$. This is analogue to the case of a spin in a static magnetic field, whose direction must change slower than the Larmor frequency for the spin to follow adiabatically.

\subsubsection{Generalisation of magnetic resonance theory to any rf polarization}
\label{sec:polarization}
We know discuss the case of an rf polarization which is no longer linear. The rf field can be written very generally as
\bea
\GG{B}_1 &=& B_x\cos(\omega t+\phi_x)\,\GG{e}_x + B_y\cos(\omega t+\phi_y)\,\GG{e}_y \nonumber\\
&+& B_z\cos(\omega t+\phi_z)\,\GG{e}_z
\eea
$z$ being the direction of the static magnetic field. In principle, the amplitudes $B_i$ (with $i=x,y,z$) and phases $\phi_i$ could depend on position. To start with, we consider a homogeneous rf field.

We now use a complex notation $\BB_1$ for the field amplitude, with $\GG{B}_1 = \BB_1\,e^{-i\omega t} + c.c.$ and
\beq
\BB_1 = \frac{B_x}{2} e^{-i\phi_x}\,\GG{e}_x + \frac{B_y}{2}e^{-i\phi_y}\,\GG{e}_y + \frac{B_z}{2}e^{-i\phi_z}\,\GG{e}_z.
\label{eq:rf_field_in_complex_notation}
\eeq

The coupling between the $\ket{m}_z$ states due to the $z$ component $B_z$ of the rf field, aligned along the quantisation axis, is extremely small if the corresponding Rabi frequency $\Omega_z$ verifies $|\Omega_z|\ll\omega$, which is the case if the RWA holds. Its effect on the resulting adiabatic energy will be discussed specifically in \secref{sec:misalignment} and be ignored elsewhere.

We introduce the spherical basis $(\GG{e}_+,\GG{e}_-,\GG{e}_z)$ where
\beq
\GG{e}_+ = -\frac{1}{\sqrt{2}}\left(\GG{e}_x+i\GG{e}_y\right), \quad\quad \GG{e}_- = \frac{1}{\sqrt{2}}\left(\GG{e}_x-i\GG{e}_y\right).
\eeq
Using this basis, we can write the rf field restricted to the $xy$ plane as
\beq
\BB_1 = B_+ \GG{e}_+ + B_- \GG{e}_-.
\label{eq:rf_field_along_sigma}
\eeq
The complex projections $B_+$ and $B_-$ are given by the scalar products
$\GG{e}^*_\pm\cdot \BB_1$:
\bea
B_+ &=& \frac{1}{2\sqrt{2}}\left(-B_x\,e^{-i\phi_x} + i B_ye^{-i\phi_y}\right) \nonumber\\
B_- &=& \frac{1}{2\sqrt{2}}\left(B_x\,e^{-i\phi_x} + i B_ye^{-i\phi_y}\right) \nonumber
\eea
and we note that $B_+ \ne B_-^\ast$.
We see that $\GG{\hat{F}}\cdot\GG{e}_+ = -\frac{1}{\sqrt{2}}\hat{F}_+$ and $\GG{\hat{F}}\cdot\GG{e}_- = \frac{1}{\sqrt{2}}\hat{F}_-$. If we define the complex coupling amplitudes as
\beq
\Omega_\pm =\mp \sqrt{2}\frac{|g_F|\mu_B}{\hbar} B_\pm,
\label{eq:complex_ampli_sigma}
\eeq
the total spin hamiltonian reads:
\bea
\hat{H} &=& \signgf\omega_0\hat{F}_z + \signgf\left[ \frac{\Omega_+}{2}\,e^{-i\omega t} \, \hat{F}_+ + \frac{\Omega_+^*}{2}\,e^{i\omega t} \, \hat{F}_-\right. \nonumber\\
&+&\left.\frac{\Omega_-}{2}\,e^{-i\omega t} \, \hat{F}_- +\frac{\Omega_-^*}{2}\,e^{i\omega t} \, \hat{F}_+\right].
\label{eq:hamil_rf_pola}
\eea

In order to emphasise the resonant terms and apply RWA, we will write the rotated Hamiltonian, in the frame rotating at frequency $\signgf\omega$. Depending on the sign of $\signgf$, either the two first terms (for $\signgf>0$) or the two last terms (for $\signgf<0$) of Eq.~\eqref{eq:hamil_rf_pola} will be static, as in the previous case of linear polarisation, see Eq.~\eqref{eq:Hrot}. We thus define the effective Rabi coupling $\Omega_1 = \left|\Omega_1\right|e^{-i\phi}$ as
\beq
\Omega_1 = \Omega_{\signgf} = \left\{
\begin{array}{l}
\Omega_+\quad\mbox{for }\signgf>0,\\
\Omega_-\quad\mbox{for }\signgf<0.\\
\end{array}
\right.
\label{eq:effective_coupling}
\eeq
Note that in any case $\Omega_1 = -\sqrt{2}g_F\mu_B B_{\signgf}/\hbar$. After a rotation around $z$ of angle $\signgf(\omega t+\phi)$, and application of the rotating wave approximation, only the term in $\Omega_{\signgf}$ remains and the effective hamiltonian is
\bea
\hat{H}\ind{eff} &=&  - \signgf\delta\hat{F}_z +\signgf\frac{|\Omega_1|}{2}\left(\hat{F}_+ + \hat{F}_-\right)\nonumber\\
&=& \signgf\left(-\delta\hat{F}_z +|\Omega_1|\hat{F}_x\right).
\label{eq:heff_sigmaplus}
\eea
We recover Eq.~\eqref{eq:Heff}, and thus the eigenenergies $m\hbar\sqrt{\delta^2+|\Omega_1|^2}$ and the eigenstates, deduced from $\ket{m}_z$ by a rotation of $\theta$ around $y$, where $\theta$ is given by Eq.~\eqref{eq:theta_dress}.

It is clear from Eq.~\eqref{eq:effective_coupling} that the relevant coupling is only the $\sigma^+$ polarised part of the rf field for $\signgf = 1$ (or the $\sigma^-$ component for $\signgf=-1$). It is related to the $x$ and $y$ projections of the rf field through
\beq
|\Omega_1| = \frac{|g_F|\mu_B}{2\hbar}\sqrt{B_x^2+B_y^2+2\signgf B_x B_y\sin(\phi_x-\phi_y)}.
\eeq

For a linearly polarised field in a plane perpendicular to $z$, for example with $B_y=0$, we recover the amplitude of Eq.~\eqref{eq:Rabi_linear}. The coupling is maximum for purely circular polarization $\sigma^{\signgf}$ with respect to $z$, which we obtain when $\phi_x-\phi_y=\signgf\pi/2$ and $B_x=B_y$. The amplitude is then twice as large as in the case of the linear field along $x$. For a linear transverse polarization with $\phi_x=\phi_y$ and $B_x=B_y=B_1$, the coupling is smaller by $\sqrt{2}$ than for the circular case. Finally, the coupling totally vanishes in the case of a $\sigma^{-\signgf}$ polarization ($\sigma^-$ for $\signgf=1$, and vice versa).

We must emphasise that all this reasoning has been done with the direction of the static magnetic field chosen as the quantisation axis. Should the direction of this field change in space, the relevant amplitude would be the $\sigma^{\signgf}$ component along the new, local direction of the magnetic field. This will be illustrated in \secref{sec:examples_of_potentials}.

\subsection{Dressed atom approach applied to magnetic resonance with an rf field}
\label{sec:quantum_field}
Although the rf field is classical in the sense that the mean photon number $\langle N\rangle$  interacting with the atoms is very large, and its relative fluctuations $\Delta N/\langle N\rangle$ negligible, it gives a deeper insight in the coupling to use a quantised description for the rf field \citep{Cohen1977}. This will make much clearer the interpretation of rf spectroscopy, see \secref{sec:quantum_spectro}, or the effect of strong rf coupling, beyond RWA, see \secref{sec:beyondRWA}. Instead of writing the rf field as a classical field, it can be written as a quantum field with creation and annihilation operators $\hat{a}$ and $\hat{a}^\dagger$ of a photon in the rf mode of frequency $\omega$. The effect of $\hat{a}$ and $\hat{a}^\dagger$ on a field state \ket{N+n} where $n<\Delta N$ is then approximately
\bea
\hat a \ket{N+n} &\simeq& \sqrt{\langle N\rangle} \ket{N+n-1} \nonumber\\
\hat a^\dagger \ket{N+n} &\simeq& \sqrt{\langle N\rangle} \ket{N+n+1}.
\label{eq:largeN}
\eea

We start from the expression Eq.~\eqref{eq:rf_field_along_sigma} of the classical field in the spherical basis.
The quantum rf magnetic field operator can be described as follows:
\beq
\GG{\hat{B}}_1 = \left(b_+\, \GG{e}_+ + b_-\, \GG{e}_-\right)\,\hat{a} + h.c.
\eeq
where $b_\pm = B_\pm/\sqrt{\langle N\rangle}$.  The field Hamiltonian is then
\beq
\hat{H}\ind{rf} = \hbar\omega(\hat{a}^\dagger\hat{a} + \frac{1}{2}).
\eeq
Keeping the definition of Eq.~\eqref{eq:complex_ampli_sigma} for the Rabi coupling, the one-photon Rabi coupling is $\Omega^{(0)}_\pm =\mp \sqrt{2} |g_F|\mu_B b_\pm/\hbar = \Omega_\pm/\sqrt{\langle N\rangle}$. The operator which couples the atom and the field now reads
\bea
\hat V_1 &&= \signgf \left[ \frac{\Omega_+^{(0)}}{2} \hat{a} \, \hat{F}_+ + \frac{\Omega_+^{(0)*}}{2} \hat{a}^\dagger \, \hat{F}_-\right.\nonumber\\
&& + \left. \frac{\Omega_-^{(0)}}{2} \hat{a} \, \hat{F}_- + \frac{\Omega_-^{(0)*}}{2} \hat{a}^\dagger \, \hat{F}_+\right].
\eea
The total Hamiltonian $\hat H_0 + \hat{H}\ind{rf} + \hat V_1$ for the spin and the field in interaction thus reads:
\bea
\hat{H} &=& \signgf\omega_0\hat{F}_z + \hbar\omega \hat{a}^\dagger \hat{a} \nonumber\\
&+& \signgf\frac{1}{2\sqrt{\langle N\rangle}} \left[ \Omega_+ \hat{a} \, \hat{F}_+ + \Omega_+^* \hat{a}^\dagger \, \hat{F}_- \right.\nonumber\\
&+&\left.\Omega_- \hat{a} \, \hat{F}_- + \Omega_-^* \hat{a}^\dagger \, \hat{F}_+\right],
\eea
where we have set the origin of energy so as to include the zero photon energy $\hbar\omega/2$.

\subsubsection{Uncoupled states}
\label{sec:uncoupled-states}

In the following, we will chose $\signgf=+1$ for simplicity, the other choice simply changing the role of the two polarisations. We also assume that $\Omega_+$ is real, without lack of generality. In the absence of coupling (for $\Omega_\pm=0$),  the eigenstates of the $\{$ atom + photons $\}$ system $\hat{H}_0 = \omega_0\hat{F}_z + \hbar\omega \hat{a}^\dagger \hat{a}$ are $\ket{m,N}_z=\ket{m}_z\ket{N}$, where $\ket{m}_z$ is an eigenstate of $\hat{F}_z$ and \ket{N} an eigenstate of $\hat{a}^\dagger \hat{a}$, with respective eigenvalues $m\hbar$ and $N$:
$$
\hat{H}_0\ket{m,N}_z = E^0_{m,N}\ket{m,N}_z, \quad\quad E^0_{m,N} = m\hbar\omega_0 + N\hbar\omega.
$$
Let us write this energy in terms of the detuning $\delta = \omega-\omega_0$:
$$
E^0_{m,N} = -m\hbar\delta + (N+m)\hbar\omega.
$$
From this expression, we see that for each fixed $N$ the states in the manifold $\mathcal{E}_N = \left\{\ket{m,N-m},m=-F\dots F\right\}$ have an energy
$$
E^0_{m,N-m} = -m\hbar\delta + N\hbar\omega.
$$

For a rf frequency $\omega$ close to $\omega_0$, that is if $|\delta|\ll\omega$, the energy splitting inside a manifold, of order $\hbar\delta$, is very small as compared to the energy splitting between neighbouring manifolds, which is $\hbar\omega$.

\subsubsection{Effect of the rf coupling}
\label{sec:effect-rf-coupling}

The Hamiltonian contains four coupling terms. The two first terms, proportional to $\Omega_+$, act inside a given ${\mathcal{E}}_N$ manifold, between two states split by a small frequency $\pm \delta$:
\bea
&&\langle m\pm 1, N-m\mp 1 \left|\frac{\Omega_+}{2\sqrt{\langle N\rangle}}\left(\hat{a} \, \hat{F}_+ + \hat{a}^\dagger \, \hat{F}_-\right)\right| m,N-m\rangle \nonumber\\
&&\simeq \frac{\Omega_+}{2}\sqrt{F(F+1)-m(m\pm 1)}=\Omega_+ ~_z\bra{m\pm1}F_x\ket{m}_z.\nonumber
\eea
Here we have assumed, following Eq.~\eqref{eq:largeN}, $N-m\simeq \langle N\rangle$ when applying the operators $\hat{a}$ and $\hat{a}^\dagger$, which is valid because $\langle N\rangle\gg1$.

The two last coupling terms, proportional to $\Omega_-$, couple states of the ${\mathcal E}_N$ manifold to states of the $\mathcal{E}_{N\pm 2}$ manifold, split by a much larger frequency $\omega_0+\omega$:
\bea
&&\langle m\pm 1, N-m\pm 1 \left|\frac{\Omega_-}{2\sqrt{\langle N\rangle}}\left(\hat{a}^\dagger \, \hat{F}_+ + \hat{a} \, \hat{F}_-\right)\right| m,N-m\rangle \nonumber\\
&&\simeq \Omega_- ~_z\bra{m\pm1}F_x\ket{m}_z.\nonumber
\eea
An estimation of the effect of these two terms on the energy with perturbation theory, valid in the limit $|\delta|\gg |\Omega_+|$, will lead to a shift of order $\hbar|\Omega_+|^2/\delta$ for the $\Omega_+$ terms, and $\hbar|\Omega_-|^2/(\omega_0+\omega)$ for the $\Omega_-$ term. The rotating wave approximation, which applies if $|\delta|,|\Omega_\pm|\ll \omega$, consists in neglecting the effect of the $\Omega_-$ terms, and to concentrate on the states belonging to a given multiplicity. An estimation of this effect in the case of large rf coupling is given in \secref{sec:beyondRWA}.

\subsubsection{Dressed states in the rotating wave approximation}
\label{sec:dress-states-RWA}

Within the rotating wave approximation, we just need to find the eigenstates in a given manifold. The result is given by using a generalised spin rotation, with the angle given at Eq.~\eqref{eq:theta_dress}, where the photon number is changed accordingly to stay in the ${\mathcal{E}}_N$ manifold. The eigenstates are the \textit{dressed states} $\ket{m,N}_\theta$ given by (see also \cite{Garraway2011})
\bea
\ket{m,N}_\theta &=& \sum_{m'=-F}^F \,_z\braket{m'}{m}_\theta\ket{m',N-m'}_z \nonumber\\
&=& \sum_{m'=-F}^F {}_z\bra{m'}\hat R_y(\theta)\ket{m}_z\ket{m',N-m'}_z
\label{eq:quantum_dressed_states}
\eea
with $\theta$ given by Eq.~\eqref{eq:theta_dress} and $\ket{m}_\theta=\hat{R}(\theta)\ket{m}_z$ as in Eq.~\eqref{eq:rotated_state}. Their energies are
\bea
E_{m,N} &=& E'_{m} + N\hbar\omega, \nonumber\\
\mbox{where} \qquad E'_{m} &=& m\hbar\sqrt{\delta^2+|\Omega_+|^2}.
\label{eq:quantum_dressed_energies}
\eea
The spin states are dressed by the rf field, in such a way that the eigenstates are now combining different spin and field states, and cannot be written as a product state spin$\otimes$field. The dressed states are connected to the uncoupled states for $|\delta|\gg |\Omega_+|$. The effect of the coupling is to repel the states inside the multiplicity: the frequency splitting increases from $|\delta|$ to $\sqrt{\delta^2+|\Omega_+|^2}$.

\section{Adiabatic potentials for rf dressed atoms}

 \label{sec:adiabatic_potential}
In the previous sections, we have assumed uniform static and rf fields, so that the spin Hamiltonian fully decouples from the external degrees of freedom. Here, we will consider instead that either the static field $\mathbf{B}_0(\mathbf{r}) = B_0(\mathbf{r}) \GG{u}(\GG{r}) $, or the rf field
\bea
\mathbf{B}_1(\mathbf{r},t) &=& \BB_1(\GG{r})\, e^{-i\omega t} + c.c. \nonumber\\
&=& \mathcal{B}_1(\mathbf{r}) \,\GG{\epsgras}(\GG{r})\,e^{-i\omega t} + c.c.,
\label{eq:rf_field_general_case}
\eea
or both, are space-dependent. $\mathcal{B}_1(\mathbf{r})$ is the complex amplitude\footnote{In the particular case of a linearly polarised field $\mathbf{B}_1(\mathbf{r},t) = B_1\cos(\omega t+\phi)\,\epsgras$ (with $\epsgras$ real) introduced in \secref{sec:classical_field}, the complex amplitude would read $\mathcal{B}_1=B_1/2$.} and $\GG{\epsgras}(\GG{r})$ is the complex polarization of the rf field. The local Larmor frequency is $\omega_0(\GG{r})=|g_F|\mu_BB_0(\GG{r})/\hbar$, the local detuning is $\delta(\GG{r})=\omega-\omega_0(\GG{r})$.

Within the semi-classical treatment of \secref{sec:classical_field}, the total Hamiltonian for the external variables and spin reads
\bea
\hat{H} &=& \frac{\GG{\hat{P}}^2}{2M} + \frac{g_F\mu_B}{\hbar} B_0(\GG{\hat{R}})\,\GG{\hat{F}}\cdot \GG{u}(\GG{\hat{R}}) \nonumber\\
&+&\left\{ \frac{g_F\mu_B}{\hbar} \mathcal{B}_1(\GG{\hat{R}})\,e^{-i\omega t}\GG{\hat{F}}\cdot \GG{\epsgras}(\GG{\hat{R}}) + h.c.\right\}.
\label{eq:hamil_adiab_pot}
\eea
Here, $\mathbf{\hat{R}}$ and $\mathbf{\hat{P}}$ are the atomic position and momentum operators, respectively, and $M$ is the atomic mass. We can recast this expression under the form
\beq
\hat{H} = \hat{T} + \hat{H}\ind{spin}(\GG{\hat{R}},t)
\label{eq:hamil_adiab_pot_recast}
\eeq
where the kinetic part is $\hat{T}=\GG{\hat{P}}^2/(2M)$ and $\hat{H}\ind{spin}(\GG{\hat{R}},t)$ contains all the angular momentum operators.

\subsection{Adiabatic potentials}
\label{sec:adiabatic_potentials}
The  approach of \secref{sec:classical_field} still holds to diagonalise the spin part of the Hamiltonian $\hat{H}\ind{spin}(\GG{\hat{R}},t)$, although the unitary transformation now depends on the position operator $\GG{\hat{R}}$, as $\hat{H}\ind{spin}(\GG{\hat{R}},t)$ commutes with $\GG{\hat{R}}$. For each value $\mathbf{r} = \langle \mathbf{\hat{R}} \rangle$ of the average position, the eigenstates are $\ket{m}_{\theta(\mathbf{r})}$, where $\theta(\mathbf{r})$ is given with respect to the local quantisation axis $\mathbf{u}(\mathbf{r})$. The generalised Rabi frequency $\Omega(\GG{r}) = \sqrt{\delta(\GG{r})^2+|\Omega_1(\GG{r})^2|}$ gives the eigenenergies at point \GG{r}, see Eq.~\eqref{eq:gen_Rabi_freq}. Here,  $\Omega_1(\GG{r})=|\Omega_1(\GG{r})|e^{-i\phi(\GG{r})}$ is the local component on the efficient  local circular polarization $\sigma^s$ (defined with respect to the quantisation axis \GG{u}). Its derivation will be given in \secref{sec:local_coupling}.

In principle, the unitary transformation applied to diagonalise the spin Hamiltonian doesn't commute with the kinetic energy. However, this effect can be ignored if the spatial dependence of the static and rf fields are weak enough, at small atomic velocities and large rf splitting $\Omega$. This is called the \emph{adiabatic theorem} \citep{MessiahEnglish}. In the following, we will assume that the adiabatic condition holds. Non-adiabatic transitions between the adiabatic states will be discussed in \secref{sec:LZlosses}.

In this limit, an atom prepared in a state $\ket{m}_{\theta(\mathbf{r})}$ will follow adiabatically this local eigenstate, called an \emph{adiabatic state}, as \GG{r} varies due to the atomic motion. The local eigenenergy thus acts as a potential for the atomic motion, indexed by $m$:
\beq
V_{m}(\GG{r}) = m\hbar\Omega(\GG{r}) = m\hbar\sqrt{\delta(\GG{r})^2 + |\Omega_1(\GG{r})|^2}.
\eeq

\begin{figure}[t]
\centering
\subfigure[Bare potentials.]{\includegraphics[width=0.8\linewidth]{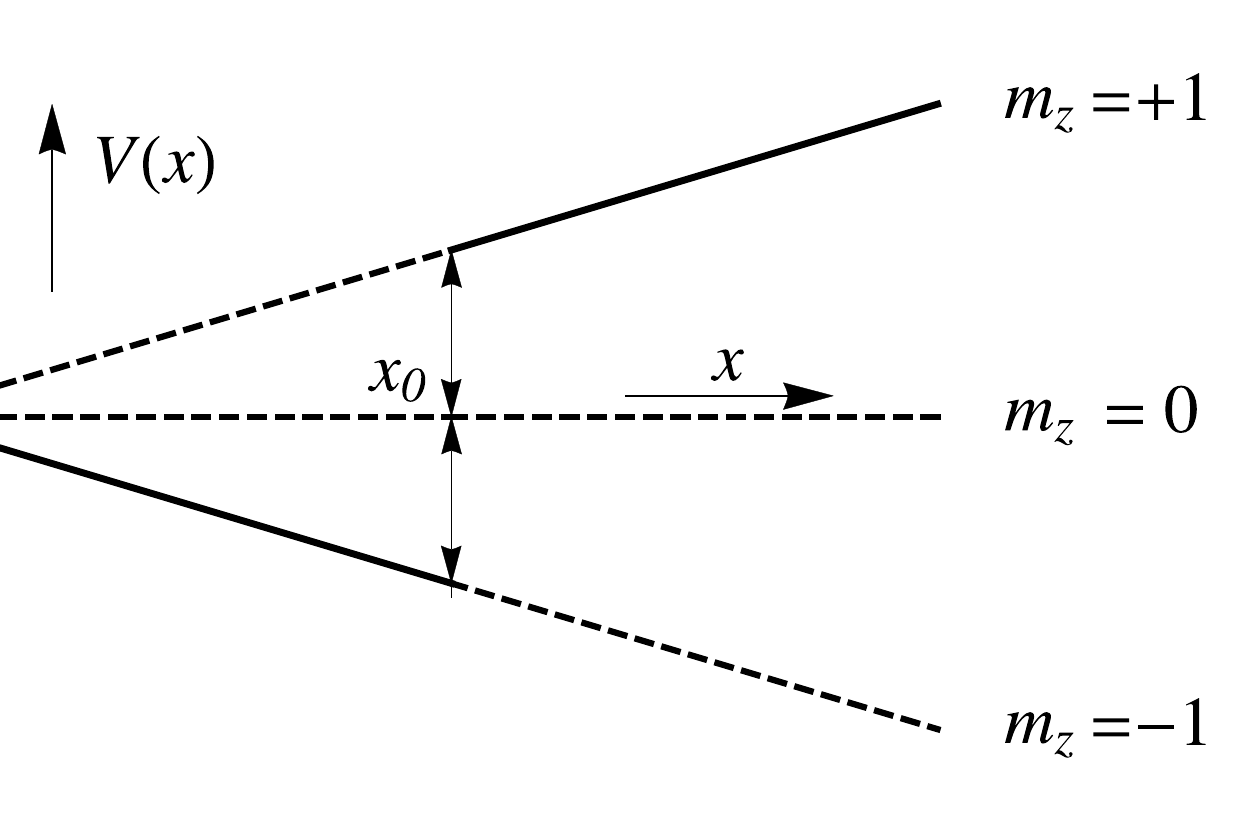}}\\
\subfigure[Adiabatic potentials.]{\includegraphics[width=0.8\linewidth]{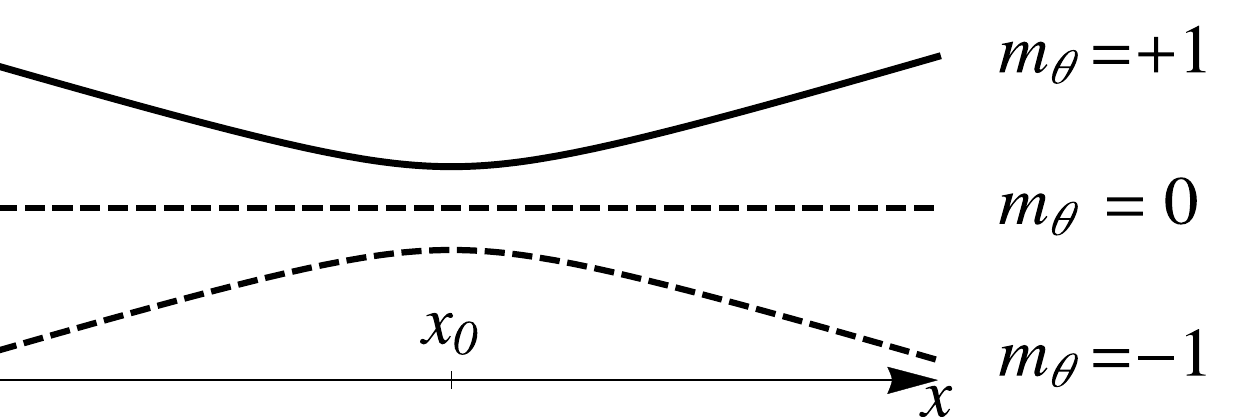}}
 \caption{\label{fig:APprinciple}
Simple example of an adiabatic potential in the case $F=1$, where the static magnetic field varies linearly with position $x$, the Rabi frequency $\Omega_1$ is position independent and the rf resonance occurs at $x=x_0$. (a) Magnetic potentials, corresponding to the energies of the uncoupled (or bare) states $\ket{m}_z$. (b) Adiabatic potentials, corresponding to the energies of the three dressed states $\ket{m}_\theta$. The frequency splitting at the avoided crossing is $\Omega_1$.
}
\end{figure}
\Figref{fig:APprinciple} illustrates the adiabatic potentials (for an angular momentum $F=1$) in the simple case where the magnetic field increases linearly with the spatial coordinate $x$, such that $\omega_0(x)=\alpha x$ where $\alpha$ is a constant gradient. The rf field is resonant at a position $x_0=\omega/\alpha$, where the adiabatic potentials present an avoided crossing, with a frequency splitting equal to the Rabi coupling. Far from this region, the adiabatic state $\ket{m}_\theta$ connects to one of the bare states $\ket{\pm m}_z$, with opposite sign on both ends, as illustrated by the bold lines of \figref{fig:APprinciple}a, which correspond to the two far ends of the bold line in \figref{fig:APprinciple}b.

In general, the maximally polarised state $\ket{m=F}_{\theta(\mathbf{r})}$ is used to trap atoms. For a spin 1 or $1/2$, this is the only trappable state around the resonance point. For $F>1$, this choice allows one to suppress inelastic collisions: the spin of the atoms initially in a non fully polarised state can flip in a collision, and cause losses, whereas such a spin-flip is prevented by conservation of total angular momentum when both atoms are fully polarised \citep{Ketterle1999}. The same is also true in a magnetic trap. One may wonder if this stays true for dressed spin, as the direction of the quantisation field rotates at frequency $\omega$. However, all the spins at the same location rotate in phase, such that they can be considered as polarised in the same extremal state. Polarising the atoms in the extremal state $\ket{m=F}_\theta$ is thus an efficient way to prevent inelastic collisions in an adiabatic potential~\citep{Moerdijk1996}.
In this adiabatic state $\ket{F}_\theta$, the potential energy is always positive

\begin{equation}
V_F(\GG{r}) = F\hbar\sqrt{\delta(\GG{r})^2 + |\Omega_1(\GG{r})|^2}.
\label{eq:VFPotential}
\end{equation}

\subsection{Local coupling}
\label{sec:local_coupling}

The rf field frequency $\omega$ is chosen in the typical range of the Larmor frequency $\omega_0(\GG{r})$. At some particular positions, the Larmor frequency is exactly $\omega$. The locus of the points such that $\omega_0(\GG{r})=\omega$ is a surface in space, characterised by a given value of the magnetic field $B_0=\hbar\omega/(|g_F|\mu_B)$, which we refer to as the \textit{resonance surface}. This surface is determined by the choice of $\omega$. It is often topologically equivalent to a sphere.

The system is described by the Hamiltonian of Eq.~\eqref{eq:hamil_adiab_pot}. 
Assuming that the adiabatic approximation is valid, we will take a semi-classical approach to describe the atom and replace the positions and momentum operators by their average value \GG{r} and \GG{p}. At each fixed position \GG{r}, we can apply the procedure described in \secref{sec:classical_field} to get the spin eigenstates and the eigenenergies: apply a rotation of angle $\signgf\omega t$ around the axis $\GG{u}(\GG{r})$, apply the rotating wave approximation to remove counter-rotating terms, and diagonalise the effective, time-independent hamiltonian.

An important point is that the $\sigma^\signgf$ polarization, which is the only efficient component to couple the spin states in the RWA, must be defined with respect to the local quantisation axis $\GG{u}(\GG{r})$. Using the \textit{local} spherical basis $\left(\GG{e}_+(\GG{r}), \GG{e}_-(\GG{r}), \GG{u}(\GG{r})\right)$, the relevant rf coupling is
\beq
\Omega_1(\GG{r}) = \Omega_{\signgf}(\GG{r}) = - \sqrt{2}\frac{g_F\mu_B}{\hbar} \mathcal{B}_1(\GG{r}) \,\GG{e}^*_{\signgf}(\GG{r})\cdot \GG{\epsgras}(\GG{r}).
\eeq
It is clear from this expression that, even if the rf complex amplitude $\mathcal{B}_1$ and the polarization $\epsgras$ are homogeneous, the Rabi frequency is position dependent, because of the position dependent direction $\GG{u}(\GG{r})$ of the static magnetic field.

Using the properties of the spherical basis, in particular $\GG{e}_\pm \times \GG{u} = \pm i \GG{e}_\pm$, or equivalently $\GG{e}_\signgf \times \GG{u} = is \GG{e}_\signgf$ where $\signgf=\pm 1$ is the sign of $g_F$, we find $\GG{e}^*_{\signgf}\cdot \GG{\epsgras}$ as follows:
\bea
\left(\epsgras - is \epsgras\times \GG{u}\right)\times \GG{u} &=&\epsgras\times\GG{u} + i\signgf \,\GG{u}\times \left(\epsgras\times\GG{u}\right)\nonumber\\
&=& 2 i\signgf (\GG{e}^*_{\signgf}\cdot \GG{\epsgras}) \GG{e}_\signgf,\\
\left| \GG{e}^*_{\signgf}\cdot \GG{\epsgras} \right| &=& \frac{1}{2}\left|\epsgras\times\GG{u} + i\signgf \,\GG{u}\times \left(\epsgras\times\GG{u}\right)\right|.
\eea
This expression makes clear that the rf coupling comes from the projections $\epsgras\times\GG{u}$ and $\GG{u}\times \left(\epsgras\times\GG{u}\right)$ of the rf field in a plane orthogonal to the direction \GG{u} of the static magnetic field. From this expression comes finally, for any complex rf field $\mathcal{B}_1(\GG{r})\epsgras(\GG{r})$ and any non vanishing static field $\GG{B}_0(\GG{r})$ such that $\GG{u}(\GG{r}) = \GG{B}_0(\GG{r})/B_0(\GG{r})$,
\bea
\left| \Omega_1(\GG{r})\right| &=& \frac{\sqrt{2}\left|g_F\mu_B \mathcal{B}_1(\GG{r})\,\GG{e}^*_{\signgf}(\GG{r})\cdot \GG{\epsgras}(\GG{r})\right|}{\hbar} \nonumber\\
&=& \frac{\left|g_F\mu_B \mathcal{B}_1\right|}{\hbar\sqrt{2}}\left|\epsgras\times\GG{u} + i\signgf\,\GG{u}\times \left(\epsgras\times\GG{u}\right)\right|.
\label{eq:rf_coupling_for_any_field}
\eea
Another convenient expression for the calculation of the rf coupling reads:
\bea
\left| \Omega_1(\GG{r})\right| &=&\frac{\left|g_F\mu_B \mathcal{B}_1\right|}{\hbar\sqrt{2}}\\ \label{eq:convenient_coupling}
&\times&\sqrt{1 - \left|\epsgras\cdot\GG{u}\right|^2 + \left|\epsgras\times\GG{u}\right|^2+2i\signgf\,\GG{u}\cdot\left(\epsgras\times\epsgras^*\right)}.\nonumber
\eea

Let us discuss two important particular choices of the rf polarization, namely circular or linear. For a polarization $\epsgras$ which is circular $\sigma^\signgf$ with respect to some given axis $z$, we remark that $(\epsgras,\epsgras^*,\GG{e}_z)$ form a basis. The last cross product reads $\epsgras\times\epsgras^* = -is\,\GG{e}_z$, such that the argument of the square root in  Eq.~\eqref{eq:convenient_coupling} simplifies to $1-\left|\epsgras\cdot\GG{u}\right|^2+\left|\epsgras\times\GG{u}\right|^2+2u_z$ with $u_z = \GG{u}\cdot\GG{e}_z$. Moreover, using the decomposition of \GG{u} in the $(\epsgras,\epsgras^*,\GG{e}_z)$ basis
$$
\GG{u} = \left(\GG{u}\cdot\epsgras^*\right)\epsgras + \left(\GG{u}\cdot\epsgras\right)\epsgras^* + u_z\GG{e}_z,
$$
we have
$$
\epsgras\times\GG{u} = \left(\GG{u}\cdot\epsgras\right)\epsgras\times\epsgras^* + u_z\epsgras\times\GG{e}_z = is\left[u_z\epsgras-\left(\GG{u}\cdot\epsgras\right)\GG{e}_z\right].
$$
As $\epsgras$ and $\GG{e}_z$ are orthogonal, the modulus square of the two components add and we get finally
\bea
&&1 -\left|\epsgras\cdot\GG{u}\right|^2 + \left|\epsgras\times\GG{u}\right|^2+2i\signgf\,\GG{u}\cdot\left(\epsgras\times\epsgras^*\right) \nonumber\\
&&= 1 - \left|\epsgras\cdot\GG{u}\right|^2+2u_z+u_z^2+\left|\GG{u}\cdot\epsgras\right|^2\nonumber\\
&&= (1+u_z)^2.\nonumber
\eea
The local coupling for a rf polarization $\sigma^\signgf$ circular of sign $\signgf$ along the axis $z$ is
\beq
\left| \Omega_1(\GG{r})\right| =\frac{\left|g_F\mu_B \mathcal{B}_1(\GG{r})\right|}{\hbar\sqrt{2}}\left[1+u_z(\GG{r})\right].
\label{eq:local_coupling_circular_polarization}
\eeq
It is maximum and equal to $\sqrt{2}\left|g_F\mu_B \mathcal{B}_1(\GG{r})\right|/\hbar$ where the field is aligned along $+z$ and vanishes at the places where the static field points towards $-z$.

In the case of a linear polarization $\epsgras$, the cross product $\epsgras\times\epsgras^*$ vanishes. Moreover, $1 -\left|\epsgras\cdot\GG{u}\right|^2$ and $\left|\epsgras\times\GG{u}\right|^2$ are equal and correspond to the square of the projection of \GG{u} in the direction orthogonal to $\epsgras$. The factor under the square root is then $2(1 -\left|\epsgras\cdot\GG{u}\right|^2)$. The local coupling for a linear rf polarization along some axis $z$ is thus
\beq
\left| \Omega_1(\GG{r})\right| =\frac{\left|g_F\mu_B \mathcal{B}_1(\GG{r})\right|}{\hbar}\sqrt{1-u_z(\GG{r})^2}.
\label{eq:local_coupling_linear_polarization}
\eeq
It is maximum and equal to $\left|g_F\mu_B \mathcal{B}_1(\GG{r})\right|/\hbar$ at the positions in space where the field is orthogonal to the $z$ axis, and vanishes where the static field and the rf field are parallel.

\subsection{Trap geometry: role of the isomagnetic surface}
\label{sec:bubble}
In order to understand better the spatial shape of the adiabatic potential, let us first assume that the rf effective coupling $|\Omega_1(\GG{r})|$ is homogeneous and equal to $\Omega_1$
while $\delta$ varies with position in Eq.~\eqref{eq:VFPotential}. 
This is relevant if the direction of the static field varies only slightly close to the potential minimum of the adiabatic potential. It is then clear that, in the absence of gravity, the potential is minimum where the detuning vanishes, $\delta(\GG{r})=0$, that is on the isomagnetic resonant surface defined by $\omega_0(\GG{r})=\omega$. This provides us with very anisotropic traps, where one direction transverse to the isomagnetic surface is confined, and may be strongly confined by the avoided crossing of the adiabatic potential, while the directions parallel to the surface are free to move.

How much are the atoms indeed confined to an isomagnetic surface $\delta(\GG{r})=0$? By definition, the direction normal to the surface is given by the gradient of the Larmor frequency $\nablagras\omega_0$. Along this direction, locally, the variations of $\omega_0$ are linear, and so are the variations of $\delta$: $\delta(\GG{r}+\GG{dr}) \simeq \delta(\GG{r}) + \GG{dr}\cdot \nablagras\omega_0$. We can expand $V_F$ around $\GG{r}$ to determine the oscillation frequency in the harmonic approximation in the direction $\nablagras\omega_0$ normal to the surface. We find:
\beq
\omega\ind{transverse} = \alpha\sqrt{\frac{F\hbar}{M\Omega_1}},
\label{eq:general_transverse_frequency}
\eeq
where $\alpha = |\nablagras\omega_0|$ is the local magnetic gradient in units of frequency. This formula is similar to the one giving the largest of the two frequencies of a cigar-shape Ioffe-Pritchard (IP) trap, provided the Larmor frequency at the trap bottom is replaced by the Rabi frequency. As the Rabi frequency is in general much smaller than the Larmor frequency, as required for the RWA, for a given magnetic gradient the confinement to the isomagnetic surface in an adiabatic potential is thus significantly larger than what is obtained in a IP trap. The transverse confinement frequency is typically in the range of a few hundred Hz to a few kHz. We can thus have a good idea of the trap geometry by assuming that the atoms will be confined to the isomagnetic surface. The effect of the position dependence of $|\Omega_1(\GG{r})|$ or of gravity is essentially to shape a refined landscape inside this surface. This will be discussed in more detail in \secref{sec:examples_of_potentials}, where we will consider various examples of interesting configurations.

\begin{figure}[t]
\begin{center}
\includegraphics[width=\linewidth]{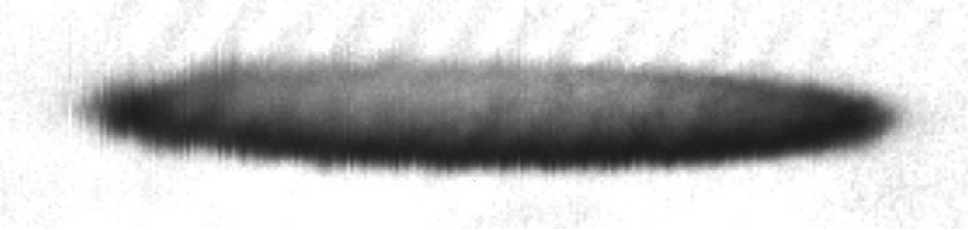}
\caption{Spatial density distribution of cold rf-dressed atoms confined in an adiabatic potential. The static magnetic field comes from a cigar shape Ioffe-Pritchard trap, see \secref{sec:examples_IP}. Atoms are spread around an ellipsoid of radii $160~\mu$m$\times 1.3$~mm, isomagnetic surface of the magnetic static field. The atomic density is higher at the bottom of the ellipsoid, where gravity causes the atoms to collect. Figure from O. Morizot's thesis~\citep{MorizotThese}, reprinted by courtesy of the author.}
\label{fig:bubble}
\end{center}
\end{figure}

If the static magnetic field $B(\GG{r})$ has a local minimum $B\ind{min}$, as is the case in a magnetic trap, the isomagnetic surfaces close to this minimum are typically ellipsoids (a trap is generally harmonic close to its minimum). This is the basic idea for a bubble trap, as proposed by Zobay and Garraway~\citep{Zobay2001,Zobay2004} and first realised by Colombe et al.~\citep{Colombe2004a}, see \figref{fig:bubble}.

When gravity is included, the isomagnetic surface is no longer an isopotential of the total potential $V\ind{tot}(\GG{r}) = F\hbar\sqrt{\delta(\GG{r})^2 + \Omega_1^2} + Mgz$. There is a single minimum, at the bottom of the ellipsoid. Depending on the energy of the cloud, which for thermal atoms is set by their temperature, the atoms will fill the bubble up to a certain height given by the barometric energy: $h\ind{max}\sim k_B T/(Mg)$. For a Bose-Einstein condensate, the relevant energy scale is the chemical potential $\mu$ and $h\ind{max}\sim \mu/(Mg)$.

\begin{figure}[t]
\begin{center}
\includegraphics[width=\linewidth]{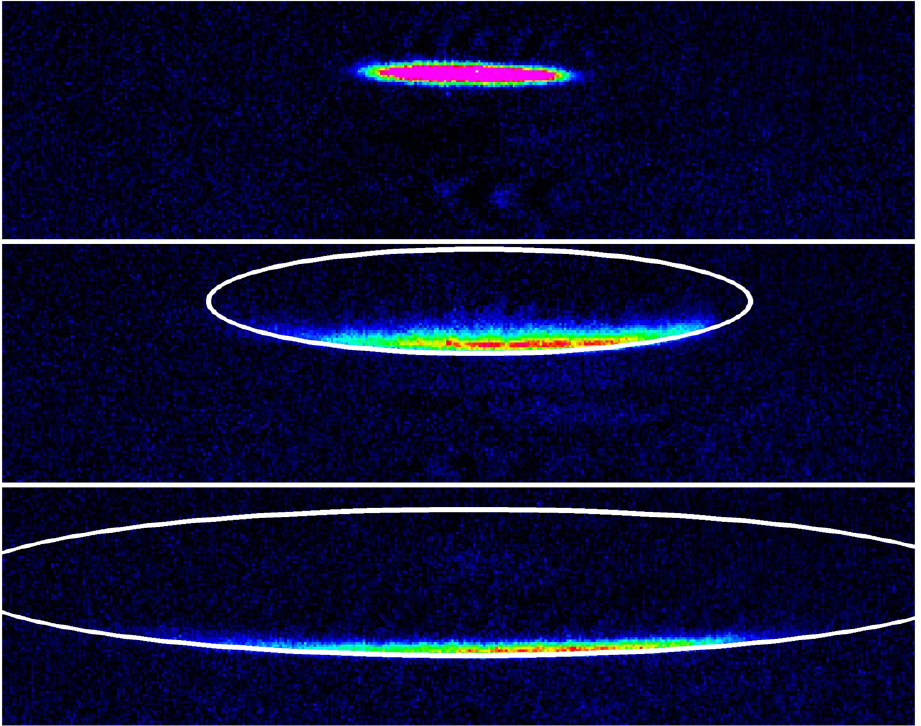}
\caption{Cold rf-dressed rubidium atoms in hyperfine state $F=2$ confined in an adiabatic potential, for different values of the dressing frequency~\citep{Colombe2004a}. Top: no rf dressing, atoms are trapped in a Ioffe-Pritchard trap with a cigar shape. The Larmor frequency at the trap bottom is 1.3 MHz. Middle: rf dressing field at 3~MHz. The atoms occupy the lower part of a bubble. The field of view is 4.5 mm (horizontally) $\times$ 1.2 mm (vertically). The centre of mass is shifted vertically by $160~\mu$m. Lower picture: dressing frequency 8~MHz. The isomagnetic bubble is larger, the atoms are more shifted vertically (by $490~\mu$m) and the cloud is even more anisotropic. The resonant isomagnetic surfaces are marked with a white line.}
\label{fig:dressedIP}
\end{center}
\end{figure}

The radii of the resonant ellipsoid can be adjusted with the choice of the rf frequency $\omega$.  Increasing $\omega$ from the minimum Larmor frequency in the trap centre $\omega\ind{$0$,min}$ makes the bubble inflate, so that the $z$ radius becomes larger than $h\ind{max}$ at some point. Above this frequency, the atoms are confined to a curved plane. The pictures of \figref{fig:dressedIP} show ultra cold atoms confined in such an anisotropic trap, for various values of the dressing frequency $\omega$.

\subsection{Loading from a magnetic trap}
\label{sec:loading}

\begin{figure}[b!]
\begin{center}
\begin{tabular}{@{}rr@{}}
\includegraphics[height=0.43\linewidth]{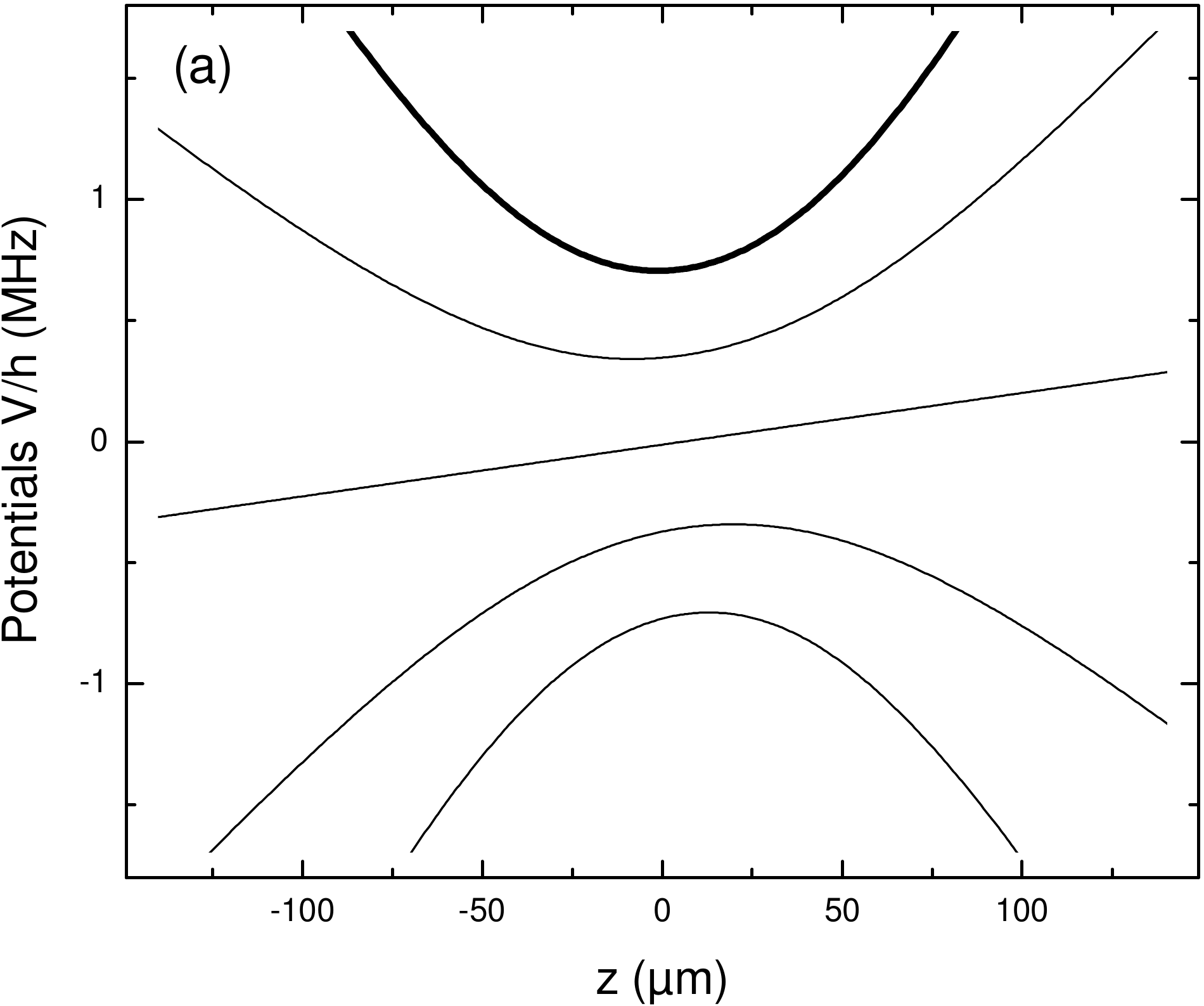}&
\includegraphics[height=0.43\linewidth]{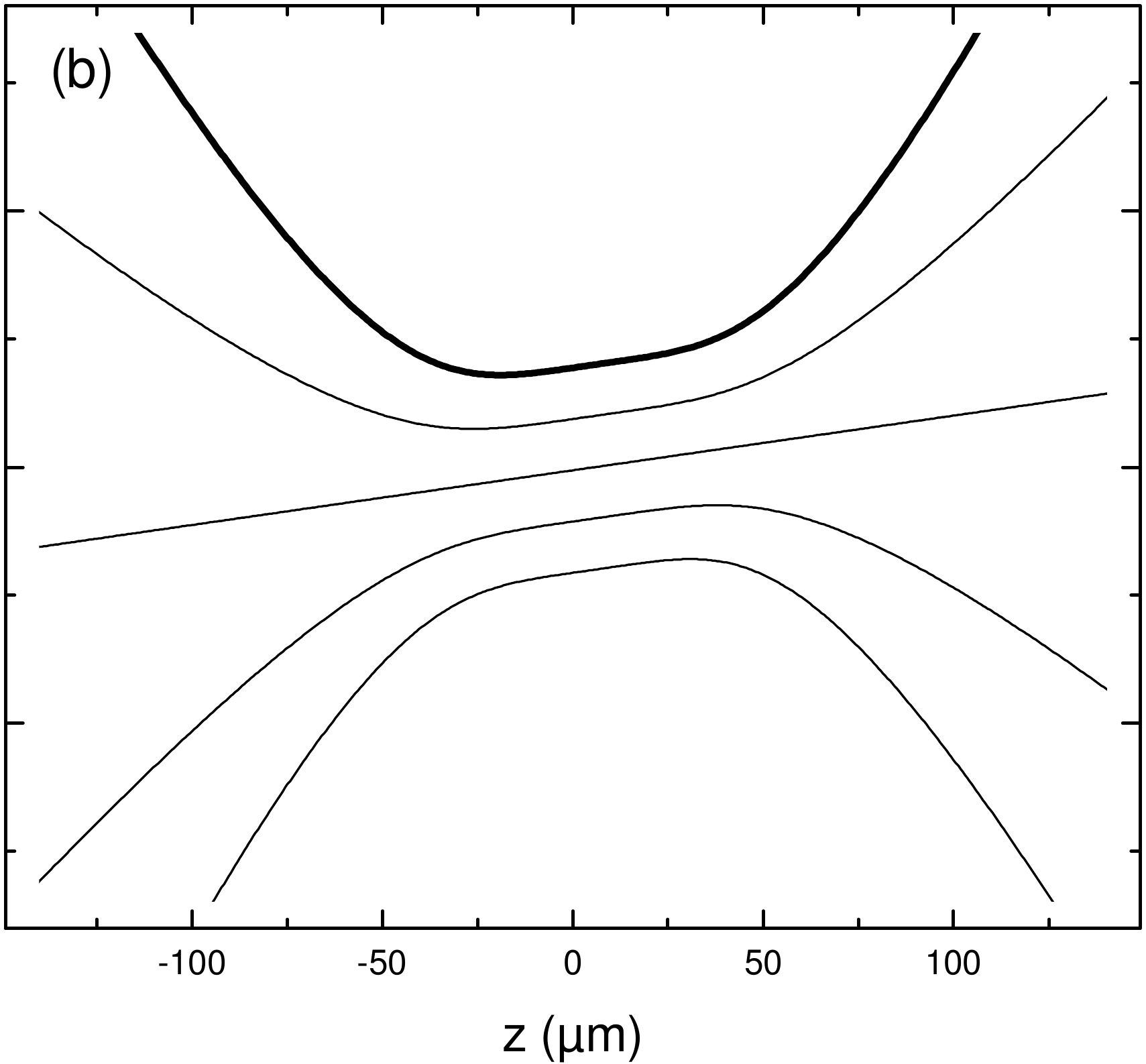}\\
\includegraphics[height=0.43\linewidth]{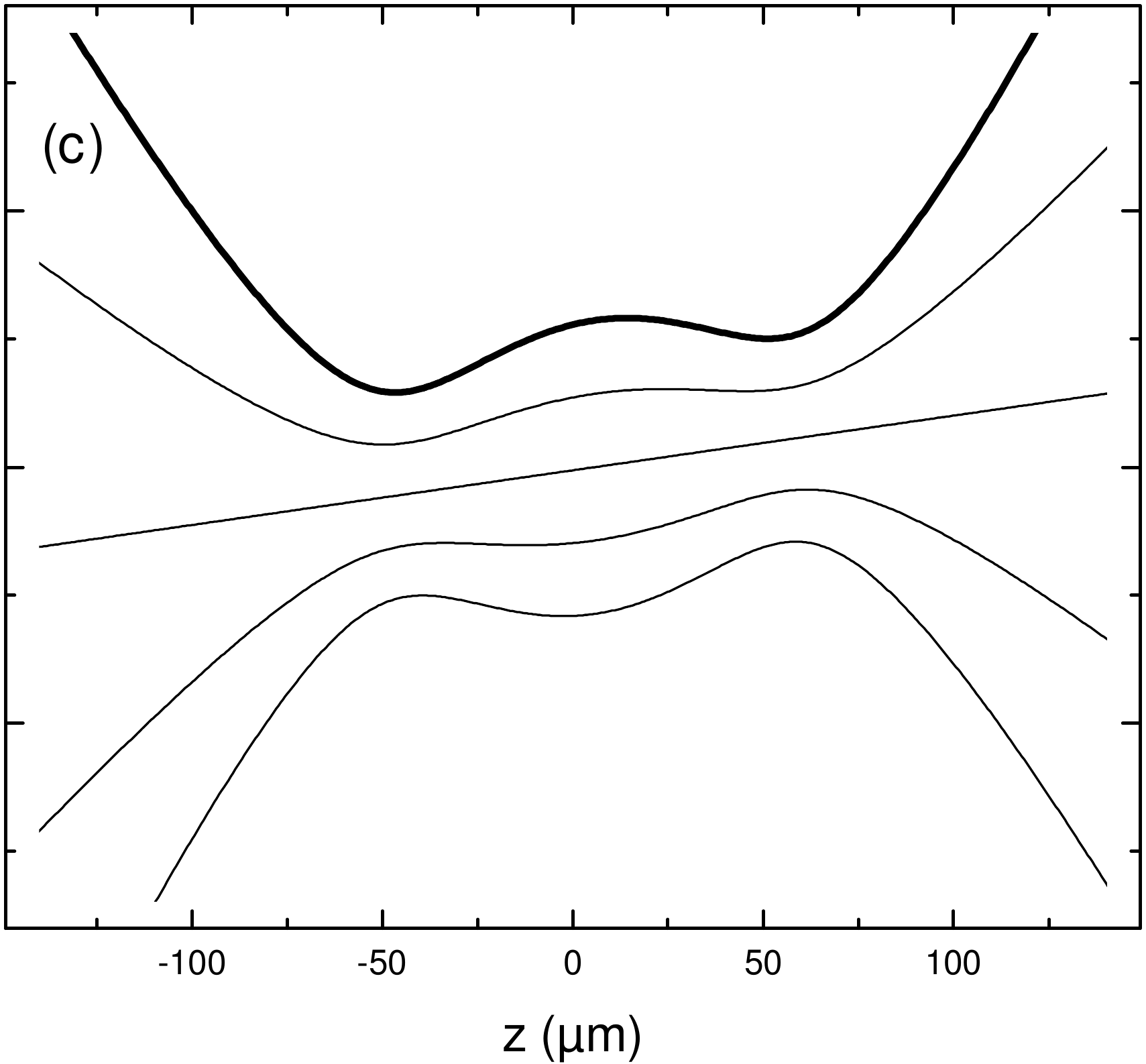}&
\includegraphics[height=0.43\linewidth]{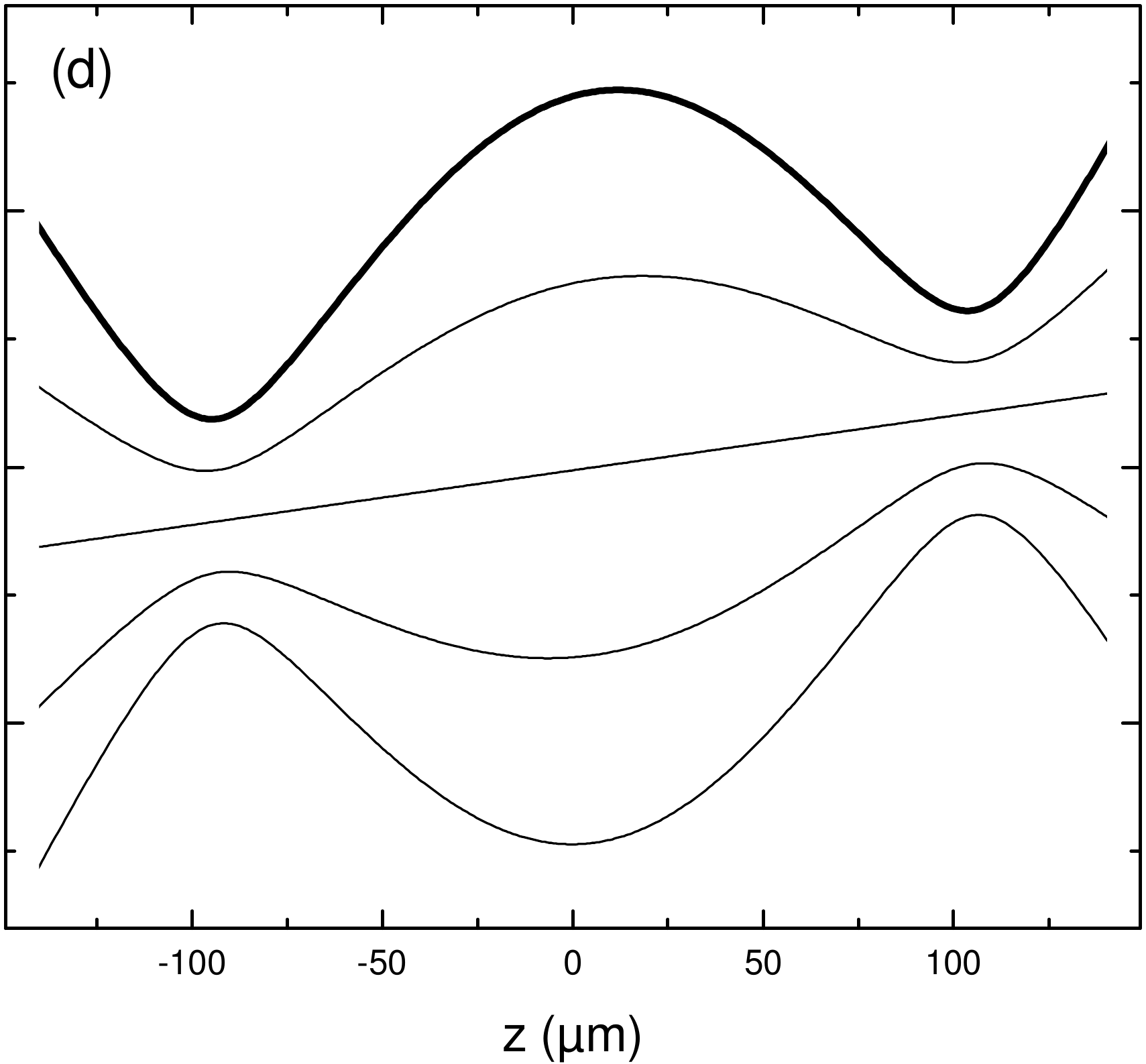}
\end{tabular}
\caption{Loading procedure from a magnetic trap with non-zero minimum, corresponding to the experimental situation of \figref{fig:dressedIP}. The adiabatic potentials are represented for the five states of a $F=2$ spin, along the vertical axis. Gravity is included. From (a) to (d), the rf frequency is ramped from below resonance to the final value. The atoms start in the upper potential, which is the same as the magnetic trapping potential in (a). (b) corresponds to the resonance at the bottom of the trap, $\omega = \omega\ind{$0$,min}$. Above $\omega\ind{$0$,min}$, see (c) and (d), there is a single trap minimum at the bottom of the ellipsoid, due to gravity, which position $z$ changes with the dressing frequency. Figure from Y.~Colombe's thesis~\citep{ColombeThese}, reprinted by courtesy of the author.}
\label{fig:loading}
\end{center}
\end{figure}

Adiabatic potentials could in principle be loaded from a magneto-optical trap, like magnetic traps. However, their trapping volume is in general much smaller, as the interesting feature of the large anisotropy also comes with a small volume. Moreover, spin flips can occur at high temperature (a few tens of $\mu$K) because the relevant effective splitting frequency is $\Omega_1$, of order 100 kHz, in general smaller than the minimum Larmor frequency $\omega\ind{0,min}$ in a magnetic trap, of order 1 MHz. Adiabatic potentials are very well adapted to trap ultra cold atoms or condensates, pre-cooled by evaporative cooling in a magnetic trap.

The loading procedure from a magnetic trap presenting a non-zero Larmor frequency $\omega\ind{$0$,min}$ at its bottom is straightforward, and is sketched on \figref{fig:loading}. The idea is to operate a frequency sweep from an initial frequency below $\omega\ind{$0$,min}$, up to the desired final value above $\omega\ind{$0$,min}$. 
The rf field is switched on at a fixed negative detuning $\delta = \omega - \omega\ind{$0$,min}$, in a time sufficiently long to ensure the adiabatic condition Eq.~\eqref{eq:adia_thetadot}, which for a variation in $\Omega_1$ only writes $|\dot\Omega_1|\ll\delta^2$. The initial, trapped, upper magnetic state $\ket{\signgf F}_z$ is connected to the upper dressed state $\ket{F}_\GG{u}$. When the rf frequency is subsequently increased, the atoms stay in this upper adiabatic state if the ramp is slow, so that $|\dot\delta|\ll\Omega_1^2$ and we used again Eq.~\eqref{eq:adia_thetadot}. Once $\omega$ reaches $\omega\ind{$0$,min}$, the atoms have reached the resonant surface, and they remain at this surface, which is now a potential minimum, as the rf frequency is further increased.

We note that apart from a condition on the adiabatic following of the adiabatic \emph{spin} state, the loading procedure involves a trap deformation, which could lead to heating if this deformation is not adiabatic with respect to the \emph{mechanical} degrees of freedom. In case of rapid loading resulting in unwanted heating, evaporative cooling can be readily applied in the final adiabatic trap, see \secref{sec:cooling}.

\section{Examples of adiabatic potentials}
\label{sec:examples_of_potentials}
We will now present two important examples of adiabatic potentials. The reasoning followed in this section can then be generalised to any kind of trapping configuration with rf-induced adiabatic potentials. Other trap configurations are reviewed in \citep{Garraway2016}, including lattices \citep{Courteille2006,Lin2010,Morgan2011}, adiabatic traps combined with optical lattices allowing to reach sub wavelength periods \citep{Lundblad2008,Shotter2008,Lundblad2014,Yi2008}, or using inductively coupled currents \citep{Griffin2008,Pritchard2012,Vangeleyn2014,SinucoLeon2014}.

Here, we discuss two configurations based on Ioffe-Pritchard and quadrupole static magnetic traps, respectively, for various choices of the rf polarization, which can substantially impact on the resulting adiabatic potential. We start with traps based on a static Ioffe-Pritchard configuration, which were the first to be implemented \citep{Colombe2004a,Schumm2005b}. We then discuss how a quadrupole field can be used to tailor a very anisotropic adiabatic trap for confining quasi two-dimensional gases \citep{Merloti2013a,Merloti2013b,DeRossi2016}, and eventually lead to an annular trap in combination with an optical potential \citep{Morizot2006,Heathcote2008}.

In this section, we will consider for simplicity a uniform rf magnetic field, in amplitude and polarization, which reads
$$
\GG{B}_1 = \mathcal{B}_1\, \epsgras \, e^{-i\omega t} + c.c.
$$
where $\epsgras$ is a complex polarization of unit modulus and $\mathcal{B}_1$ a complex amplitude.

\subsection{The dressed Ioffe-Pritchard trap}
\label{sec:examples_IP}
The first rf-dressed trap was obtained from dressing atoms initially confined in a Ioffe-Pritchard magnetic trap \citep{Colombe2004a}. We will first recall the static field configuration in this case, then discuss the effect of the rf field depending on the rf polarization and amplitude.
\subsubsection{Isomagnetic surfaces}
\label{sec:IPisoB}

The static magnetic field we start from is the one of a Ioffe-Pritchard magnetic trap \citep{Ketterle1999}. Let us write explicitly the magnetic field in this trap near the trap minimum, up to second order in the coordinates:
\bea
&&\GG{B}_0(\GG{r}) = B_z(z)\,\GG{e}_z + b'(x\,\GG{e}_x - y\,\GG{e}_y) - \frac{b''}{2}z(x\,\GG{e}_x + y\,\GG{e}_y)\nonumber\\
&&\mbox{where}\qquad B_z(z)=B\ind{min}+\frac{b''}{2}z^2
\eea
Its modulus up to second order in $x,y,z$ is
\bea
B_0(\GG{r}) &=& B\ind{min}\left[1 + \frac{b''}{2B\ind{min}}z^2 + \frac{b'^2}{2B\ind{min}^2}(x^2+y^2)\right] \nonumber\\
&=& B_0(\rho,z)\nonumber
\eea
giving rise to a cylindrically symmetric cigar-shape harmonic trapping near the centre, with frequencies $\omega_x=\omega_y\gg\omega_z$. Away from the $z$ axis, specifically when $\rho=\sqrt{x^2+y^2}\gg B\ind{min}/b'$, the trap is closer to a linear potential, with $B_0(\GG{r})\simeq b'\rho$. The quantisation axis is defined as
\beq
\GG{u} = \frac{B_z(\GG{r})}{B_0(\GG{r})}\,\GG{e}_z + \frac{b'x - b''xz/2}{B_0(\GG{r})}\,\GG{e}_x - \frac{b'y + b''yz/2}{B_0(\GG{r})}\,\GG{e}_y.
\label{eq:quantization_IP}
\eeq

For discussing adiabatic potentials, frequency units are often more convenient. The Larmor frequency is denoted $\omega_0(\rho,z)$, with a minimum value $\omega\ind{$0$,min}$ at the centre, and the gradient in frequency units is $\alpha = |g_F|\mu_Bb'/\hbar$. We also define $\eta = |g_F|\mu_B b''/\hbar$, the curvature in units of frequency. We can then also write, up to second order in the coordinates,
$$
\omega_0(\rho,z) = \omega_0(0,z) + \frac{\alpha^2}{2\omega\ind{$0$,min}}\rho^2,
$$
with
$$
\omega_0(0,z) = \omega\ind{$0$,min} + \frac{1}{2}\eta z^2.
$$

The resonant surface is defined by $\omega_0(\rho\ind{res}(z),z)=\omega$. At each longitudinal position $z$, the resonant radius $\rho\ind{res}$ is given by
\beq
\rho\ind{res}(z) = \frac{1}{\alpha}\sqrt{\omega^2 - \left(\omega\ind{$0$,min} + \frac{\eta z^2}{2}\right)^2}.
\label{eq:rho_resIP}
\eeq
It depends only slowly on $z$. For $\omega\sim\omega\ind{$0$,min}$, the maximum radius $\rho_0 = \rho\ind{res}(0)$ scales like $\sqrt{\omega-\omega\ind{$0$,min}}$. For rf frequencies much larger than $\omega\ind{$0$,min}$, the dependence becomes linear:
$$
\rho_0 \simeq \frac{\omega}{\alpha} \quad\quad\mbox{for}\quad \omega\gg\omega\ind{$0$,min}.
$$
Tuning the frequency is thus a natural and efficient way to increase the radius of the isomagnetic surface.

\subsubsection{Circular rf polarization in the Ioffe-Pritchard case}
\label{sec:IP_trap_circ_pol}
We consider here a circular rf polarization along $z$, in a direction chosen according to the sign of $g_F$, $\epsgras = -\frac{1}{\sqrt{2}}\left(\signgf\GG{e}_x+i\GG{e}_y\right)$. This choice corresponds to a maximum coupling on the $z$ axis of the trap.
Defining $\Omega_0$ as the maximum coupling, corresponding to $\hbar\Omega_0 = \sqrt{2}|g_F|\mu_B |\mathcal{B}_1|/\hbar$,  we deduce the local coupling from Eq.~\eqref{eq:local_coupling_circular_polarization}
\bea
|\Omega_1(\GG{r})| &=& \frac{\Omega_0}{2}\left(1+u_z\right) = \frac{\Omega_0}{2}\left[1+\frac{B_z(z)}{B_0(\rho,z)}\right] \nonumber\\
&=& \frac{\Omega_0}{2}\left[1+\frac{\omega_0(0,z)}{\omega_0(\rho,z)}\right].
\eea

The rf coupling reaches its maximum $\Omega_{\rm max}$ on the $z$ axis, and is reduced as $\rho$ increases. The coupling does not depend on the polar angle around $z$, and the rotational invariance of the Ioffe-Pritchard potential is preserved. On the resonance surface $\omega_0=\omega$, the rf coupling is
$$
|\Omega_1(\rho\ind{res}(z),z)| = \frac{\Omega_0}{2}\left[1+\frac{\omega_0(0,z)}{\omega}\right].
$$

\begin{figure}[t]
\begin{center}
\includegraphics[width=0.47\linewidth]{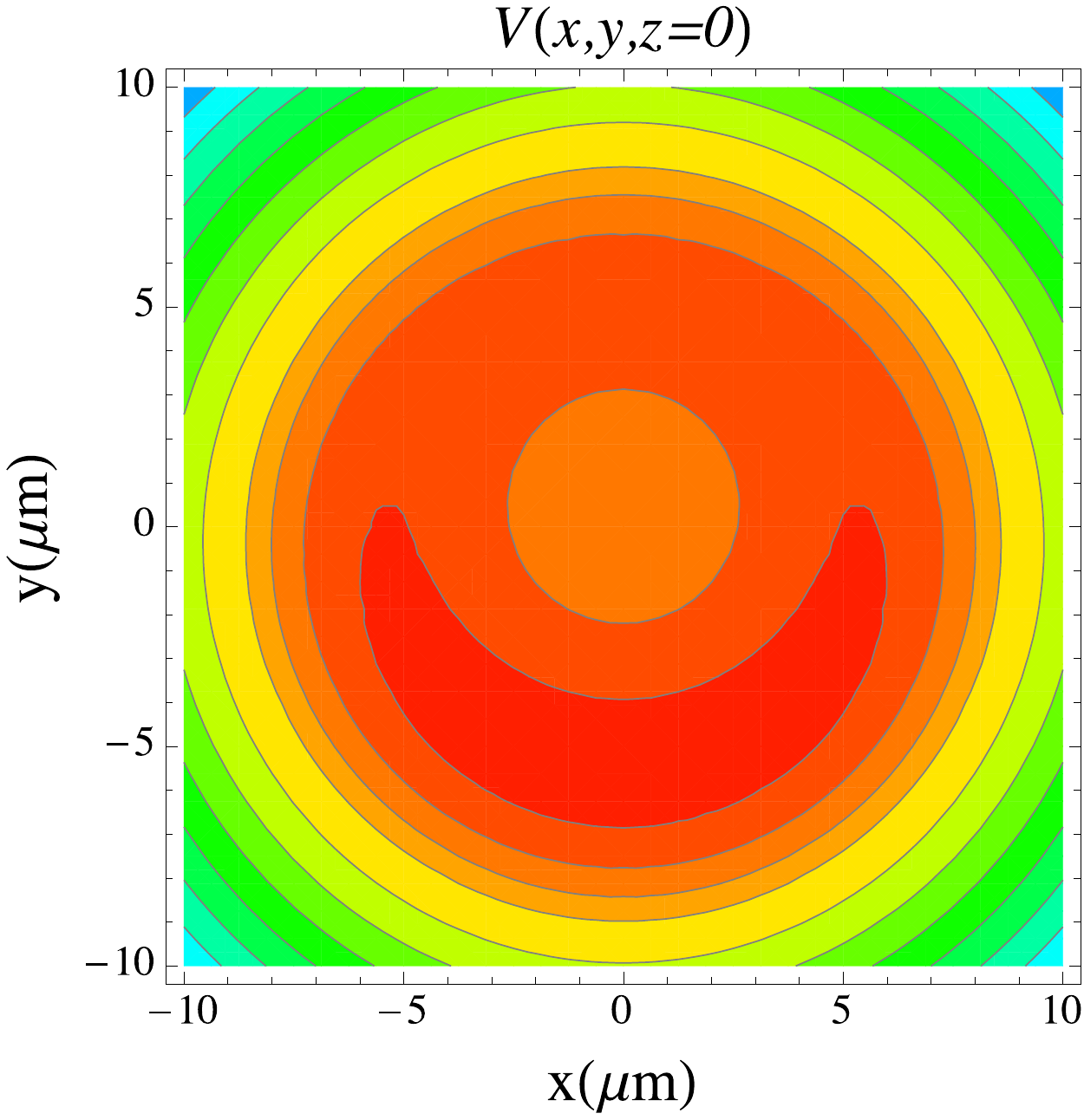}\quad
\includegraphics[width=0.47\linewidth]{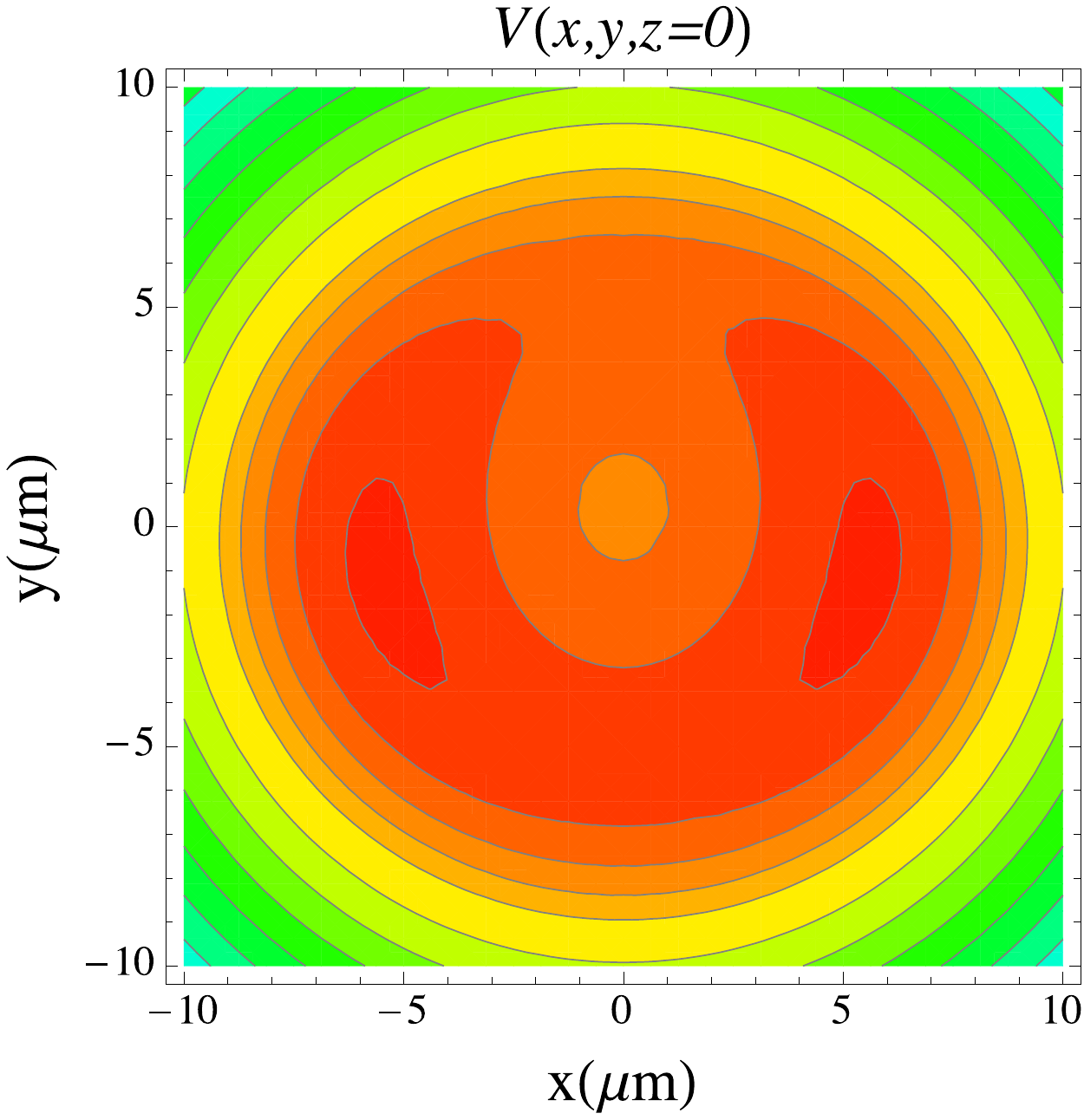}
\caption{Isopotential lines in the `ring' trap (left) and the double well trap (right), seen in the $xy$ plane. Gravity is along $y$ and is taken into account. These traps are elongated in the $z$-direction leading to a tubular trap on the left and two long traps on the right. Parameters for both plots: $\omega\ind{$0$,min}/(2\pi)=1000$~kHz, $\alpha/(2\pi)=100$~kHz$\cdot\mu$m$^{-1}$, 
$\omega/(2\pi)=1120$~kHz, $\Omega_0/(2\pi)=200$~kHz, $F=1$. Left: circular polarisation. Right: linear polarisation.
}
\label{fig:double_well}
\end{center}
\end{figure}

The bubble geometry described at \secref{sec:bubble} is not much changed, see \figref{fig:double_well}, left. Because of the reduction of the coupling with increasing $\rho$, the potential minimum is simply shifted very slightly to a larger radius with respect to the resonant radius $\rho\ind{res}(z)$. Along $z$, on the resonance surface, the minimum coupling is obtained for $z=0$.

In the presence of gravity, the trap minimum is at the bottom of the resonant surface, as was the case in the first demonstration \citep{Colombe2004a}. If however one can manage to compensate for gravity, the region of minimum total potential is ring shaped, or more precisely tubular, because it is very elongated along $z$~\citep{Lesanovsky2006a}. Such a gravity compensation can be approximately obtained by using a position varying rf amplitude, as is the case with atom chips producing large magnetic gradients, including for the rf field, as recently demonstrated \citep{Kim2016}.

\subsubsection{Linear rf polarization}
\label{sec:IP_polar_lin}
The situation is quite different for a linear polarization. If gravity is along the $y$ axis, let us consider a rf field polarised along $x$: $\epsgras = \GG{e}_x$. From Eq.~\eqref{eq:local_coupling_linear_polarization}, we get for the local coupling
\bea
|\Omega_1(\GG{r})|&=& \Omega_0\sqrt{1-u_x^2}= \Omega_0\left[1-\frac{b'^2x^2}{B_0^2}\right]^{1/2} \nonumber\\
&=& \Omega_0\left[1-\frac{\alpha^2x^2}{\omega_0(\rho,z)^2}\right]^{1/2}
\eea
up to second order in the coordinates. $\Omega_0 = |g_F|\mu_B |\mathcal{B}_1|/\hbar$ is the maximum coupling, reached in the $x=0$ plane. Note that, for identical rf power, it is smaller than the maximum in the case of a circular polarization by a factor $\sqrt{2}$. The coupling is lower on the $x$ axis than on the $y$ axis, by a factor $B_z/B_0$. On the resonant surface $\rho= \rho\ind{res}(z)$, the effective Rabi coupling  is thus
$$
|\Omega_1| = \Omega_0\left[1-\frac{\alpha^2x^2}{\omega^2}\right]^{1/2}, \quad |x|\leq \rho\ind{res}(z).
$$
$|\Omega_1|$ is minimum for $y=0$ and $x=\rho\ind{res}(z)$, for which $|\Omega_1| = \Omega_0\,\omega_0(0,z)/\omega$. The absolute coupling minimum within the resonant surface is thus $|\Omega_1| = \Omega_0\,\omega\ind{$0$,min}/\omega$, obtained at the two equatorial positions $(\pm\rho_0,0,0)$ where $\rho_0=\rho\ind{res}(z=0)=\alpha^{-1}\sqrt{\omega^2-\omega\ind{$0$,min}^2}$.

Let us look for the potential minimum inside the resonant surface. Because of the reduced coupling at these points, in the absence of gravity, the potential has two minima, located at $x=\pm \rho_0, y=0,z=0$, see \figref{fig:double_well}, right. The energy at these points is $F\hbar\Omega_0 \sqrt{1 - (\alpha\rho_0/\omega)^2} = F\hbar\Omega_0\,\omega\ind{$0$,min}/\omega$, to be compared to the highest energy points at $x=0,y=\pm \rho\ind{res}(z)$ where the energy is $F\hbar\Omega_0$. The energy difference due to the inhomogeneity in the rf coupling is equal to $F\hbar\Omega_0 \left(1-\omega\ind{$0$,min}/\omega\right)$.

\begin{figure}[t]
\begin{center}
\includegraphics[width=0.23\linewidth,height=0.23\linewidth]{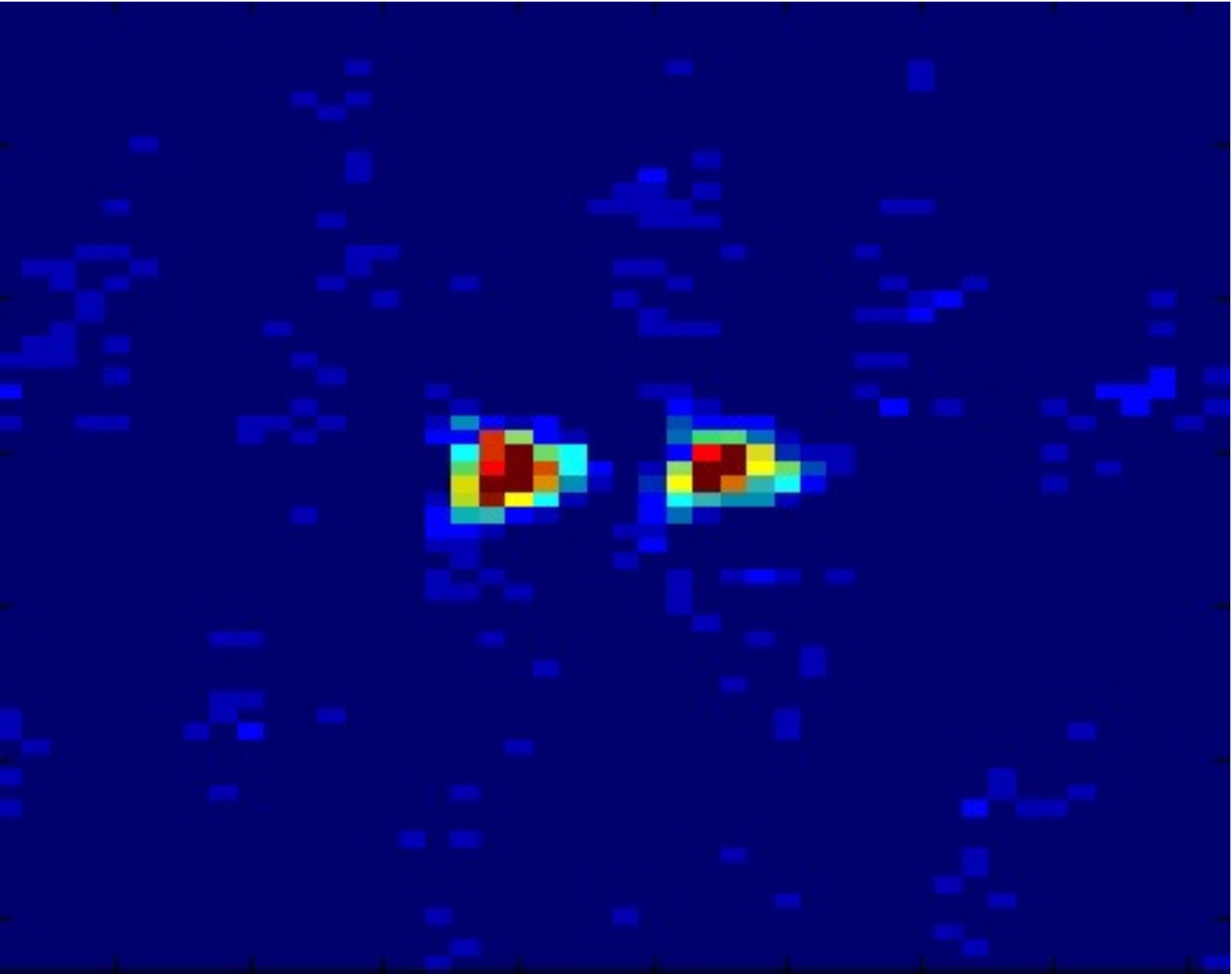}\,
\includegraphics[width=0.23\linewidth,height=0.23\linewidth]{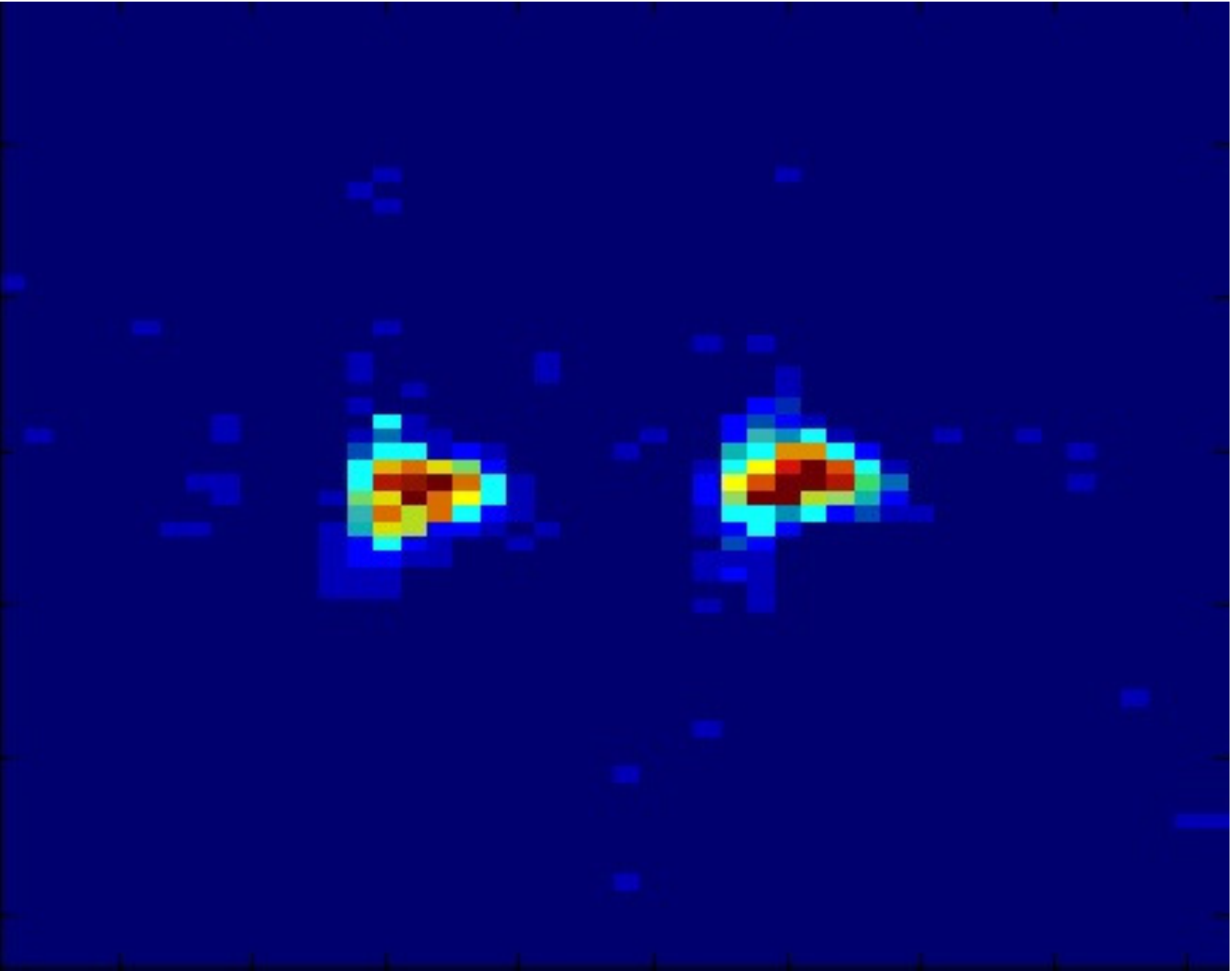}\,
\includegraphics[width=0.23\linewidth,height=0.23\linewidth]{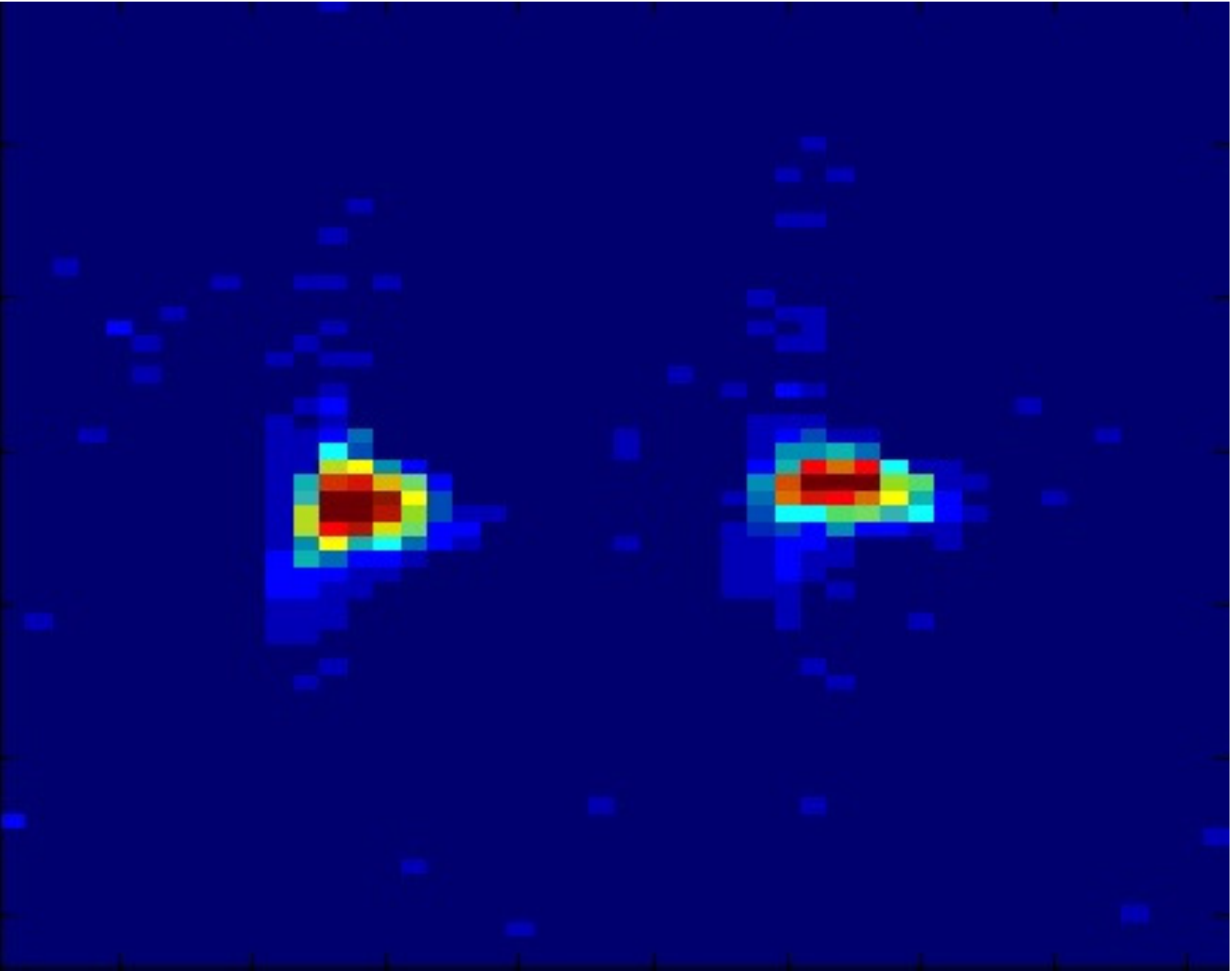}\,
\includegraphics[width=0.23\linewidth,height=0.23\linewidth]{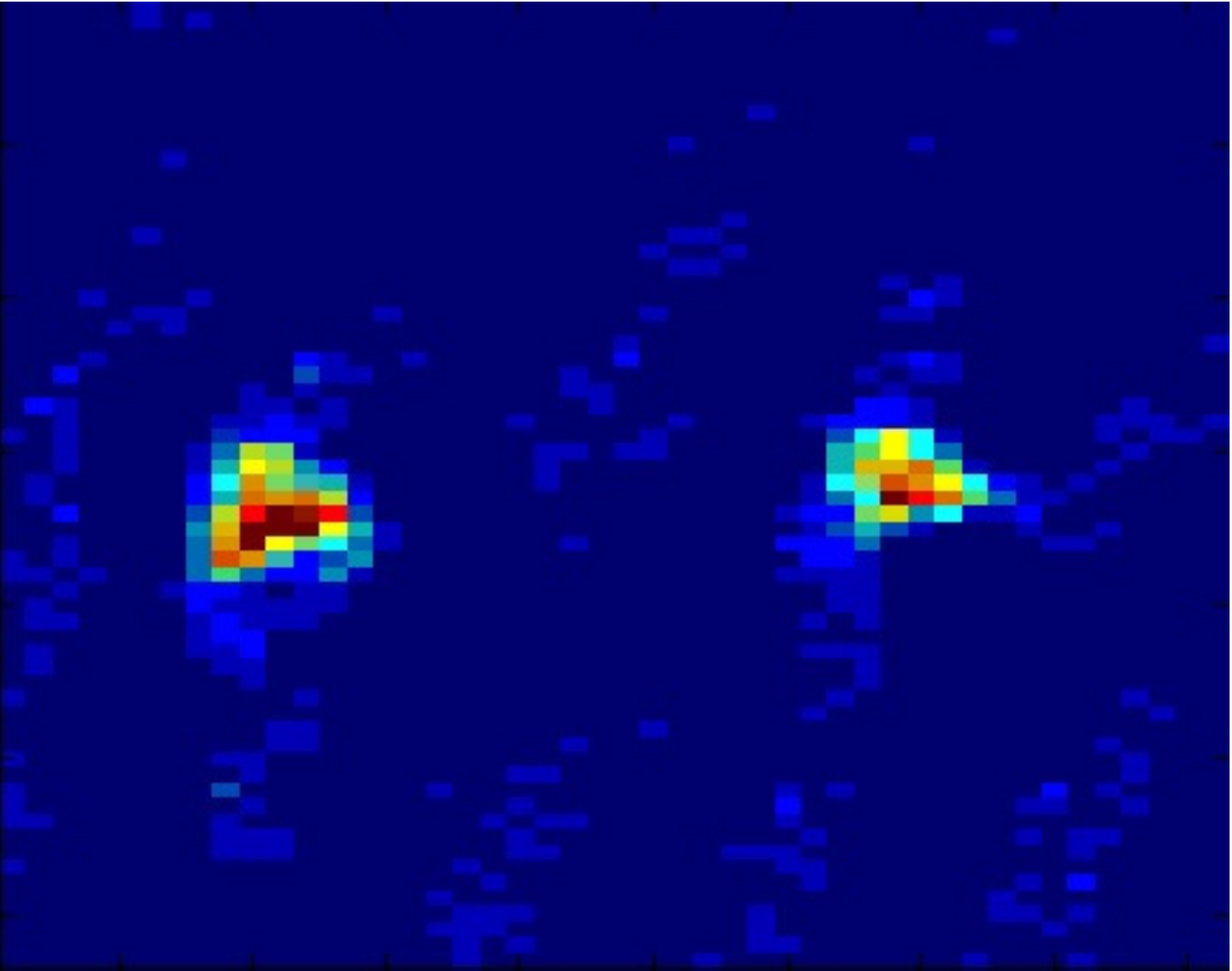}
\caption{First experimental realisation of the rf double well trap~\citep{Schumm2005b}. The distance between two Bose-Einstein condensates in elongated traps is adjusted with the rf frequency. Distance between wells, from left to right: 20 $\mu$m, 36 $\mu$m, 45 $\mu$m, 60 $\mu$m. Unpublished figure, courtesy of Thorsten Schumm.}
\label{fig:Schumm}
\end{center}
\end{figure}

As discussed in \secref{sec:IPisoB}, the resonant radius $\rho_0$ is tuned by changing the rf frequency. For $\omega\sim\omega\ind{$0$,min}$, the maximum radius $\rho_0$ scales like $\sqrt{\omega-\omega\ind{$0$,min}}$. For rf frequencies much larger than $\omega\ind{$0$,min}$, $\rho_0 \simeq \omega/\alpha$. It is then straightforward to adjust at will the distance $2\rho_0$ between the two wells by tuning the rf frequency, which has been done to perform atom interferometry on a chip~\citep{Schumm2005b,Jo2007a}, see \figref{fig:Schumm}, or to realise species selective trapping \citep{Extavour2006}. These double waveguides have also been used to split a cloud in two parts during its propagation \citep{vanEs2008}.

We must now discuss the effect of gravity, which in the case of the tubular potential significantly changes the shape of the minimum if it is not compensated. Here, gravity won't affect qualitatively the double well trap if the gravitational energy difference between the equator and the bottom of the resonant surface is less pronounced than the rf coupling difference, that is if
\beq
Mg\rho_0< F\hbar\Omega_0 \left(1-\omega\ind{$0$,min}/\omega\right).
\label{eq:cond_gravity}
\eeq
To discuss if this relation is fulfilled, we introduce the parameter $\beta = F\frac{\hbar\alpha}{Mg}$, which is the ratio between the magnetic force and the gravity force. It has to be larger than one for the Ioffe-Pritchard trap to confine atoms against gravity. The condition given by Eq.~\eqref{eq:cond_gravity} can be written using $\beta$ as:
$$
1<\beta\frac{\Omega_0}{\omega}\sqrt{\frac{\omega-\omega\ind{$0$,min}}{\omega+\omega\ind{$0$,min}}}.
$$

The ratio under the square root is always less than one, and reaches one in the limit where $\omega\gg\omega\ind{$0$,min}$. A necessary condition for Eq.~\eqref{eq:cond_gravity} to be fulfilled for all double well distances is thus
\beq
1<\beta\frac{\Omega_0}{\omega}, \quad \mbox{or} \quad \omega<\beta\Omega_0.
\label{eq:condition_beta_IP}
\eeq
If we want to stay in the RWA where $\Omega_0\ll\omega$, the magnetic gradient must be much stronger than gravity, $\beta\gg 1$. This explains why atom chips are ideal devices for these double well traps, as they provide us with large magnetic gradients.

\subsubsection{Trapping away from resonance}
\label{sec:out_of_resonance}
The existence of a resonant surface is not necessary for an adiabatic trapping to occur. If, as is the case in a Ioffe-Pritchard configuration, the magnetic field presents a local minimum, the rf frequency can be less than the Larmor frequency everywhere, provided that $\omega<\omega\ind{$0$,min}$. The magnetic potential in the vicinity of the magnetic minimum is nonetheless modified if $|\omega-\omega\ind{$0$,min}|\lesssim |\Omega_1|$. A double-well potential with a small separation between the wells, enabling tunnel coupling between the wells, has been produced with the Ioffe-Pritchard magnetic field and a linear rf polarization discussed in the previous section but with a frequency below the minimum Larmor frequency \citep{Schumm2005b}.

Let us write the condition to obtain a double-well. The adiabatic potential for the adiabatic state $\ket{m=F}_\theta$ reads
$$
V_F(\GG{r}) = F\hbar\sqrt{(\omega_0(\rho,z) - \omega)^2 + \Omega_0^2\left(1-\frac{\alpha^2x^2}{\omega_0(\rho,z)^2}\right)}.
$$
We ignore gravity here for simplicity. It only has a small effect on the atom chip experiment where the magnetic gradients are large ($\beta\gg 1$). The minima will lie on the $x$ axis by symmetry. Along this axis, the expression of the potential simplifies as
\bea
&&V_F(x) = \\ \label{eq:double_well_non_res}
&&F\hbar\omega\ind{$0$,min}\sqrt{\left(\frac{X^2}{2} - \tilde\delta\ind{min}\right)^2 + \frac{1-X^2/2}{1+X^2/2}\,\tilde\Omega_0^2}\nonumber
\eea
where $X=\alpha x/\omega\ind{$0$,min}$, $\tilde\Omega_0 = \Omega_0/\omega\ind{$0$,min} < 1$ and $\tilde\delta\ind{min} = (\omega-\omega\ind{$0$,min})/\omega\ind{$0$,min} < 0$.

For small values of $X$, the Taylor expansion of the potential is of the form
$$
V_F(x)\simeq V_0 + a_2X^2+a_4X^4+O(X^6).
$$
We can show that $a_4$ is always positive. A double-well occurs as soon as $a_2<0$. As $a_2\propto -(\tilde\Omega_0^2+\tilde\delta\ind{min})$, see Eq.~\eqref{eq:double_well_non_res}, the condition on the rf coupling for a double-well to appear is
$$
\tilde\Omega_0^2>-\tilde\delta\ind{min}\qquad\mbox{or}\qquad \Omega_0^2 > \omega\ind{$0$,min}|\delta\ind{min}|.
$$
The separation between the two wells scales as $\Delta x \propto \sqrt{\Omega_0^2 - \omega\ind{$0$,min}|\delta\ind{min}|}$ and can thus be adjusted by a control on the rf amplitude at fixed detuning.

The trap loading is simplified in this case, as the rf detuning can remain fixed at $\delta\ind{min}<0$, while the rf amplitude is ramped up to its final value. The single initial minimum of the Ioffe-Pritchard trap splits into two minima as the rf amplitude increases and approaches $|\omega-\omega\ind{$0$,min}|$. This method allows in particular to prepare two copies of an initial Bose-Einstein condensate, which can then interfere after a given waiting time, enabling the study of the phase diffusion in elongated BECs with interferometric methods \citep{Schumm2005b,Betz2011}.

\subsection{The dressed quadrupole trap}
\label{sec:dressed_quad}
After the dressed Ioffe-Pritchard trap, another interesting case, which allows very flat traps for 2D quantum gases, is the adiabatic potential obtained from dressing atoms in a quadrupole field~\citep{Morizot2007,Merloti2013a}. We will discuss this trap here.

\subsubsection{Magnetic field geometry in the quadrupole trap}
\label{sec:magnetic-field-quadrupole}

In a quadrupole field, which can be used as a magnetic trap \citep{Migdall1985}, the magnetic field is linear in the position, for example:
\beq
\GG{B}_0(\GG{r}) = b'(x\,\GG{e}_x + y\,\GG{e}_y - 2z\,\GG{e}_z).
\label{eq:quadrupole_field}
\eeq
The factor 2 in the $z$ gradient ensures a vanishing divergence of the magnetic field. The corresponding Larmor frequency is
\beq
\omega_0(\GG{r}) = \alpha\sqrt{x^2+y^2+4z^2},
\label{eq:Larmor_quad_field}
\eeq
where $\alpha = |g_F|\mu_B b'/\hbar$.

The isomagnetic surfaces is this case are ellipsoids, with a radius smaller in the vertical direction by a factor of two. For a given rf frequency $\omega$, the equation of the resonant ellipsoid is
$$
x^2+y^2+4z^2 = r_0^2,
$$
where the ellipsoid horizontal radius is related to the frequency through
\beq
r_0 = \frac{\omega}{\alpha}.
\eeq
In the following, we will use the generalised distance to the centre of the quadrupole,
\beq
\ell(\rho,z) = \sqrt{\rho^2+4z^2},
\eeq
where $\rho$ is the polar coordinate in the $xy$ plane, $\rho\, \GG{e}_\rho = x\,\GG{e}_x + y\,\GG{e}_y$.

The direction of the magnetic field is given by
\beq
\GG{u} = \frac{\rho\, \GG{e}_\rho - 2z\,\GG{e}_z}{\ell(\rho,z)}.
\eeq
The naive potential, forgetting polarization issues, is thus a bubble defined by $\ell(\rho,z)=r_0$, the atoms being attracted to its bottom by gravity. This explains why this configuration is well adapted to the trapping of two-dimensional gases.

\subsubsection{Circular polarization in the quadrupole trap}
The magnetic field at the bottom of the ellipsoid $(0,0,-r_0/2)$ is aligned along the $+z$ axis. It is then natural to consider first a circularly polarised field. 
To maximise the coupling at the bottom, we will hence chose a circular polarization $\sigma^\signgf$ of sign $\signgf$ aligned with $z$:
$$
\epsgras = -\frac{1}{\sqrt{2}}\left(\signgf\,\GG{e}_x + i\GG{e}_y\right).
$$
Using Eq.~\eqref{eq:local_coupling_circular_polarization}, the efficient Rabi component is then
\beq
|\Omega_1(\GG{r})| = \frac{\Omega_0}{2}\left(1+u_z\right) = \frac{\Omega_0}{2}\left(1-\frac{2z}{\ell}\right).
\eeq
$\Omega_0$ is the maximum Rabi frequency, obtained as expected on the negative side of the $z$ axis, where $\ell=-2z$. On the other hand, the effective coupling vanishes on the positive side of the $z$ axis, where $\ell=2z$. The polarization here is $\sigma^{-\signgf}$ with respect to the local orientation of the magnetic field, which points downwards. This results in a half axis of zero coupling. The trap minimum must lie away from this $+z$ axis, in order to prevent spin flips.

The total potential, including gravity, in the extreme adiabatic state $\ket{m=F}_\theta$ is
\beq
V_F(\GG{r}) = F\hbar\sqrt{\left[\alpha\ell(\rho,z)-\omega\right]^2+\frac{\Omega_0^2}{4}\left[1-\frac{2z}{\ell(\rho,z)}\right]^2}+Mgz.
\eeq
We can immediately notice that the potential is rotationally invariant around $z$, and depends only on $\rho$ and $z$. The first term will be minimum when both the detuning term and the coupling term vanish, at the point $(0,0,r_0/2)$. However, this is the point of the resonant surface where gravity is maximum, and gravity will help bringing the atoms to the bottom of the ellipsoid, away from the region of vanishing Rabi frequency.

Again, it is reasonable to assume that the minimum will lie on the resonance surface, which allows one to set $\delta$ to zero. Let us write the value of the potential for atoms living on the two-dimensional resonance surface $\ell(\rho,z)=r_0$. It depends only on $z$ now, as $\rho$ is given by $z$ and $r_0$:
$$
V_{F,\rm surf}(z) = F\hbar\frac{\Omega_0}{2}\left(1-\frac{2z}{r_0}\right) + Mgz, \quad|z|\leq\frac{r_0}{2}.
$$
The potential on the surface is linear in $z$.

From this expression, it is clear that there will be a competition between the gradient of rf coupling and gravity. Let us use the same parameter $\beta = F\frac{\hbar\alpha}{Mg}$ as in \secref{sec:IP_polar_lin}. It has to be larger than $1/2$ if the quadrupole field itself is supposed to confine atoms against gravity. If $Mgr_0>F\hbar\Omega_0$, that is for $\omega>\beta\Omega_0$, gravity is dominant and the minimum is at $z=-r_0/2$, at the bottom of the ellipsoid where the coupling is maximum. On the other hand, if $\omega<\beta\Omega_0$, the atoms are pushed upwards to the point of the ellipsoid where the coupling vanishes, and the trap is unstable with respect to Landau-Zener losses. For a given coupling, this sets a minimum frequency which should be used:
\beq
\omega > \beta\Omega_0.
\label{eq:cond_bottom}
\eeq
In contrast to the requirement for a double well, this is now easily compatible with RWA, which requires $\omega\gg\Omega_0$. The trap is thus normally at the bottom of the ellipsoid, even for moderate gradients.

\subsubsection{Isotropic trap for a 2D gas  using quadrupole fields}
\label{sec:2Dtrapcirc}
Let us assume that we have indeed $\omega > \beta\Omega_0$. In the vicinity of the bottom of the resonant surface, we can develop the full potential to find the oscillation frequencies. In the vertical direction, the trap is similar to the radial direction of a Ioffe Pritchard magnetic trap:
\bea
V_F(0,0,z) &=& F\hbar\sqrt{\alpha^2\left(r_0+2z\right)^2 + \Omega_0^2} + Mgz\nonumber\\
&\simeq& F\hbar\Omega_0 + Mgz + F\hbar\frac{4\alpha^2}{2\Omega_0}\left(z+\frac{r_0}{2}\right)^2.\nonumber
\eea
The vertical oscillation frequency is thus \citep{Merloti2013a}
\bea
\omega_z &=& 2\alpha \sqrt{\frac{F\hbar}{M\Omega_0}}\left(1-\frac{1}{4\beta^2}\right)^{3/4},\label{eq:nuz}\\
\omega_z &\simeq& 2\alpha\sqrt{\frac{F\hbar}{M\Omega_0}}.
\label{eq:nuz_approx}
\eea
The $1/\beta^2$ correction arises because the minimum does not strictly belong to the resonant surface, due to gravity~\citep{Merloti2013a}. The analogy with the Ioffe Pritchard trap is immediate: the Larmor frequency is just replaced by the Rabi frequency.

In the horizontal direction, the trap is isotropic. The oscillation frequency is imposed by the geometry of the ellipsoid and by gravity: the motion is pendulum-like, with an oscillation frequency in the harmonic approximation of order $\sqrt{g/(2r_0)}$. More precisely, defining the absolute value of the vertical equilibrium position as
\beq
R = \frac{r_0}{2}\left( 1 + \frac{1}{\sqrt{4\beta^2-1}}\frac{\Omega_0}{\omega} \right),
\label{eq:eqpos}
\eeq
the expression of the horizontal frequency is \citep{Merloti2013a}
\bea
\omega_\rho &=& \sqrt{\frac{g}{4R}}\left[ 1 - \frac{F\hbar\Omega_0}{2Mg R}\sqrt{1-\frac{1}{4\beta^2}} \right]^{1/2} ,\label{eq:nurho}\\
\omega_\rho  &\simeq& \sqrt{\frac{g}{2r_0}}\left[ 1 - \frac{\beta\Omega_0}{\omega}\right]^{1/2}.\label{eq:nurho_approx}
\eea
The correction in $\beta\Omega_0/\omega$ comes from the vertical dependence of the Rabi frequency. The exact expression Eq.~\eqref{eq:nurho} is slightly affected by the gravitational sag. The two values given by Eqs.~\eqref{eq:nuz_approx} and \eqref{eq:nurho_approx} however are a very good estimate of the oscillation frequencies. The trap is very anisotropic, and the aspect ratio is approximately
\beq
\frac{\omega_z}{\omega_\rho} \simeq 2\alpha\sqrt{\frac{2F\hbar r_0}{Mg\Omega_0}} = 2\sqrt{\frac{2F\hbar\alpha\omega}{Mg\Omega_0}} = 2\sqrt{\frac{2\beta\omega}{\Omega_0}}.
\eeq
It is bounded from above and from below:
$$
2\sqrt{2}\,\beta < \frac{\omega_z}{\omega_\rho} < 2\sqrt{2}\,\frac{\omega}{\Omega_0}.
$$
The last inequality is the condition Eq.~\eqref{eq:cond_bottom} for the trap minimum to be at the bottom of the ellipsoid. We see here that as soon as the underlying quadrupole trap compensates gravity, the trap is naturally anisotropic. Moreover, if the rf field fulfils the RWA $\omega\gg\Omega_0$, the upper bound is very high. In a regime where the aspect ratio reaches about 100, this trap has been used to prepare quasi two-dimensional quantum gases~\citep{Merloti2013a}.

\subsubsection{Linear polarization}
If the polarization is chosen to be linear, the best choice is a horizontal polarization, to ensure a maximum coupling at the bottom. Let us chose a polarization $\GG{e}_x$ along the $x$ axis. From Eq.~\eqref{eq:local_coupling_linear_polarization}, the rf coupling is
$$
|\Omega_1(\GG{r})| = \Omega_0\sqrt{1-u_x^2} = \Omega_0\sqrt{1-\frac{x^2}{\ell^2}}.
$$
$\Omega_0$ is defined as the maximum coupling.
If we again assume that the potential minimum lies on the resonance surface, we end up with
\bea
V_{F,\rm surf}(\GG{r}) = F\hbar\Omega_0\sqrt{1-\frac{x^2}{r_0^2}} + Mgz, \nonumber\\
|z|\leq\frac{r_0}{2}, \quad |x|\leq r_0, \quad \ell(\rho,z)=r_0.\nonumber
\eea

Now, the rotational symmetry is broken by the choice of the rf polarization. Two `holes' with a vanishing coupling appear on the resonant surface, at positions $(\pm r_0,0,0)$ opposite on the equator of the resonant ellipsoid. Spin-flips can occur if the atoms reach these points \citep{Morizot2007}, which can however be prevented thanks to gravity.

Indeed, there is again a competition between gravity and the coupling gradient, in the $y=0$ plane where $V\ind{surf}(z) = 2F\hbar\Omega z/r_0 + Mgz$. Gravity wins if $\omega>2\beta\Omega_0$. This condition is almost the same as in the previous case, and is fulfilled within the RWA for moderate magnetic gradients. For large magnetic gradients, the rf frequency must be increased to fulfil the condition on gravity.

The vertical oscillation frequency at the trap bottom is unchanged with respect to the previous case of Eq.~\eqref{eq:nuz}. As the coupling strength is now uniform in the $yz$ plane, the oscillation frequency along $y$ is the bare pendulum frequency:
\beq
\omega_y =  \sqrt{\frac{g}{4R}} \simeq \sqrt{\frac{g}{2r_0}}.
\eeq

The $x$ oscillation frequency is lowered by the attraction to the holes \citep{Merloti2013a}:
\bea
 \omega_x &=& \sqrt{\frac{g}{4R}}\left[ 1 - \frac{F\hbar\Omega_0}{Mg R}\sqrt{1-\frac{1}{4\beta^2}} \right]^{1/2} \\
 &\simeq& \sqrt{\frac{g}{2r_0}}\left[ 1 - \frac{2\beta\Omega_0}{\omega}\right]^{1/2}.\nonumber
\eea
This effect is twice as large as in the circularly polarised case, because the holes are at half the height. The condition on gravity is clear from the expression of the oscillation frequency, which would vanish at the limit $\omega=2\beta\Omega_0$ where the single minimum at the bottom disappears, and the two minima at the equator appear --- with zero coupling and strong spin flips.

Here, the important point is that, while the trap is still extremely aniso\-tropic in the vertical versus horizontal directions, it also becomes anisotropic in-plane, with an approximate aspect ratio $\sqrt{1 - 2\beta\Omega_0/\omega}$ controlled by the rf amplitude, opening the way to trapping two-dimensional anisotropic gases. The direction of this anisotropy is controlled by the polarization axis. In short, modulating the direction of the polarization axis or the rf amplitude allows one to set the gas into rotation or to excite quadrupole oscillations, respectively. This has been used recently to study collective modes of a two-dimensional superfluid and probe its superfluidity, with an anisotropy of about 4/3~\citep{Dubessy2014,MerlotiThese,DeRossi2016}. The smooth character of the adiabatic potential~\citep{vanEs2008,Merloti2013a} is very well suited for such studies.

\begin{figure}[t]
\begin{center}
\includegraphics[width=0.5\linewidth]{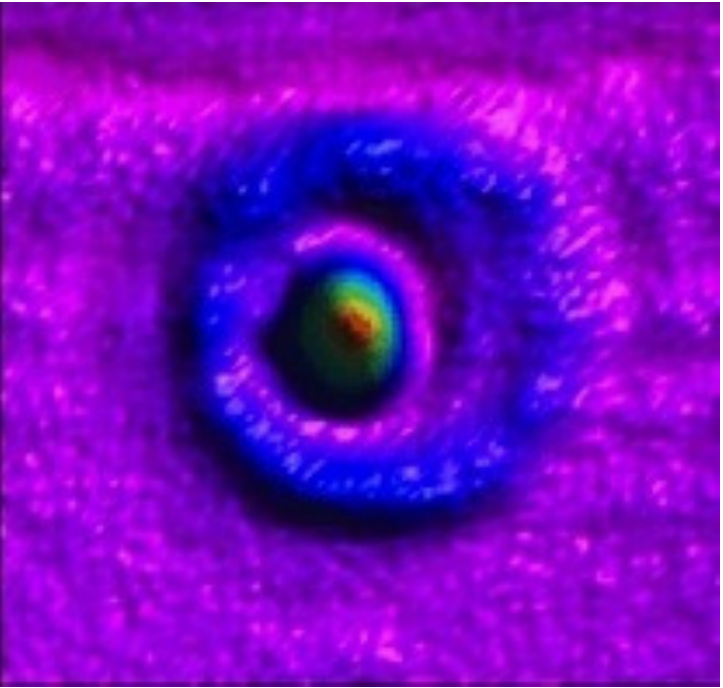}
\caption{Rubidium atoms confined in an annular adiabatic potential analogue to \figref{fig:double_well}, left, but with a negative detuning. The central spot corresponds to atoms trapped at the minimum of the static magnetic field in the adiabatic states with $m_\theta<0$. The experimental parameters are: $\omega\ind{$0$,min}/2\pi=700$~kHz, $\alpha=75$~kHz$\cdot\mu$m$^{-1}$, $\omega/(2\pi)=650$~kHz, $\Omega_0/(2\pi)\simeq 550$~kHz, $F=2$. Note that the rf amplitude is beyond the RWA regime. Picture size: $4\times 4~\mu$m$^2$. Figure adapted with permission from \citep{Kim2016}. Copyrighted by the American Physical Society.}
\label{fig:ring_trap_Dana}
\end{center}
\end{figure}

How much can we increase the in-plane anisotropy while keeping a strong vertical-horizontal anisotropy necessary for a 2D gas? We remark that the two aspect ratios $\omega_z/\omega_y$ and $\omega_x/\omega_y$ are related:
\bea
\frac{\omega_y}{\omega_z} &\simeq& \sqrt{\frac{\Omega_0}{8\beta\omega}}\nonumber\\
 \quad\Rightarrow \quad\frac{\omega_x}{\omega_y} &\simeq& \left[ 1 - \frac{2\beta\Omega_0}{\omega}\right]^{1/2} \simeq \left[ 1 - 16\beta^2\left(\frac{\omega_y}{\omega_z}\right)^2\right]^{1/2}.\nonumber
\eea
For an important in-plane anisotropy, $\omega_y/\omega_z$ should approach $1/(4\beta)$. In order to stay in the 2D regime, which requires in particular a strong anisotropy $\omega_z\gg\omega_y$, very large gradients $4\beta\gg 1$ are necessary. Experimentally, the maximum in-plane anisotropy which can be obtained in a 2D gas thus depends on the maximal feasible gradient.

\subsection{Ring traps}
\label{sec:ring-trap-quad}
A proper choice of inhomogeneous static and rf fields can lead to an annular confining potential, as proposed by \cite{Lesanovsky2006a} and demonstrated by \cite{Kim2016} with an atom chip in the regime of negative detuning ($\omega<\omega\ind{$0$,min}$) as in \secref{sec:out_of_resonance}: see \figref{fig:ring_trap_Dana}. This geometry has been discussed briefly at \secref{sec:IP_trap_circ_pol}, and is presented in more detail in \cite{Garraway2016}. The gravitational sag can be compensated by a gradient in the rf amplitude. This trap can provide a strong radial annular confinement (with frequencies in the kHz range), as it relies on the confinement to a resonant surface. The axial confinement is typically weaker, and the resulting geometry is tubular.

\begin{figure}[t!]
\begin{center}
\subfigure[]{\includegraphics[width=0.32\linewidth]{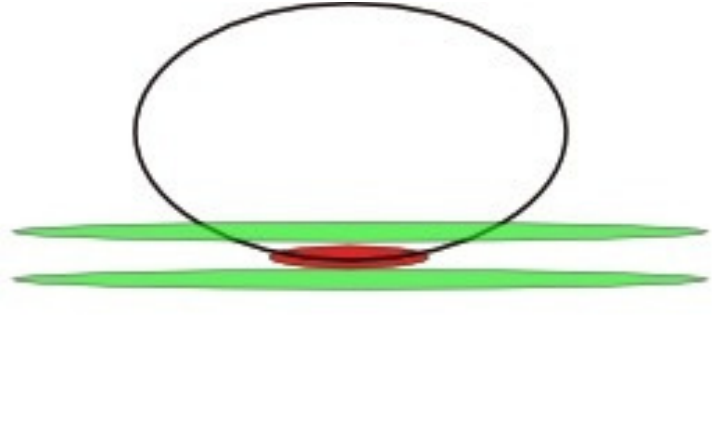}}
\subfigure[]{\includegraphics[width=0.32\linewidth]{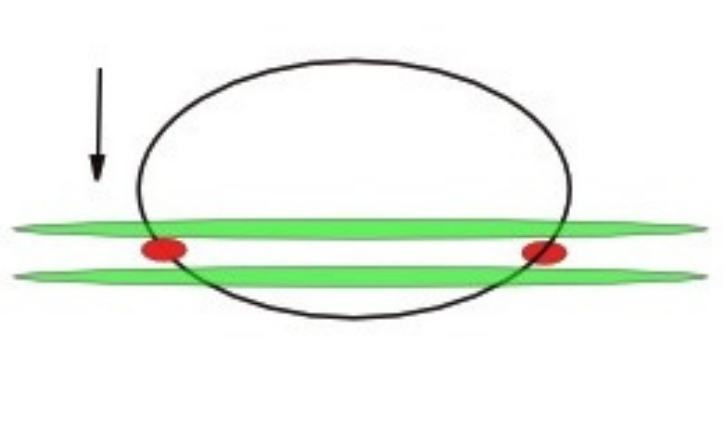}}
\subfigure[]{\includegraphics[width=0.32\linewidth]{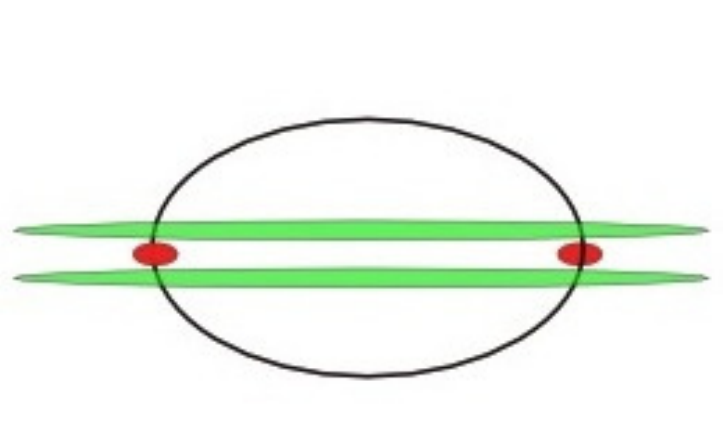}}\\
\subfigure[]{\raisebox{-0.028\linewidth}{\includegraphics[width=0.445\linewidth]{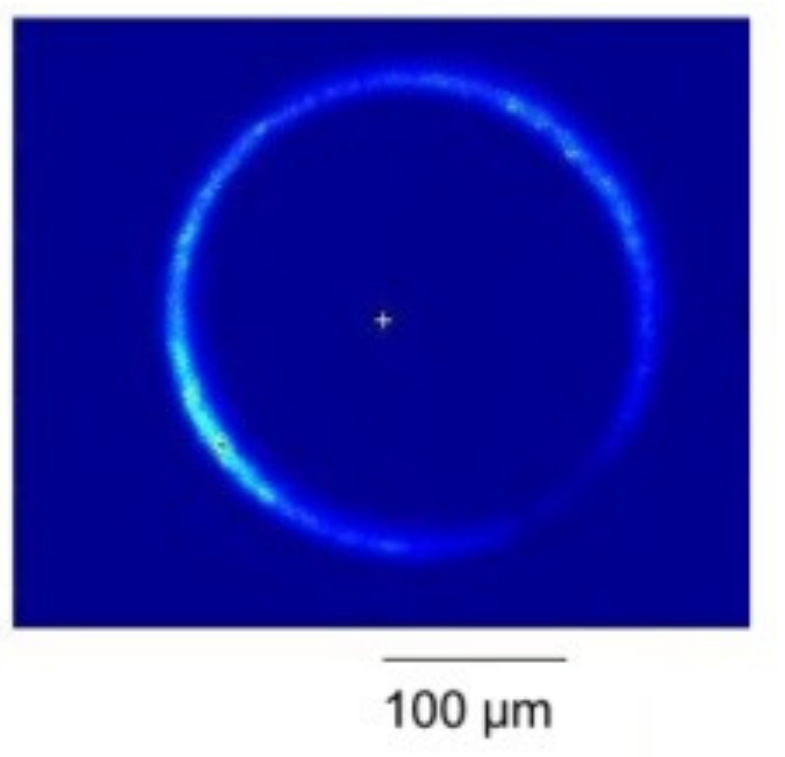}}}
\subfigure[]{\includegraphics[width=0.4\linewidth]{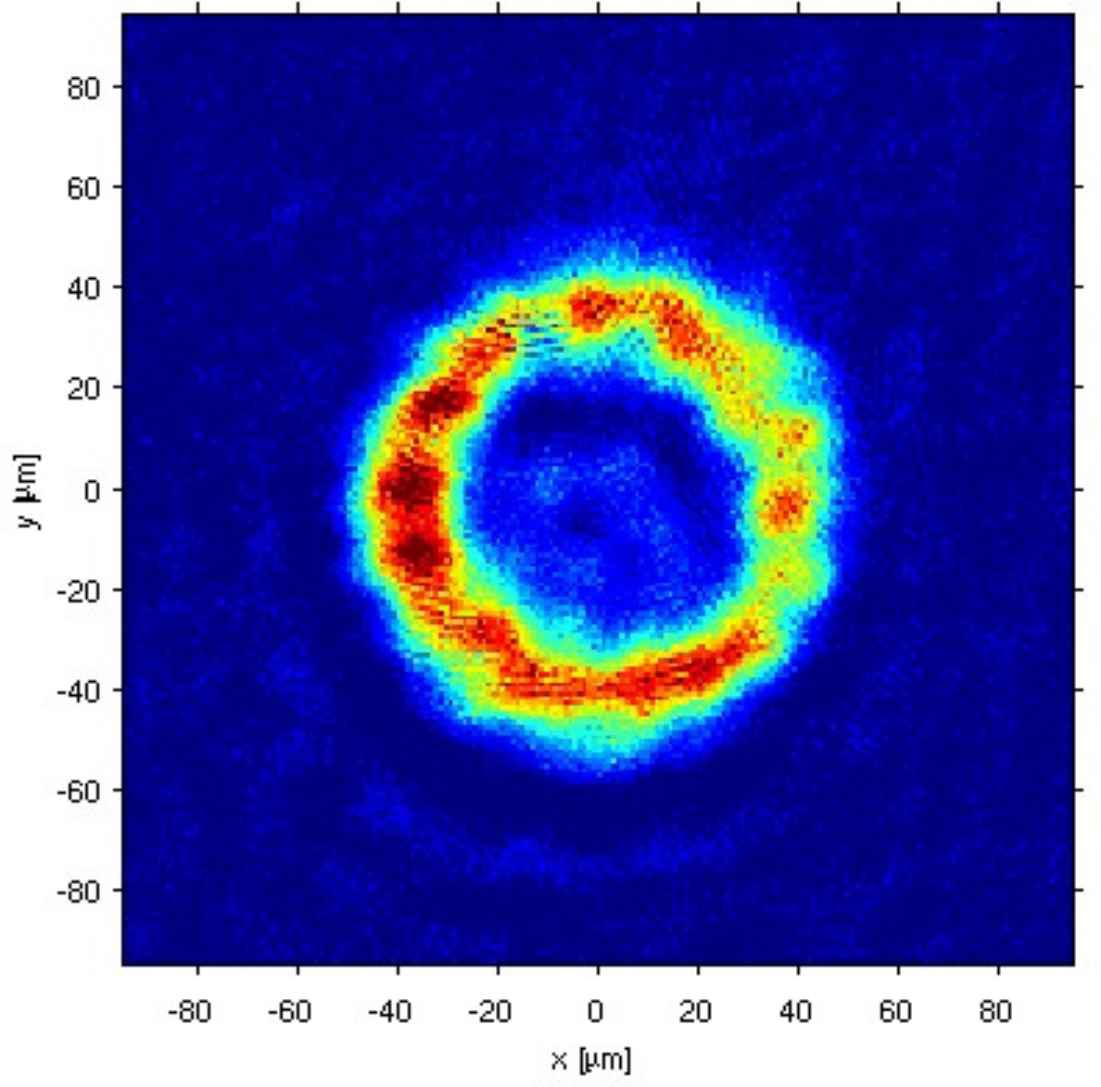}}
\caption{Top row: Loading of a ring trap starting from atoms confined to a dressed quadrupole trap. (a) The two blue-detuned light sheets (in green) are switched on in order to confine the initial oblate atomic cloud (in red) between the two light maxima. (b) The resonant surface in then translated downwards thanks to an additional static magnetic field aligned along the vertical axis. The atoms stay at fixed height due to the light sheets and a ring forms. (c) The maximum ring radius -- and maximum radial trapping frequency -- is obtained when the atoms are in the equatorial plane. Bottom row: atoms confined in the ring trap. (d) Radius 130~$\mu$m obtained for a low magnetic gradient of $b'=55.4$~G$\cdot$cm$^{-1}$ and a rf frequency of 1~MHz. (e) Radius 40~$\mu$m corresponding to a larger magnetic gradient $b'=218$~G$\cdot$cm$^{-1}$ and a dressing rf frequency of 600~kHz (image LPL). (a-d): Figure adapted from~\citep{DeRossiThese}, reprinted by courtesy of the author.\label{fig:ring_loading}}
\end{center}
\end{figure}

In order to increase the axial confinement and get a ring with comparable trapping frequencies in all directions, adiabatic potentials can be combined with optical dipole potentials \citep{Grimm2000} which widens the range of possible trapping geometries. Starting from a dressed quadrupole trap with a circular polarization, see \secref{sec:2Dtrapcirc}, and taking advantage of the strong confinement to the ellipsoidal resonant surface, a ring trap is obtained by cutting the ellipsoid by a horizontal plane. The confinement to a given horizontal plane can be done with a standing laser wave~\citep{Morizot2006} or of a pair of blue detuned light sheets~\citep{Heathcote2008}, as illustrated in \figref{fig:ring_loading}a-c where a possible loading mechanism is sketched.

The ring radius is easily adjustable dynamically, for example with the rf frequency $\omega$ or the magnetic gradient of the static quadrupole field (see \figref{fig:ring_loading}d-e). If the confinement plane is at the equator of the ellipsoid, the radial confinement is purely due to the adiabatic potential and is given by Eq.~\eqref{eq:nuz_approx} (provided that the vertical gradient $2\alpha$ is replaced by the horizontal gradient $\alpha$), while the vertical confinement is purely set by the optical trap, allowing an independent tuning of these two parameters.

A circular polarization should be used to ensure the rotational symmetry of the adiabatic potential. On the other hand, using a non circular polarization, with an additional component on $x$ for example, induces a deformation of the ring along $x$. Rotating the direction of this additional rf component allows one to induce a rotation of the atoms in the ring and create a superflow \citep{Heathcote2008}.

\section{Time-averaged adiabatic potentials}
\label{sec:taaps}

Time-averaged potentials have ben used for some time in atomic physics. An example, in a field which is close to this review, is the Time Orbiting Potential, otherwise known as a TOP trap \citep{Petrich1995}. This is a pure magnetic trap (with no rf radiation) which is often based on a quadrupole configuration. The quadrupole field (e.g.\ $[b'x,b'y,-2b'z]$, \secref{sec:magnetic-field-quadrupole}) has a field zero at its centre where the spin direction is undefined. As a result atoms are easily lost from the trap. This problem can be solved for the magnetic trap by adding an additional potential (e.g.\ an optical `plug' using different, light induced forces, as done by \cite{Davis1995,Naik2005,Dubessy2012a}). Alternatively, the time-orbiting potential approach can be used to make the TOP trap. In this case the centre of the quadrupole field is rotated about another point by using a rotating bias field. Provided the rotation of the field is fast compared to the atomic dynamics, the atoms `see' an average field: the resulting time-averaged potential is a harmonic trapping potential. This requires that the harmonic trap frequency is less than the TOP rotation frequency which is less than the Larmor frequency $\omega_0$, Eq.~\eqref{eq:Larmor}.

The same approach can be used with adiabatic traps, with the only major complication being that although the atoms should experience a time-averaged potential, requiring fast changes in the potential, the changes should not be so fast as to cause non-adiabatic transitions. The first example was proposed in a theoretical paper by \cite{Lesanovsky2007b} who proposed a double-well and a ring trap by this method.

\begin{figure}[t]
  \centering
  \includegraphics[width=\linewidth]{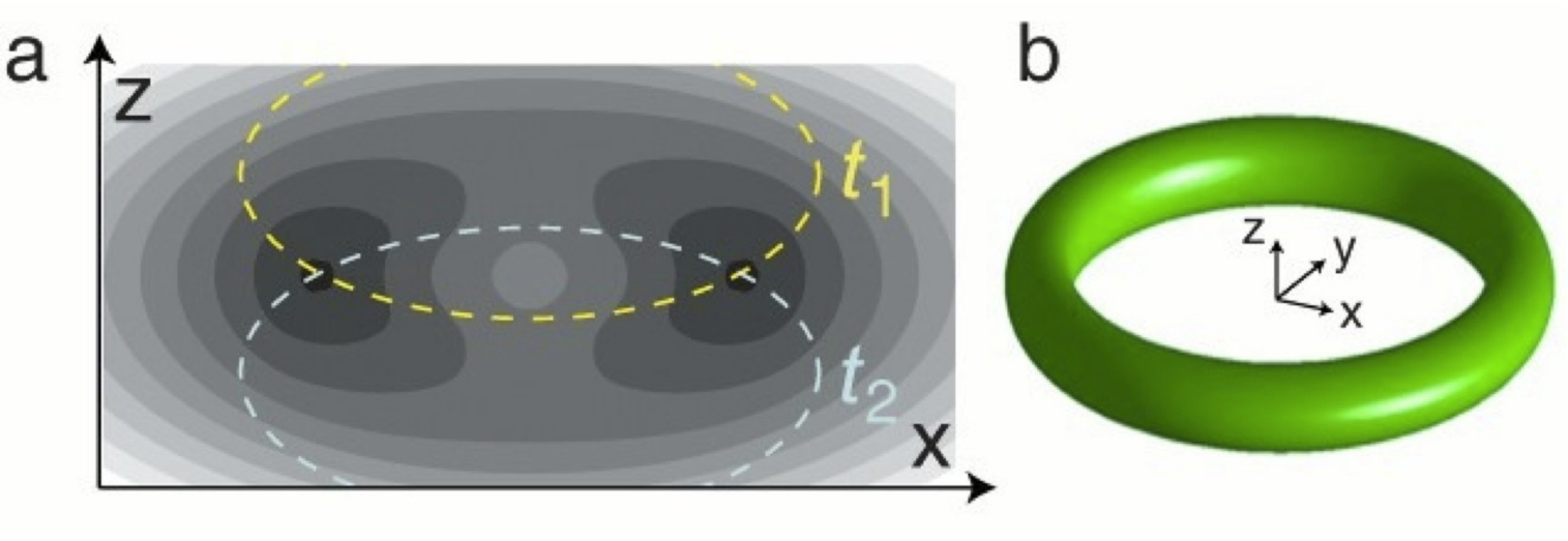}
  \caption{Use of time-averaged adiabatic potential, or TAAP, to change a bubble potential to a ring potential: (a) Vertical slice through the time-averaged ring potential with the averaged adiabatic potential shown with contours. Two extreme positions of the instantaneous adiabatic potential are shown with the dashed lines which indicate the instantaneous minimum in that potential. (b) A 3D visualisation of an equipotential shown in section in (a). From \citep{Lesanovsky2007b}. Copyrighted by the American Physical Society.}
  \label{fig:TAAP-Klitzing}
\end{figure}

Although several different types of time-averaged adiabatic potentials (TAAP) are possible, and have now been realised \citep{Gildemeister2010,Sherlock2011,Gildemeister2012,Navez2016}, we will here give one of the original proposals of Lesanovsky and von Klitzing as a particular example.
We have already seen in \secref{sec:bubble} how resonant dressing with a quadrupolar field tends to make a `bubble' potential with the atoms trapped on the resonance surface. Lesanovsky and von Klitzing add an oscillating bias field which shifts the centre of the bubble up and down (see \figref{fig:TAAP-Klitzing}). In this example they also oscillate the rf frequency and the strength of the rf to achieve the ring trap potential after time-averaging over all these additional oscillations.

We start by imposing an oscillating bias field on the quadrupole trap of \secref{sec:magnetic-field-quadrupole}, that is on Eq.~\eqref{eq:quadrupole_field} so that 
\begin{equation}
  \label{eq:BG:TAAP-osc-bias-field}
  \GG{B}_0(\GG{r},t) = b'(x\,\GG{e}_x + y\,\GG{e}_y - 2z\,\GG{e}_z )
   +  B_m \sin (\omega_m t) \GG{e}_z 
\,,
\end{equation}
where $B_m$ is the amplitude of the oscillating (modulated) field which is taken to have a vector direction along $z$ and an oscillation frequency $\omega_m$. The immediate effect of this oscillation is that the centre of the quadrupole field can be regarded as oscillating up and down with a new centre given by
\begin{equation}
  \label{eq:BG:TAAP-z0}
  z_0(t) =  \frac{ B_m }{ 2 b' } \sin (\omega_m t)
\,.
\end{equation}
At a fixed location (in cylindrical co-ordinates where $\rho\equiv\sqrt{x^2+y^2}$) the Larmor frequency, see Eq.~\eqref{eq:Larmor_quad_field}, is now given by 
\begin{equation}
\omega_0(\rho,z,t) = \alpha\sqrt{\rho^2+4[z-z_0(t)]^2},
\label{eq:BG:TAAP-Larmor}
\end{equation}
where, as before, we have the frequency gradient $\alpha = |g_F|\mu_B b'/\hbar$ in a radial direction of the magnetic field.
The system interacts with a radio-frequency field which is taken to be uniform and with a linear polarisation in $z$-direction, i.e.\
\begin{equation}
  \label{eq:BG:TAAP-Brf}
    \GG{B}_1(t) 
       =  B_1 \cos(\omega_\mathrm{rf} t) \,\GG{e}_z 
\end{equation}
which is also 
$
    \GG{B}_1(t) 
       = \mathcal{B}_1\,e^{-i\omega_\mathrm{rf} t}\,\GG{e}_z + \mathcal{B}_1^\ast\,e^{i\omega_\mathrm{rf} t}\,\GG{e}_z
$, Eq.~\eqref{eq:rf_field_general_case} with a real value of the complex vector amplitude $\mathcal{B}_1 = B_1/2$.
In Eq.~\eqref{eq:BG:TAAP-Brf} we use the notation $\omega_\mathrm{rf}$ for the frequency of the rf, rather than the usual $\omega$, as in the development below this frequency will itself become time-dependent.

The linear polarisation could be replaced with a circular polarisation around $z$; the main point is that at the location of the proposed ring the interaction of the rf radiation with the atoms should be strong, given the approximately radial field at that point, and that the rf coupling should not break the rotational symmetry around the $z$-axis. Two aspects of the rf field have to be considered carefully: the amplitude and the frequency. Both of these quantities can also be time-dependent as we see below.

The instantaneous dressed potential is given by a modified (i.e.\ time-dependent) version of Eq.~\eqref{eq:VFPotential}
\begin{equation}
  \label{eq:BG-TAAP-Vdress}
  V_F(\rho,z,t) = F\hbar\sqrt{\left[
\omega_0(\rho,z,t)-\omega_\mathrm{rf}(t)
\right]^2+\Omega_1(\rho,z,t)^2 }
\,,
\end{equation}
where the local Rabi frequency $\Omega_1$ is given by Eq.~\eqref{eq:local_coupling_linear_polarization}
\bea
  \label{eq:BG-TAAP-Rabi}
  \Omega_1(\rho,z,t) &=& 
\left|\frac{g_F \mu_B}{2\hbar}\right| B_1\sqrt{1-u_z(\rho,z,t)^2} \nonumber\\
&=& \Omega_{0} \sqrt{1-u_z(\rho,z,t)^2}\,.
\eea
The maximum Rabi frequency for a linear polarisation $\Omega_{0} = |g_F \mu_B B_1|/(2\hbar) $ is also given in Eq.~\eqref{eq:Rabi_linear}, while, for rf polarised in the $z$-direction,
\bea
u_z(\rho,z,t) &=& \frac{\GG{B}_0(\rho,z,t)\cdot\GG{e}_z}{|\GG{B}_0(\rho,z,t)|} \nonumber\\
&=& \frac{-2[z-z_0(t)]}{\sqrt{\rho^2+4[z-z_0(t)]^2}}
\label{eq:uz_quad}
\eea
reduces the Rabi frequency as the fields become less orthogonal. Note that we neglect the effect of gravity in this calculation for simplicity. The gravitational term will simply cause some sag in the final ring potential.

Now we suppose that we will form a ring trap centred on the origin and with a ring radius $\rho_0$, and we examine the time dependent fields acting on an atom located at $\rho=\rho_0$, $z=0$. The aim is to make this location a stable and static circle so that the surrounding fields, when time-averaged, form a trapping potential. The stable ring would naturally be a circle on the magnetic iso-surface given by $\omega_0=\omega_\mathrm{rf}$, as this would already have a trapping potential in the radial direction. However, we note that the oscillation in $\omega_0$ would make the iso-surface expand and contract at twice the modulation frequency $\omega_m$. This would not imply a static potential point at $\rho=\rho_0$, $z=0$, but we can compensate for this by simultaneously modulating the rf frequency so that $\omega_0(\rho_0,z{=}0,t)-\omega_\mathrm{rf}(t)=0$. This implies that
\begin{equation}
  \label{eq:BG-TAAP_wrf-modulation}
  \omega_{\mathrm{rf}}(t)= \alpha\sqrt{\rho_0^2+[2z_0(t)]^2},
\end{equation}
and then defining the lowest rf frequency as $\omega_\mathrm{rf}^0=\alpha\rho_0$ we
can also write
this time-dependent rf frequency as 
\begin{equation}
  \label{eq:BG:TAAP-wrf-result}
  \omega_\mathrm{rf}(t) =\omega_\mathrm{rf}^0 
    \sqrt{ 1 +\beta_m^2  \sin^2(\omega_m t )}
\,,
\end{equation}
where $\beta_m = B_m /( b' \rho_0)$ gives the scale of vertical shaking relative to the size of the resonance surface.

This nearly leads to a stable ring in Eq.~\eqref{eq:BG-TAAP-Vdress}, but we still have oscillations in energy at $\rho=\rho_0$, $z=0$ because the oscillating magnetic field $\GG{B}_0(\rho,z,t)$ causes the Rabi frequency, Eq.~\eqref{eq:BG-TAAP-Rabi}, to oscillate. This is because as the quadrupole field centre moves out of the $z=0$ plane, the field changes its angle in the $z=0$ plane so that it points slightly out of the plane with its magnitude increasing. If the rf field is in the fixed direction $z$, as above, the Rabi frequency around the ring circle will oscillate as 
\begin{equation}
  \Omega_1(\rho_0,0,t) = 
  \frac{ \Omega_0}{    \sqrt{ 1 + \beta_m^2  \sin^2(\omega_m t )}  }
\label{eq:BG-TAAP-RabiOsc}
\end{equation}
from Eq.~\eqref{eq:BG-TAAP-Rabi} and using $2 z_0(t) = \rho_0\beta_m \sin(\omega_m t)$.
This Rabi frequency would not oscillate at the proposed ring location $\rho=\rho_0, z=0$ if the rf magnetic field amplitude $B_1$ is itself modulated by replacing
\begin{equation}
  \label{eq:BG:TAAP-rf-amplitude-modulation}
    B_1 \longrightarrow B_1  \sqrt{ 1 + \beta_m^2  \sin^2(\omega_m t )}
\,,
\end{equation}
so that the Rabi frequency around the trapping circle is fixed at $\Omega_{0} = |g_F \mu_B B_1|/(2\hbar) $ although it will oscillate at other locations (away from $\rho=\rho_0$, $z=0$). In this case, i.e.\ away from the trapping circle, using Eq.~\eqref{eq:uz_quad} we find
\begin{equation}
  \label{eq:BG:TAAP-Rabi-in-space}
   \Omega_1(\rho,z,t) = 
 \Omega_0
\frac{ (\rho/\rho_0) \sqrt{ 1 + \beta_m^2  \sin^2(\omega_m t )}
}{\sqrt{ (\rho/\rho_0)^2 +   [ 2 z/\rho_0   - \beta_m \sin(\omega_m t) ]^2 } }
\,.
\end{equation}

So far we have engineered the three oscillating field parameters to produce a non-time-oscillating point around $\rho=\rho_0, z=0$. To demonstrate that it is stable we must expand the dressed potential, Eq.~\eqref{eq:BG-TAAP-Vdress}, about this point and then time-average it to obtain the TAAP. 

Thus we expand Eq.~\eqref{eq:BG-TAAP-Vdress}, firstly in $\rho$, and we let $\rho=\rho_0+\Delta\rho$. Then for the first part in the square root of Eq.~\eqref{eq:BG-TAAP-Vdress} we find that 
\begin{equation}
  \omega_0(\rho,z{=}0,t)-\omega_\mathrm{rf}(t) = 
\omega_\mathrm{rf}^0 \frac{\Delta\rho}{\rho_0}
\frac{1}{\sqrt{ 1 +  \beta_m^2  \sin^2(\omega_m t )}}
\,.
\end{equation}
This expression is only needed in first order in $\Delta\rho$ to expand $V_F(\rho,z,t)$ to second order in $\Delta\rho$, which then becomes
\bea
&&  V_F(\rho_0+\Delta\rho,z,t) \simeq
F \hbar \Omega_{0}\label{eq:VF-expanded}\\
&&\times \left\{
1 + \frac{1}{2} 
\left(\frac{\omega_\mathrm{rf}^0}{\Omega_0}\right)^2
\left(\frac{\Delta\rho}{\rho_0}\right)^2 
 \frac{1}{ 1 +  \beta_m^2  \sin^2(\omega_m t ) }
  + \ldots
\right\}
\,.\nonumber
\eea
Here we have neglected terms from the expansion of Eq.~\eqref{eq:BG:TAAP-Rabi-in-space} about  $\rho=\rho_0$, $z=0$. This is because they are smaller by a factor of  $(\Omega_{0}/\omega_\mathrm{rf}^0)^2$ which makes only a slight difference to the second order term in Eq.~\eqref{eq:VF-expanded}, which will be the main result. There is also a first order term from the spatial dependence of the Rabi frequency, Eq.~\eqref{eq:BG:TAAP-Rabi-in-space}, but this simply slightly shifts the radial location of the ring trap.

Finally, for the radial direction, we time-average the approximate potential given by Eq.~\eqref{eq:VF-expanded}.
By using 
\begin{equation}
\int_0^{2\pi/\omega_m}
 \frac{1}{ 1 +  \beta_m^2  \sin^2(\omega_m t )}
\, dt =   \frac{1}{\sqrt{ 1 +  \beta_m^2 }}
\,,
\label{eq:TAAP-time-average-integral}
\end{equation}
we obtain the approximate second order potential, expanded in the $\rho$ direction, as 
\bea
  \label{eq:VF-final-rho}
&& V_F(\rho_0+\Delta\rho, z ) \simeq  F \hbar \Omega_{0} \\
&&\times\left\{
1 + \frac{1}{2} \left(\frac{\omega_\mathrm{rf}^0}{\Omega_0}\right)^2
\left(\frac{\Delta\rho}{\rho_0}\right)^2 
  \frac{1}{\sqrt{ 1 +  \beta_m^2 }}
\right\} \,,\nonumber
\eea
where, again, we have neglected the slight effects of the spatial variation of the Rabi frequency in Eq.~\eqref{eq:BG:TAAP-Rabi-in-space}.  The potential of Eq.~\eqref{eq:VF-final-rho} implies a vibrational frequency through 
\begin{equation}
V_F = V_0 + \frac{1}{2} M \omega_\rho^2 (\rho-\rho_0)^2
\,,
\label{eq:TAAP-harmonic}
\end{equation}
where $V_0$ is the potential at the ring trap bottom (where $\rho=\rho_0$, $z=0$), $M$ is the mass of the atom and $\omega_\rho$ is the vibrational frequency. By comparing Eq.~\eqref{eq:VF-final-rho} to Eq.~\eqref{eq:TAAP-harmonic} we find the radial trapping frequency
\begin{equation}
  \label{eq:TAAP-vib-rho}
  \omega_\rho = \omega\ind{trans}
  ( 1 +  \beta_m^2 )^{-1/4}
\,.
\end{equation}
This result has been referred to the standard transverse trapping frequency $\omega\ind{trans}$, Eq.~\eqref{eq:general_transverse_frequency}, which would be the dressed trapping frequency in the radial direction if there was no modulation $B_m$ and with the atoms confined by other means to the plane $z=0$. In this situation we can obtain $\omega\ind{trans}$ from the limit of Eq.~\eqref{eq:TAAP-vib-rho} as $\beta_m\longrightarrow 0$, or from
Eq.~\eqref{eq:general_transverse_frequency} as $\omega\ind{trans}=|g_F \mu_B b'|\sqrt{F/(M\hbar\Omega_0)}$. We note that, as expected, this result, Eq.~\eqref{eq:TAAP-vib-rho}, does not depend on the modulation frequency $\omega_m$.

A similar calculation can be made in the $z$ direction. 
This time, the position is set to $(\rho,z)\longrightarrow(\rho_0,\Delta z)$, which has a small displacement $\Delta z$ in the $z$ direction. From Eqs.~\eqref{eq:BG:TAAP-Larmor} and \eqref{eq:BG-TAAP_wrf-modulation} we find the approximate detuning
\begin{equation}
  \omega_0(\rho_0,\Delta z,t)-\omega_\mathrm{rf}(t) \simeq 
-2 \omega_\mathrm{rf}^0 \frac{\Delta z}{\rho_0}
\frac{\beta_m \sin(\omega_m t)}{\sqrt{ 1 +  \beta_m^2  \sin^2(\omega_m t )}}
\,.
\end{equation}
This becomes squared inside $V_F$, Eq.~\eqref{eq:BG-TAAP-Vdress}, so that if we make the same approximation of a constant Rabi frequency, because of the modulation given by Eq.~\eqref{eq:BG:TAAP-rf-amplitude-modulation} and small displacements, we can obtain the approximated adiabatic potential
\bea
&& V_F(\rho_0,\Delta z,t) \simeq
F \hbar \Omega_{0} \label{eq:VF-expanded-z}\\
&&\times\left\{
1 + 2
\left(\frac{\omega_\mathrm{rf}^0}{\Omega_0}\right)^2
\left(\frac{\Delta z}{\rho_0}\right)^2 
 \frac{\beta_m^2  \sin^2(\omega_m t )}{ 1 +  \beta_m^2  \sin^2(\omega_m t )}
  + \ldots
\right\}\,.\nonumber
\eea
Then after time-averaging over a modulation period $2\pi/\omega_m$ we will find the $z$-direction vibrational frequency:
\begin{equation}
  \label{eq:TAAP-vib-z}
  \omega_z =  2 \omega\ind{trans}
\sqrt{ 1-  ( 1 +  \beta_m^2 )^{-1/2} } \,.
\end{equation}
Equation~\eqref{eq:TAAP-vib-rho} shows that the effect of increased modulation depth is to slowly weaken the confinement of the atom in the radial direction. However, in the $z$-direction, Eq.~\eqref{eq:TAAP-vib-z} shows that the opposite happens. With no modulation, there is no confinement in this direction, but as the modulation depth is increased the trapping frequency monotonically increases (in the harmonic approximation). The choice $\beta_m=3/4$ gives equal oscillation frequencies along $z$ and $\rho$.

An argument based on the symmetry of the vertical oscillation, see Eq.~\eqref{eq:BG:TAAP-osc-bias-field}, and the expansion oscillation, see Eq.~\eqref{eq:BG:TAAP-wrf-result}, which has half the period, indicates that when neglecting the spatial variation of the Rabi frequency in Eq.~\eqref{eq:BG-TAAP-Rabi}, the local $\rho$ and $z$ axes form the principal axes in the potential.

The full 3D TAAP calculation (\cite{Lesanovsky2007b}) produces the ring trap seen in \figref{fig:TAAP-Klitzing}. Here the potentials are calculated exactly, i.e.\ beyond the harmonic regime. The resonance surface is shown at two extreme positions as dashed lines in \figref{fig:TAAP-Klitzing}a, where it has an expanded size due to Eq.~\eqref{eq:BG:TAAP-wrf-result}.

As mentioned above, the TAAP ring was realised by \cite{Sherlock2011} and \cite{Navez2016}. The TAAP concept has been applied to other trapping situations involving adiabatic traps. \cite{Lesanovsky2007b} proposed `dumbbell' traps and other exotic structures. \cite{Vangeleyn2014} proposed an rf ring trap based on fields induced in a metal ring. The bias field required for rf dressing necessarily introduced an asymmetry which was removed by rotating the bias potential at a low frequency and time-averaging it (see discussion in \cite{Garraway2016}): i.e.\ by essentially forming a TAAP. 
In another example, \cite{Garraway2010} obtained a TAAP by rotating a polarisation direction which affected the couplings in the magnetic resonance problem.

\section{Multiple rf fields}
\label{sec:multipleRF}

\subsection{Well separated rf frequencies}
Up to this point in this review we have focussed on single frequency rf fields, although it could be argued that the time-dependent fields of the previous section on TAAPs (\secref{sec:taaps}) present a spread of frequencies to the atom. In this section we examine in detail the physics and features of multiple discrete rf frequencies with a particular emphasis on two fields with frequencies $\omega_1$ and $\omega_2$.

It would seem clear that if an atom interacts with two radio-frequency fields which are `well' separated the fields will not interfere with each other and at a particular point in space the atom will interact with the dominant field at that location. (Here we consider that, because of experimental preparation, only a single hyperfine multiplet $F$ is involved.) That is, given a spatially inhomogeneous Larmor frequency $\omega_0(\mathbf{r})$, the conditions $\omega_0(\mathbf{r})=\omega_1$ and $\omega_0(\mathbf{r})=\omega_2$ define different regions of space when $\omega_1$ and $\omega_2$ are different.  This means that, for resonant rf trapping, we could have atoms trapped in adiabatic traps in two locations with the two rf frequencies defining the two distinct traps.  But then we have the question of what is meant by `well' separated, and what happens if the regions of space overlap? A plausible criterion is that the frequency separation $\Delta=\omega_2-\omega_1$ should be much larger than a characteristic Rabi frequency, $\Omega_1$ or $\Omega_2$, at all places visited by the atom for this simple, single field, picture to be valid. An interesting aspect of even this simple picture is that two such adiabatic traps are not simply connected if the magnetic field strength changes monotonically from one place to another. Following the adiabatic potential from the minimum at $\omega_0(\mathbf{r})=\omega_1$ we will find we arrive at a potential maximum at $\omega_0(\mathbf{r})=\omega_2$. Essentially, this is because adiabatic potentials cannot cross.

\begin{figure}[t]
\centering
\begin{minipage}{0.45\linewidth}
  \centering
  \subfigure[Bare states.]{\includegraphics[width=\linewidth]{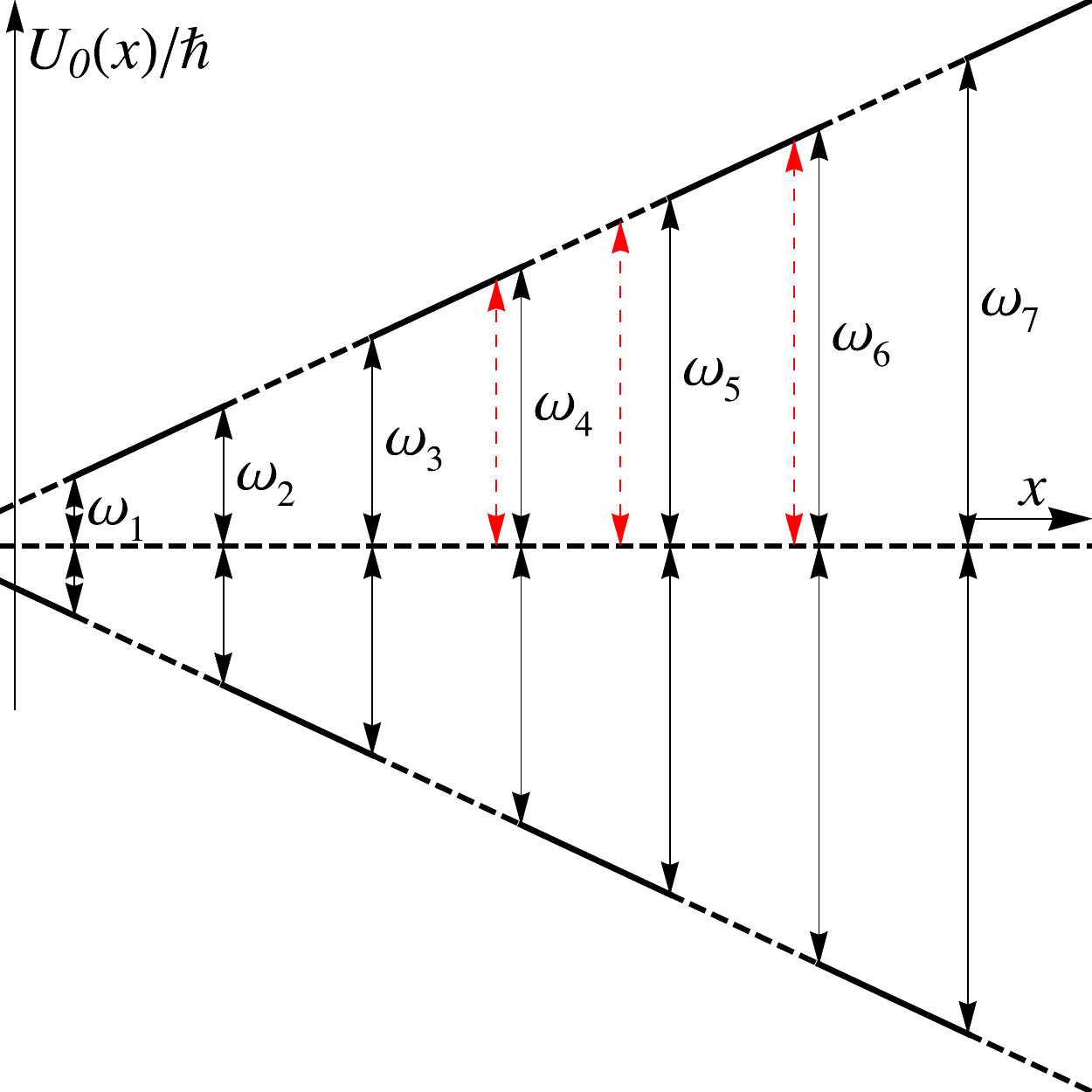}}
\end{minipage}
\hspace{0.05\linewidth}
\begin{minipage}{0.45\linewidth}
  \centering
\subfigure[Adiabatic potentials for a rf comb.]{\includegraphics[width=\linewidth]{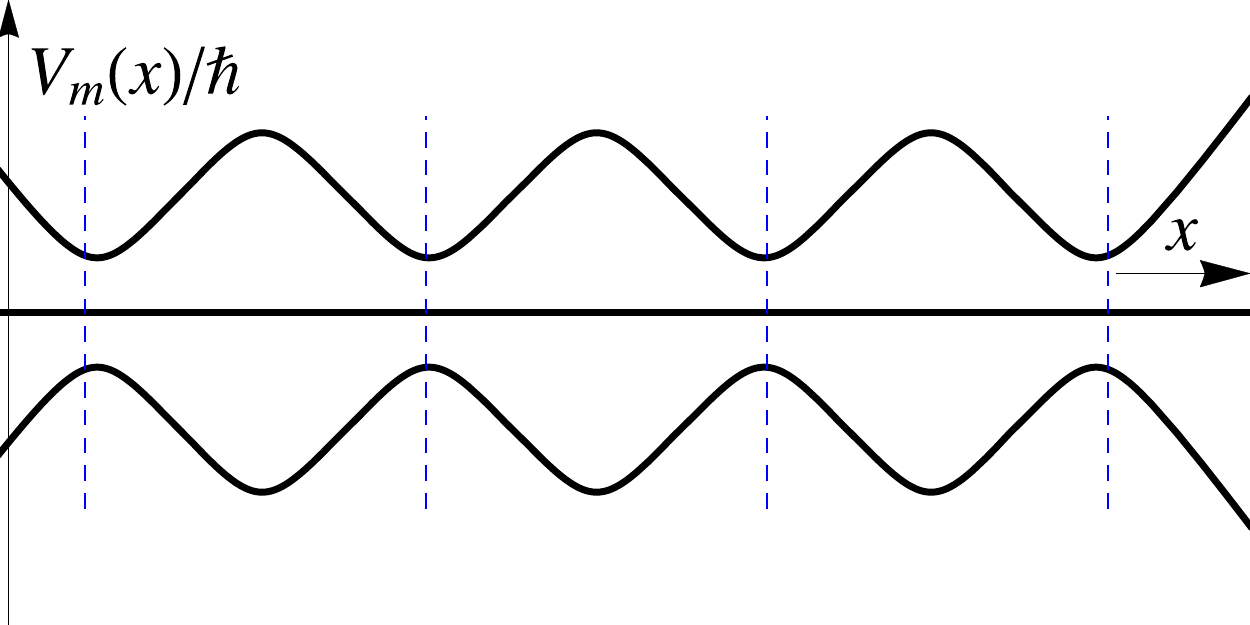}}\\
\subfigure[Frequency shifted $\omega_4$, $\omega_5$ and $\omega_6$.]{\includegraphics[width=\linewidth]{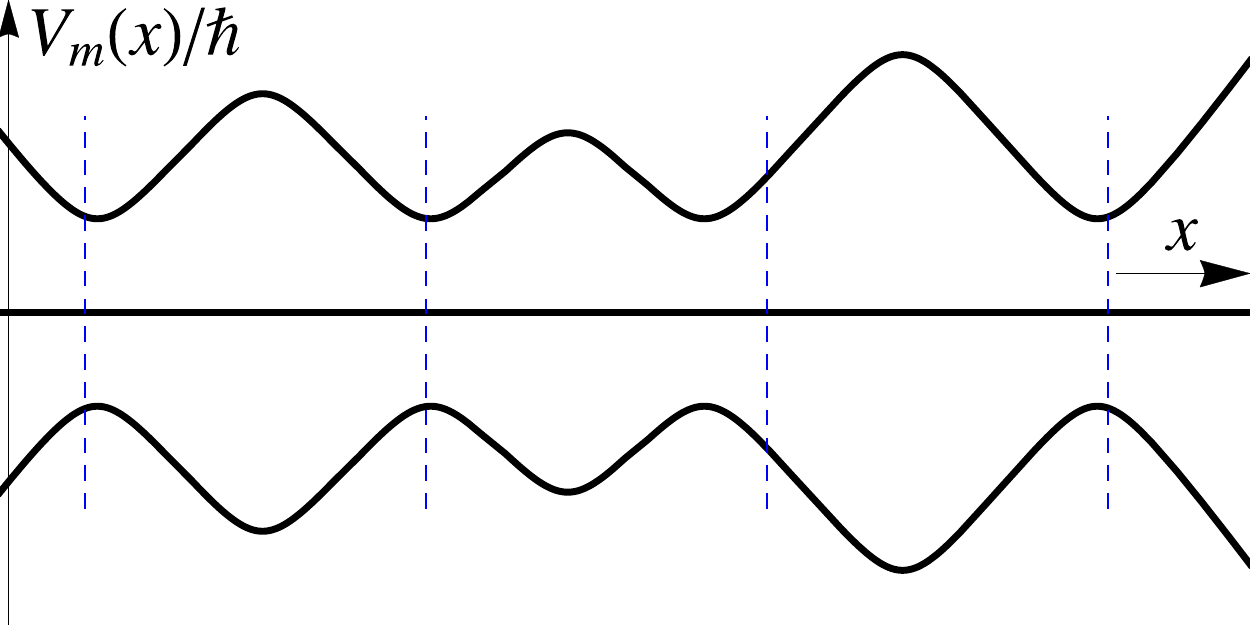}}\end{minipage}
\caption{A one-dimensional lattice obtained with a rf frequency comb (here an example with seven frequencies such that $\omega_n=n\omega_1,$ $n=1\dots 7$ for a $F=1$ state). (a) The bare magnetic potential $U_0(x)$ is linear, and the resonance points (full line arrows) occur at evenly spaced places. (b) Resulting adiabatic potentials $V_m(x)$. The correction given by Eq.~\eqref{eq:VF-large-delta-limit} has been included. (c) Adiabatic potential where $\omega_5$ is shifted to a lower value, while $\omega_4$ and $\omega_6$ are shifted twice as less, see the red dashed arrows in (a). As a consequence, the third well is shifted to the left.
\label{fig:dressed_lattice}
}
\end{figure}

This is not to preclude the formation of double-well potentials with a single frequency, or more, because of spatial variations in magnetic field strength, or rf polarisation, which are not monotonic. However, by using three rf fields it is possible to create two connected resonant rf atom traps, and with five rf fields three traps can be created, and so on. This approach has been suggested to make an rf dressed atom lattice by \cite{Courteille2006}. Such a lattice would be highly controllable as individual lattice sites could have their position adjusted by changing their associated rf frequency and the height of the barriers between lattice sites could be adjusted, to some extent, by controlling the strength of the Rabi coupling at those locations (whilst maintaining adiabaticity). \Figref{fig:dressed_lattice} presents an example with seven frequencies forming evenly separated wells. By modifying three of these frequencies, it is possible to shift spatially one of the wells amongst the others in a controlled way, for example to modify the tunneling between sites.

However, even this picture of well separated traps is not so simple as the atom trap at one location will be perturbed by the field of another even if we have the condition $|\Delta|  \gg  \Omega_1$. For example, Eq.~\eqref{eq:VFPotential} can be considered in the limit where  $|\omega_1 - \omega_0(\GG{r})| \gg \Omega_1$ in order to examine perturbatively the effect of the field $\omega_1$ at the resonant location of the field $\omega_2$ (where $\omega_2 = \omega_0(\GG{r})$) under the condition $|\Delta|  \gg  \Omega_{1,2}$. Thus, letting $\omega_1-\omega_0(\GG{r})\rightarrow\Delta$ we find
\bea
V_{F,1}(\GG{r}) &=& F\hbar\sqrt{[ \omega_1 - \omega_0(\GG{r})]^2 + |\Omega_1(\GG{r})|^2}\nonumber\\
&\rightarrow& F\hbar|\Delta| + \frac{1}{2}F\hbar \frac{ |\Omega_1(\GG{r})|^2 }{|\Delta|}\,.
\label{eq:VF-large-delta-limit}
\eea
In perturbation theory this acts as a shift in the resonant frequency required for $\omega_2$, or alternatively, if a perfect spatial lattice is required then offsets such as Eq.~\eqref{eq:VF-large-delta-limit} must be compensated for to obtain resonance in the correct location. The effects are pervasive because the correction, Eq.~\eqref{eq:VF-large-delta-limit}, depends on the inverse of the frequency difference $\omega_1-\omega_2$ which vanishes slowly with frequency difference. Furthermore, if many rf frequencies are involved, such as in the rf lattice proposed by \cite{Courteille2006}, then the shifts of all other rf frequencies have to be included, as a sum, at any particular lattice site. An example with seven frequencies is shown on \figref{fig:dressed_lattice}. In addition, from a practical point-of-view, care has to be taken over the presence of harmonics from any of the chosen frequencies as harmonic contamination could produce couplings in unexpected places (see \secref{sec:noise}).

The above picture is, however, essentially one-dimensional and may be much more complex in three-dimensions. For example, using an atom chip lattice of wires one of us has been able to design a 2D rf dressed lattice with a single field where the spatial variation is produced by near-fields to the chip rather than different frequencies \citep{SinucoLeon2015}.

Despite these complexities, trapping with well separated multi-frequency fields has been analysed using the quadrupole trap \citep{Bentine2017}. In this case, two rf fields which are well separated in frequency space create two well separated resonance surfaces (see \secref{sec:dressed_quad}). The same work uses four rf frequencies to create a two-species trap with a double well for one species and a single well for the other.

Returning to the case of two rf frequencies, we realise that if one of the fields is strong, and the other is weak, we can consider the case of a probe field (the second rf field) performing spectroscopy on a dressed rf atom trap. Such a probe, which typically induces atom loss at particular locations from the atom trap can also be used as a tool for evaporative cooling of the adiabatic trap. This system is also amenable to analytic treatment, even when the two fields are not too well separated,  and we turn in the next sections to this problem in more detail.

 \subsection{Double-dressing and rf evaporative cooling}
 \label{sec:cooling}

The situation when a second, weaker rf field is added to the main dressing rf field is of important practical interest, as the second rf field can be used to evaporatively cool the trapped atoms or to probe the Rabi splitting between the dressed states.

In this section, to simplify the discussion, we take $g_F>0$ ($s=1$) and consider a homogeneous static magnetic field aligned along $z$ with a Larmor frequency $\omega_0$. The generalisation to the inhomogeneous case proceeds as in \secref{sec:adiabatic_potential}. The spin states are coupled by a first, strong rf field of frequency $\omega_1$ and Rabi frequency $\Omega_{1,+}$, real and optimally ($\sigma^+$) polarised, with $\Omega_{1,+}\ll\omega_1$ to ensure the application of RWA. The purpose of this section is to describe the effect of a second field of frequency $\omega_2$ with Rabi frequencies $\Omega_{2,z}$ and $\Omega_{2,+}$ in the $\pi$ (linear along $z$) and $\sigma^+$ (circularly polarised around $z$) directions, respectively. We do not consider a $\sigma^-$ polarization, as its effect is neglected within the rotating wave approximation.

We start with the time-dependent hamiltonian
\bea
    \label{eq:defH}
\hat  H(t) &=& \omega_0 \hat{F}_z + \Omega_{1,+}\left(\hat{F}_x \cos\omega_1 t + \hat{F}_y \sin\omega_1 t\right) \label{eq:twoclassicalfields}\\
  &+& \Omega_{2,+}\left(\hat{F}_x \cos\omega_2 t + \hat{F}_y \sin\omega_2 t\right) + 2\Omega_{2,z} \hat{F}_z \cos\omega_2 t .\nonumber
\eea
The states dressed by the first field are obtained by first applying a rotation of frequency $\omega_1$ around $z$, then a rotation of angle $\theta$ around $y$, as in \secref{sec:classical_field}. We introduce sequentially the state $\ket{\psi\ind{rot}}$ defined by Eq.~\eqref{eq:psirot} following the rotated hamiltonian of Eq.~\eqref{eq:hrot}, and the state \ket{\tilde\psi}, related to the initial state \ket{\psi} through $\ket{\psi} = \hat{R}_z(\omega_1 t)\hat R_y(\theta)\ket{\tilde\psi}$. These unitary transformations diagonalise the first line of Eq.~\eqref{eq:twoclassicalfields}, but also modify the terms describing the second field. The transformed hamiltonian for \ket{\tilde\psi} reads
\bea
&&\hat{\tilde H}=\nonumber\\
&&\left[\hat R_y(\theta)\right]^\dagger\left(- i\hbar\hat{R}_z^\dagger\left[\partial_t \hat{R}_z\right] + \left[\hat{R}_z(\omega_1 t)\right]^\dagger\hat{H}\hat{R}_z(\omega_1 t)\right)\hat R_y(\theta)\nonumber\label{eq:htilde_full}\\
&&=\Omega \hat F_z + \Omega_{2,+}\left[\cos\theta\cos\Delta t\hat F_x +\sin\theta\cos\Delta t\hat F_z + \sin\Delta t\hat F_y\right]\nonumber\\
&&+ 2\Omega_{2,z}\cos\omega_2 t\left(\cos\theta\hat F_z-\sin\theta\hat F_x\right) \label{eq:htilde_sum}
\eea
We can write $\hat{\tilde H}=\hat{H}_\parallel + \hat{H}_+ + \hat{H}_- + \hat{H}_z$, with
\bea
\hat{H}_\parallel&=&\Omega \hat F_z + \Omega_{2,+}\sin\theta\cos\Delta t\,\hat F_z\nonumber\\
&+& 2\Omega_{2,z}\cos\theta\cos\omega_2 t\,\hat F_z\label{eq:vzdress}\\
\hat{H}_+&=& \Omega_{2,+}\cos^2(\theta/2)\left(\cos\Delta t\,\hat F_x + \sin\Delta t\,\hat F_y\right)\label{eq:vp}\\
\hat{H}_-&=& -\Omega_{2,+}\sin^2(\theta/2)\left(\cos\Delta t\,\hat F_x - \sin\Delta t\,\hat F_y\right)\label{eq:vm}\\
\hat{H}_z&=&-2\Omega_{2,z}\sin\theta\cos\omega_2 t\,\hat F_x\label{eq:vz}
\eea
where $\Omega=\sqrt{\delta_1^2+|\Omega_{1,+}|^2}$, $\delta_1=\omega_1-\omega_0$ and $\Delta=\omega_2-\omega_1$.

This new hamiltonian is analogous to with the one of a spin in a static magnetic field and a rf field. The terms in $\hat{H}_\parallel$, Eq.~\eqref{eq:vzdress}, proportional to $\hat F_z$ do not couple the states dressed by the first field, but the terms along $\hat F_x$ or $\hat F_y$ do. They oscillate either at $\pm\Delta$ or at $\omega_2$, and will be resonant when one of these three frequencies approaches the splitting $\Omega$ between states.

Specifically, for a frequency $\omega_2$ close to $\omega_1+\varepsilon\Omega$ with $\varepsilon=\pm 1$, either $\hat{H}_+$, Eq.~\eqref{eq:vp}, or $\hat{H}_-$, Eq.~\eqref{eq:vm}, is nearly resonant, and we can define a detuning $\delta_2=\omega_2-\omega_1-\varepsilon\Omega=\Delta-\varepsilon\Omega$. We repeat the procedure described in \secref{sec:classical_field}: another rotation around $\hat F_z$ at a frequency $\varepsilon\Delta$ and the application of the rotating wave approximation yields the rotated hamiltonian
\bea
\hat H_{2,+} &=& -\delta_2\hat F_z+\Omega_{2,+}\cos^2(\theta/2)\hat F_x\label{eq:h2p}\quad\mbox{for }\varepsilon=1,\\
\hat H_{2,-} &=& \delta_2\hat F_z-\Omega_{2,+}\sin^2(\theta/2)\hat F_x\label{eq:h2m}\quad\mbox{for }\varepsilon=-1.
\eea
This expression is valid in the situation where $|\delta_2|,\Omega_{2,+}\ll |\Delta|,\Omega\ll\omega_1$. This ensures the application of RWA first to the main $\omega_1$ dressing field, then to the second, weaker $\omega_2$ field.

We note that the effective coupling $\Omega_{2,{\rm eff}} = \Omega_{2,+}\cos^2(\theta/2)$, or $\Omega_{2,+}\sin^2(\theta/2)$, of the second rf field is reduced by a factor $\cos^2(\theta/2)$ or $\sin^2(\theta/2)$. This reduced coupling, asymmetric with respect to the sign of $\delta_1$, is easily understood in the limit $|\delta_1|\gg\Omega_{1,+}$, where $\Omega\simeq|\delta_1|$. For $\delta_1<0$, $\theta\simeq 0$ and only the coupling at $\omega_2\simeq\omega_1+\Omega\simeq\omega_1-\delta_1=\omega_0$ is non vanishing. This simply corresponds to the direct coupling of $\omega_2$ to the initial undressed spin, at the Larmor frequency $\omega_0$. In the same way, for large and positive $\delta_1$, $\theta\simeq\pi$ and only the coupling at $\omega_2\simeq\omega_1-\Omega\simeq\omega_1-\delta_1=\omega_0$ remains. The apparition of the other coupling is always due to the dressing by the first rf field, such that $\delta_1$ should not be too large for the second resonance to occur. We come back to the interpretation of the different couplings at the end of \secref{sec:quantum_spectro}.

In the case of a low frequency $\omega_2$, close to $\Omega$, the last term $\hat{H}_z$, Eq.~\eqref{eq:vz}, is resonant. A rotation at $\omega_2$ around $z$ yields the transformed hamiltonian within RWA
\beq
\hat H_{2,z} = -\delta_2'\hat F_z - \Omega_{2,z}\sin\theta\hat F_x,\quad\mbox{with}\quad\delta_2'=\omega_2-\Omega.
\eeq
Resonance occurs at $\omega_2=\Omega$, with again a reduced coupling amplitude $\Omega_{2,{\rm eff}} = \sin\theta\,\Omega_{2,z}$, which is now symmetric with respect to the sign of $\delta_1$, and maximum for $\delta_1=0$. This shows that coupling is only relevant when the first rf field is nearly resonant, and $\theta$ is different from both $0$ and $\pi$.

In any of these three cases, if now the Larmor frequency depends on position, the application of the second rf field can lead to doubly dressed states, with a modified adiabatic potential
\beq
V_{{\rm dd},m}(\GG{r})=m\hbar\sqrt{\delta_2(\GG{r})^2+\Omega_{2,{\rm eff}}(\GG{r})^2}.
\eeq

\begin{figure}[t]
\centering
\subfigure[$\omega_2\simeq\omega_1+ \Omega_1$]{\includegraphics[width=0.32\linewidth,height=0.32\linewidth]{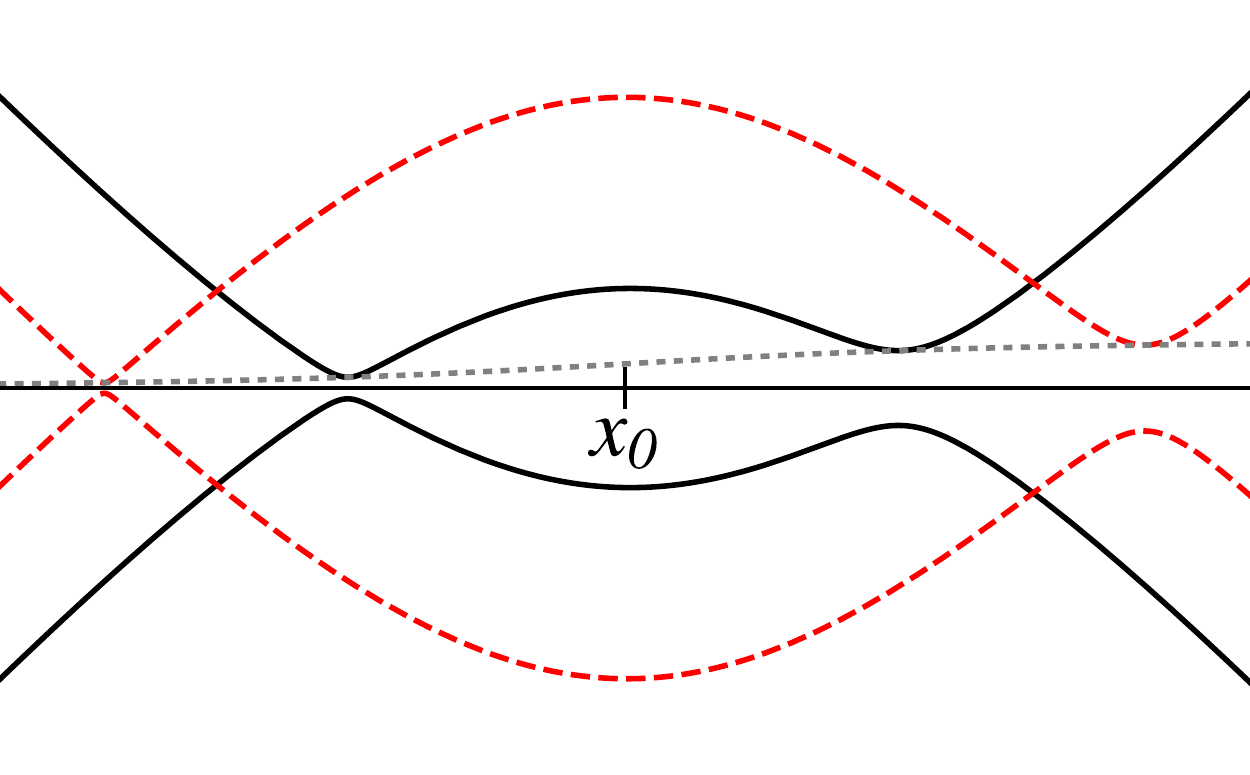}}
\subfigure[$\omega_2\simeq\omega_1- \Omega_1$]{\includegraphics[width=0.32\linewidth,height=0.32\linewidth]{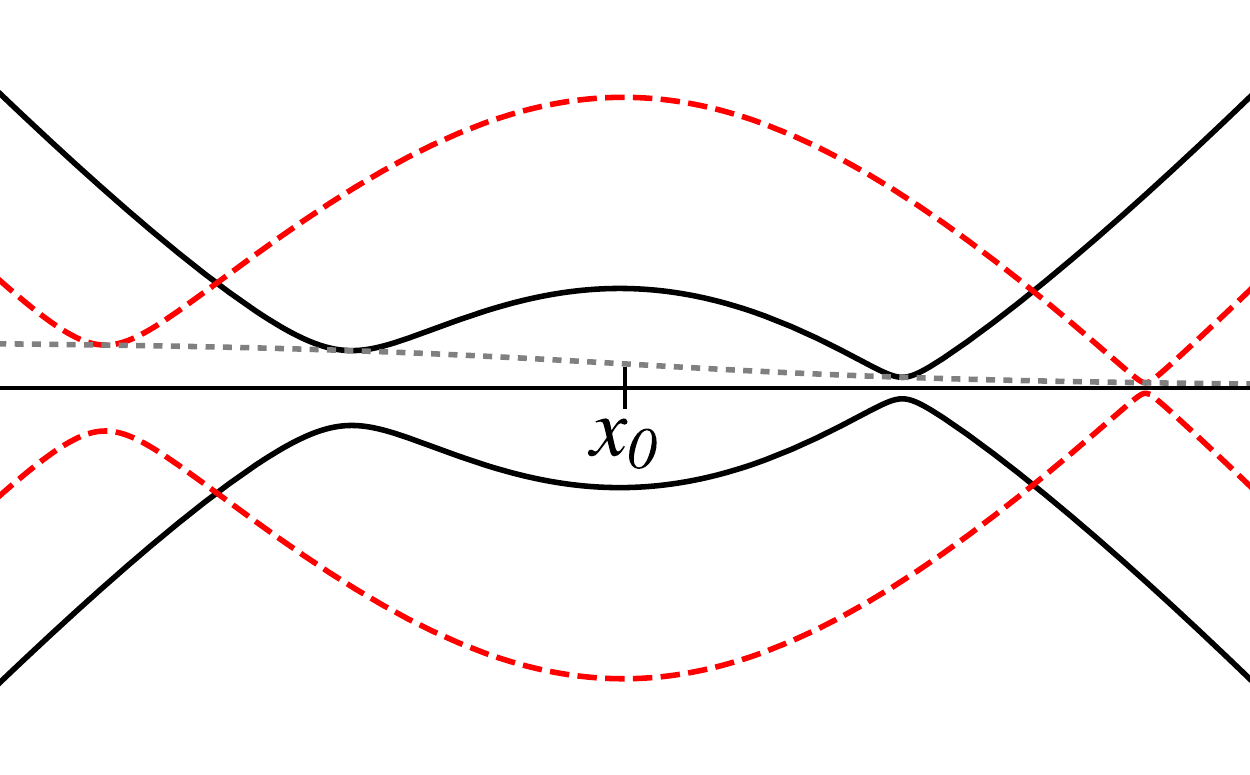}}
\subfigure[$\omega_2\simeq \Omega_1$]{\includegraphics[width=0.32\linewidth,height=0.32\linewidth]{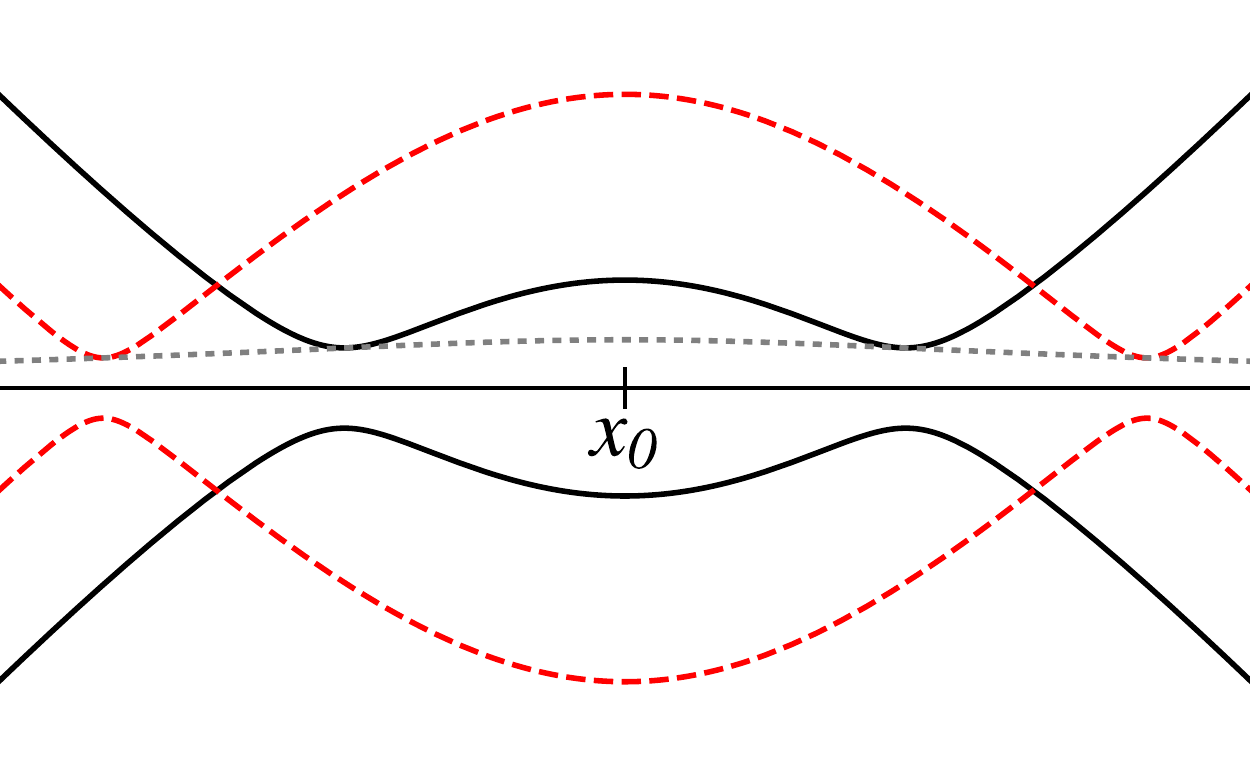}}
\caption{\label{fig:ddplots}
Effect of a second rf field on the adiabatic potentials sketched on \figref{fig:APprinciple} ($F=1$), for the three cases $\omega_2\simeq\omega_1\pm \Omega_1$ and $\omega_2\simeq\Omega_1$, each time for two values of $\omega_2$. In each plot, the dotted line represents the Rabi splitting due to the second field, which is asymmetric in (a) and (b) and symmetric in (c). (a) $\omega_2=\omega_1+1.2\,\Omega_1$ (full black line) and $\omega_2=\omega_1+1.6\,\Omega_1$ (dashed red line). (b) $\omega_2=\omega_1-1.2\,\Omega_1$ (full black line) and $\omega_2=\omega_1-1.6\,\Omega_1$ (dashed red line). (c) $\omega_2=1.2\,\Omega_1$ (full black line) and $\omega_2=1.6\,\Omega_1$ (dashed red line). 
}
\end{figure}

Due to the asymmetry in the effective coupling when $\omega_2\simeq\omega_1\pm\Omega_1$, the doubly dressed potentials also are asymmetric, as seen on \figref{fig:ddplots} for the simple case of a linear static magnetic field (compare with \figref{fig:APprinciple} sketching the adiabatic potential with a single frequency in the same situation). Atoms initially trapped in the adiabatic potential resulting from the first field $\omega_1$ now see a reduced trap depth as a result of the second rf field $\omega_2$, which allows one to perform rf evaporation in the adiabatic potential \citep{Garrido2006}. In this context, the asymmetry in the cases depicted in \figref{fig:ddplots}a and \figref{fig:ddplots}b means that, far from the trap bottom (for large $|\delta_2|$), evaporation will occur essentially on a resonance surface around (or inside) the trapping surface of the first adiabatic potential. On the other hand, the low frequency resonance at $\omega_2\simeq\Omega_1$ provides an evaporation on both sides, but is less efficient at larger detuning $|\delta_2'|$, where the effective Rabi frequency decreases.

 \subsection{Trap spectroscopy}
 \label{sec:quantum_spectro}
When the second rf field $\omega_2$ is very weak, the adiabatic condition is not fulfilled any more and the effect of the second field is to induce non adiabatic transitions between the states dressed by the first rf field \citep{Cohen1977}. The second field can thus be used as a probe, by measuring the non adiabatic losses as a function of the frequency $\omega_2$, which allows us to characterise the effect of the first field. To describe the effect of the probe, the quantum approach gives more insight in the elementary processes at work. In the quantum field approach the fields $\omega_1$ and $\omega_2$ are described by annihilation and creation operators. The full hamiltonian in the presence of the second field at $\omega_2$, with the same notations as in the previous section, reads
\beq
\hat H = \hat H_0 + \hat V_+ + \hat V_z,
\eeq
where
\bea
\hat H_0 &=& \hat H_1 + \hbar\omega_2\hat a_2^\dagger \hat a_2,\\
\hat H_1 &=& \hbar\omega_1\hat a_1^\dagger \hat a_1 + \omega_0\hat F_z + \frac{\Omega_{1,+}^{(0)}}{2}(\hat a_1\hat F_+ + \hat a_1^\dagger \hat F_-),\\
\hat V_+&=&\frac{\Omega_{2,+}^{(0)}}{2}(\hat a_2\hat F_++\hat a_2^\dagger \hat F_-),\\
\hat V_z&=&\Omega_{2,z}^{(0)}(\hat a_2+\hat a_2^\dagger)\hat F_z.
\eea
$\Omega_{i}^{(0)}$ with $i=1,2+,2z$ are the single photon Rabi frequencies, corresponding to average Rabi frequencies $\Omega_i=\sqrt{\langle N_i\rangle}$, taken as real for simplicity. The effect of the probe field $\omega_2$ is to couple the eigenstates of the `unperturbed' hamiltonian $\hat H_0$, describing the effect of the dressing field $\omega_1$.

\begin{table}[b!]
\centering
\begin{tabular}{cclcc}
\hline
res. $\omega_2$ & polar. & $\qquad\Omega\ind{eff}$ & $\theta\to 0$ & $\theta\to\pi$\\
\hline
$\omega_1+\Omega_{1,+}$ & $\sigma_+$ & $\Omega_{2,+}\cos^2(\theta/2)$ & $\Omega_{2,+}$ & $\ds\frac{\Omega_{1,+}\Omega_{2,+}^2}{2\Omega^2}$\\[3mm]
$\omega_1-\Omega_{1,+}$ & $\sigma_+$ & $\Omega_{2,+}\sin^2(\theta/2)$ & $\ds\frac{\Omega_{1,+}\Omega_{2,+}^2}{2\Omega^2}$ & $\Omega_{2,+}$\\[3mm]
$\Omega_{1,+}$ & $\pi$ & $\Omega_{2,z}\sin\theta$ & $\ds\frac{\Omega_{1,+}\Omega_{2,+}}{\Omega}$ &  $\ds\frac{\Omega_{1,+}\Omega_{2,+}}{\Omega}$\\[3mm]
\hline
\end{tabular}
\caption{Effective coupling $\Omega\ind{eff}$ of the probe for the various one-photon resonances, with the corresponding resonant frequency $\omega_2$ and probe polarisation.
\label{tab:spectro_coupling}
}
\end{table}

The states dressed by the first field, eigenstates of $\hat H_1$, are $\ket{m,N_1}_\theta$, given by Eq.~\eqref{eq:quantum_dressed_states}. The eigenstates of $\hat H_0$ are then simply $\ket{m,N_1}_\theta\ket{N_2}_2$, where $\ket{N_2}_2$ is a Fock state for the probe field $\omega_2$, with the same decomposition on the bare product states than Eq.~\eqref{eq:quantum_dressed_states}:
\beq
\ket{m,N_1}_\theta\ket{N_2}_2 = \sum_{m'=-F}^F \bra{m'}_z\hat R_y(\theta)\ket{m}_z\ket{m'}_z\ket{N_1-m'}_1\ket{N_2}_2
\label{eq:quantum_dressed_states_2}
\eeq
where $\ket{N}_i$ represents a Fock states with $N$ photons for the field $\omega_i$.

\begin{figure}[t]
\hspace*{0.15\linewidth}\includegraphics[width=0.8\linewidth]{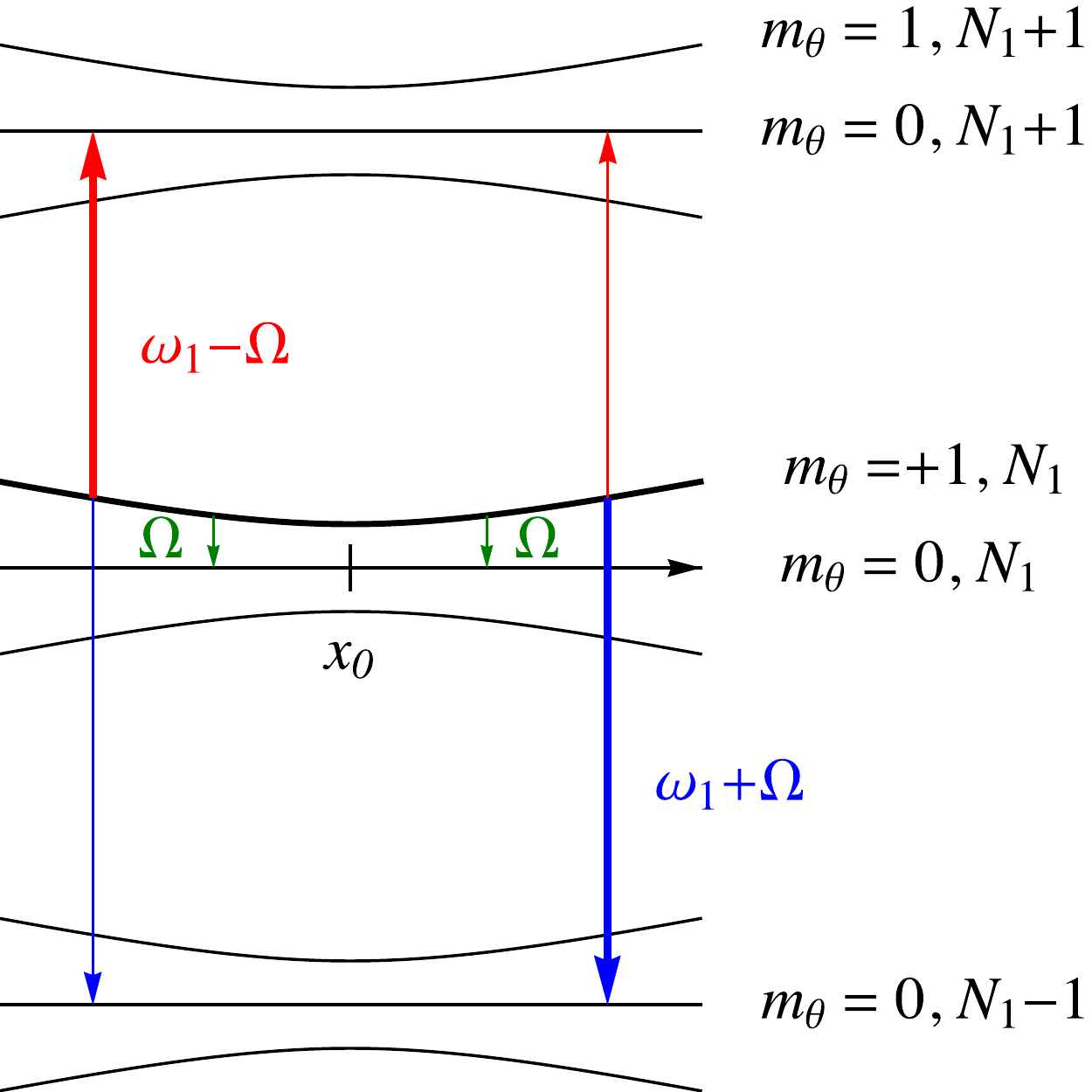}\\[-0.75\linewidth]
\hspace*{-0.85\linewidth}
$\mathcal{E}_{N_1+1}$\\[0.28\linewidth]
\hspace*{-0.85\linewidth}
$\mathcal{E}_{N_1}$\\[0.28\linewidth]
\hspace*{-0.85\linewidth}
$\mathcal{E}_{N_1-1}$\\[0.05\linewidth]
\caption{\label{fig:spectro_dressed}
Effect of a second, weak rf field on the adiabatic potentials sketched on \figref{fig:APprinciple} ($F=1$), represented in the dressed picture for three cases: $\omega_2=\omega_1+\Omega$ (blue arrows) and $\omega_2=\omega_1-\Omega$ (red arrows) with $\Omega=1.6\Omega_{1,+}$, which couple different manifolds; and $\omega_2=\Omega$ (green arrows) with $\Omega=1.2\Omega_{1,+}$, within a given manifold. The stronger resonances in the cases $\omega_2\simeq\omega_1\pm \Omega_{1,+}$ are indicated by a bolder arrow.
}
\end{figure}

These eigenstates of $\hat H_0$ are coupled by $\hat V_+$ and $\hat V_z$. When the frequency $\omega_2$ is close to $\omega_1\pm\Omega_{1,+}$, the eigenstates of $\hat H_0$ are resonantly coupled by $\hat V_+$, while they are resonantly coupled by $\hat V_z$ when $\omega_2\simeq\Omega_{1,+}$. After simple algebra, writing the matrix element between two states $\ket{m,N_1}_\theta\ket{N_2}$ and $\ket{m',N_1'}_\theta\ket{N_2'}$ induced by $\hat V_+$ or $\hat V_z$ allows us to recover the effective couplings derived in the last section. The matrix elements are given below:
\bea
&&\bra{m-1,N_1-1}_\theta\bra{N_2+1}\hat V_+\ket{m,N_1}_\theta\ket{N_2} \nonumber\\
&&\quad\quad= \frac{\Omega_{2,+}}{2}\cos^2\frac{\theta}{2}\langle m-1|_z\hat F_-|m\rangle_z\nonumber\\
&&\bra{m-1,N_1+1}_\theta\bra{N_2-1}\hat V_+\ket{m,N_1}_\theta\ket{N_2}\nonumber\\
 &&\quad\quad=- \frac{\Omega_{2,+}}{2}\sin^2\frac{\theta}{2}\langle m-1|_z\hat F_-|m\rangle_z\nonumber\\
&&\bra{m-1,N_1}_\theta\bra{N_2+1}\hat V_z\ket{m,N_1}_\theta\ket{N_2} \nonumber\\
&&\quad\quad= \frac{\Omega_{2,z}}{2}\sin\theta\bra{m-1}_z\hat F_-\ket{m}_z\nonumber
\eea
and the corresponding effective coupling $\Omega\ind{eff}$, where $\Omega\ind{eff}/2$ appears in front of $\hat F_\pm$, is given in Table~\ref{tab:spectro_coupling}. We recover the asymmetric couplings also valid for rf-dressed atoms coupled to probe fields in the optical or microwave range \citep{Cohen1969b}. The corresponding absorption or emission of probe photons are sketched on \figref{fig:spectro_dressed}.

\begin{figure}[t]
\centering
\subfigure[$\omega_2\simeq\omega_1+ \Omega_1$]{\includegraphics[height=0.65\linewidth]{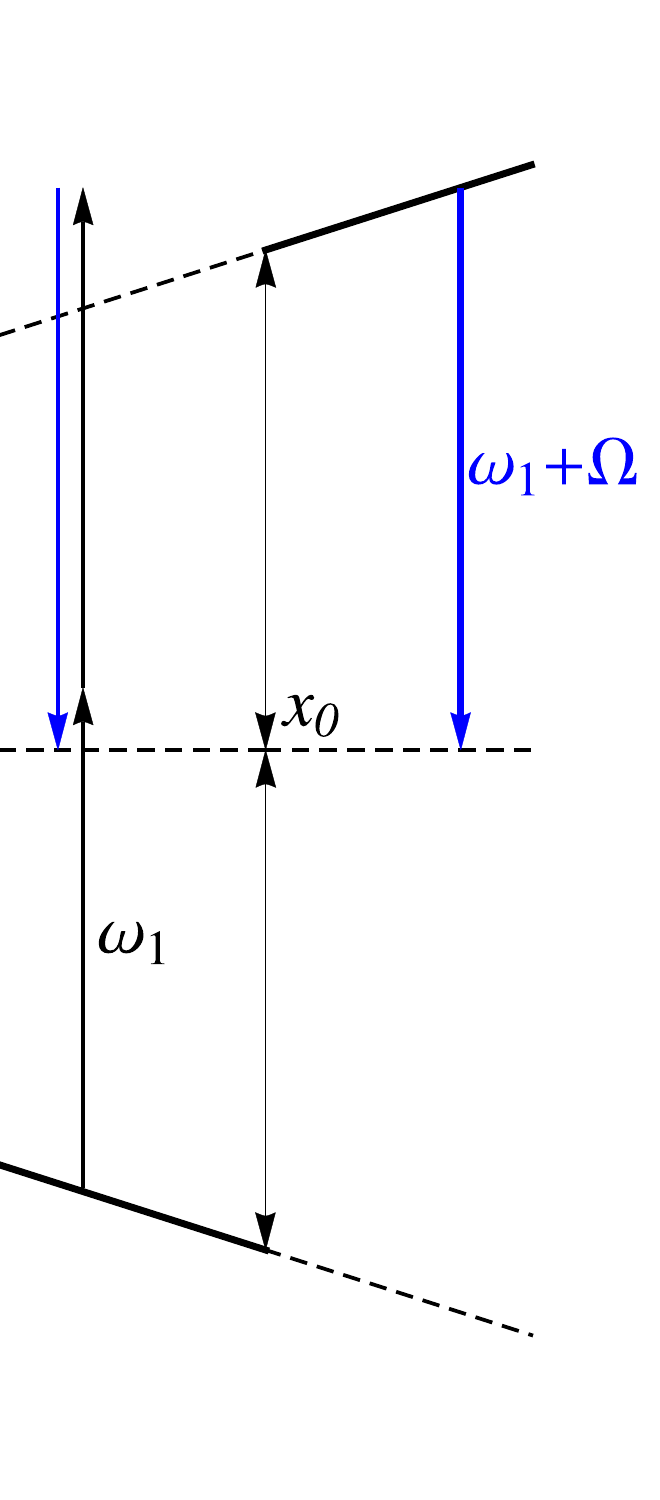}}
\hspace{0.02\linewidth}
\subfigure[$\omega_2\simeq\omega_1- \Omega_1$]{\includegraphics[height=0.65\linewidth]{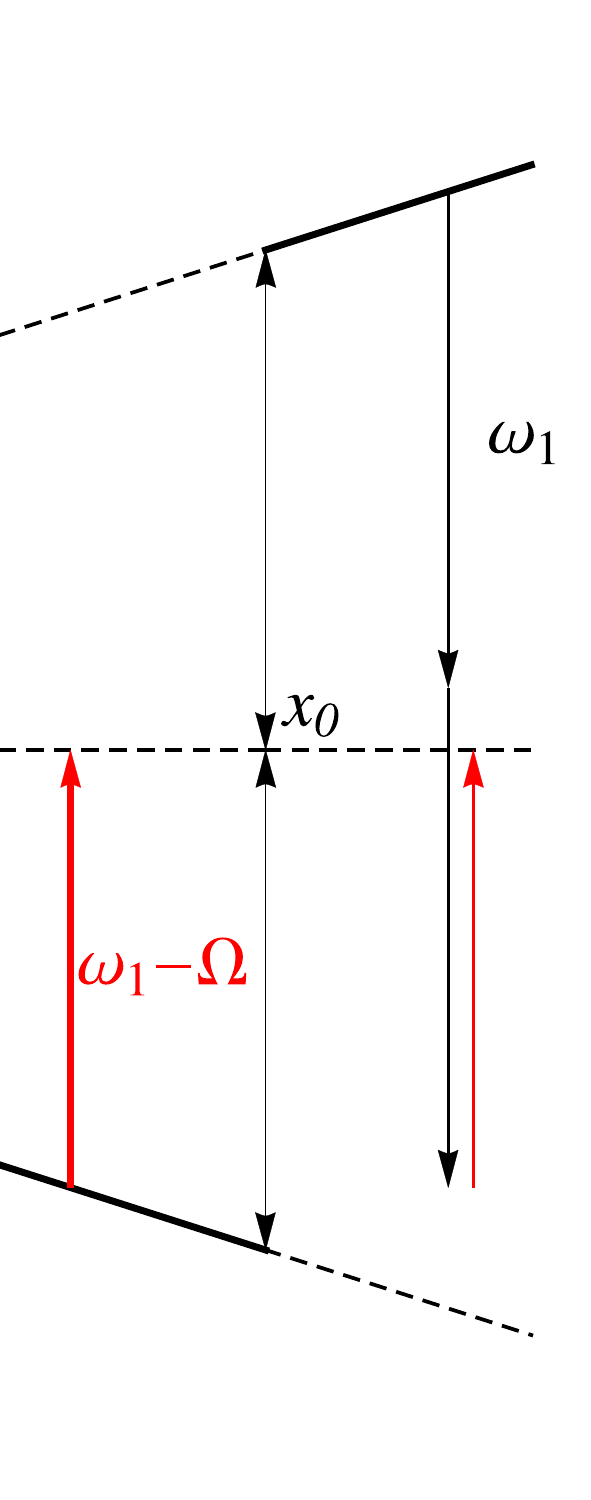}}
\hspace{0.02\linewidth}
\subfigure[$\omega_2\simeq \Omega_1$]{\includegraphics[height=0.65\linewidth]{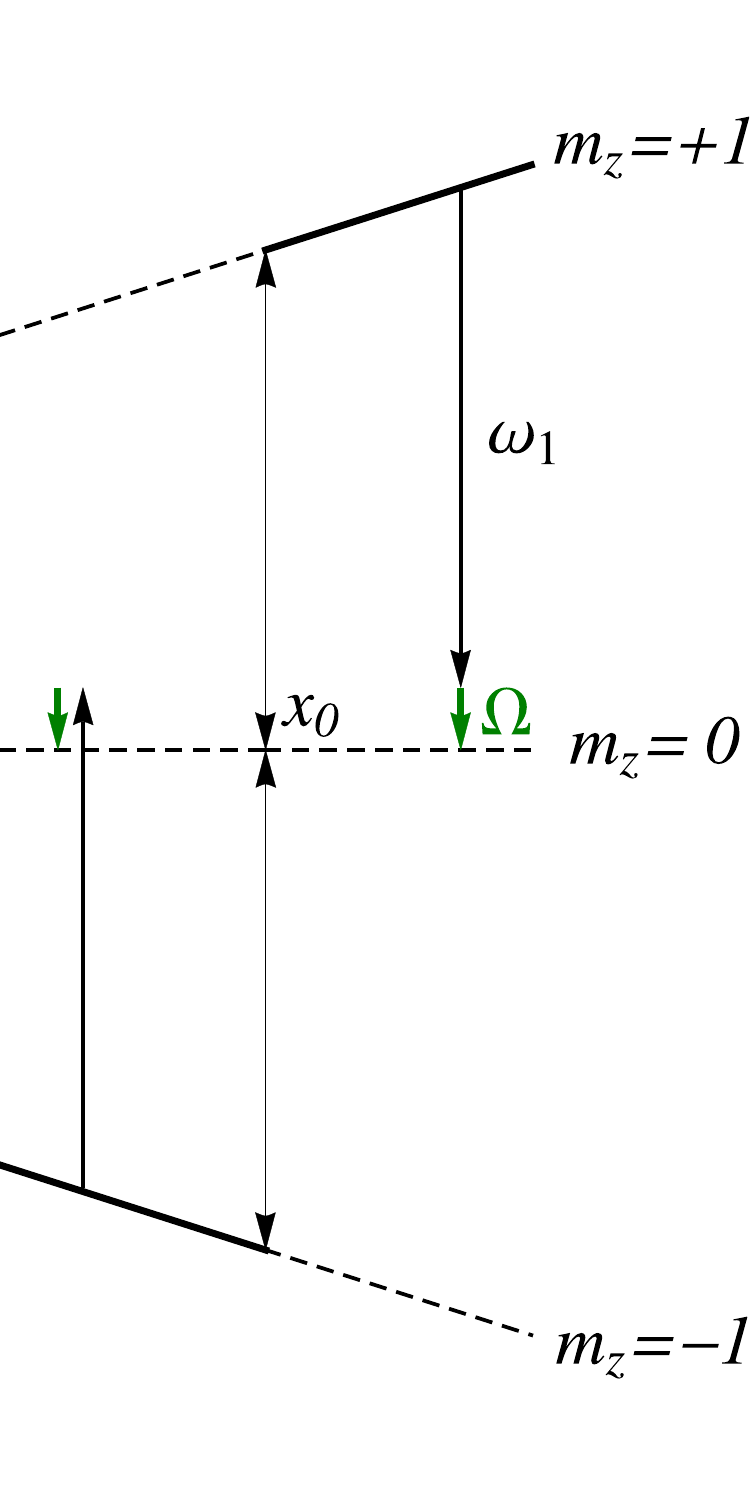}}
\caption{\label{fig:spectro_dressed_bare}
Probe resonances as seen in the basis of the bare states, for a linear static magnetic field (same field as in \figref{fig:spectro_dressed}) and away from the resonance point at $x_0$. The processes at frequency $\omega_2\simeq\omega_1\pm\Omega_{1,+}$, resonant at the two positions $x$ such that $|\omega_2-\omega_1|=\Omega(x)$, involve either one probe photon alone or three photons, with one probe photon and two photons from the dressing field. The resonances at frequency $\Omega_{1,+}$ always involve a probe photon and a dressing photon.}
\end{figure}

In the limit where $\theta$ is close to $0$ or $\pi$, which corresponds to probing regions away from the resonance of the strong dressing field ($|\delta_1|\gtrsim\Omega_{1,+}$), the coupling amplitudes are proportional to $\Omega_{1,+}^n\Omega_{2,+}$ (see Table~\ref{tab:spectro_coupling}). The transitions can then be interpreted in terms of single, two or three-photon processes, with one photon of the probe and $n$ photons of the dressing field, depending on the power $n$ of $\Omega_{1,+}$ in the expression of the coupling. \Figref{fig:spectro_dressed_bare} shows the three resonances in the basis of the bare states, which is relevant for large $\delta_1$, as a decomposition into multi-photon processes implying one or several photons of the dressing field. This picture makes apparent the reason of the asymmetry in the coupling of the resonances at $\omega_1\pm\Omega_{1,+}$, and the symmetry of the resonance at $\Omega_{1,+}$. The weak resonance is a three-photon process, whereas the strong resonance results from the direct coupling of the probe $\Omega_{2,+}$ between bare states.

\subsection{Higher order probe in the dressed atom trap}
\label{sec:multiRF}

Even higher order processes can contribute if a probe $\omega_2$ is used with a strength that is too great. For example, an $\omega_2$ excitation is possible between dressed states $\ket{m_\theta=+1, N_1+1}_\theta$ and $\ket{m_\theta=0, N_1-1}_\theta$ (see \figref{fig:spectro_dressed}) via a point half-way between $\ket{m_\theta=+1, N_1}_\theta$ and $\ket{m_\theta=0, N_1}_\theta$, under the condition $\Delta=\omega_2-\omega_1=\Omega/2$.
Therefore, unlike the probe process of the previous section, this process takes place through an intermediate point which is non-resonant (as in a non-resonant Raman process in three-level systems).
We will call this a two-photon multi-photon probe process and examine the conditions under which it can take place, and explain when it should be avoided, in this section.
The more general case of this kind of probe resonance requires 
$\Delta = \Omega/\np$,
where $\np$ is an integer ($\np\ne 0$) characterising the order of the process.

Our starting point for this analysis is the Hamiltonian in the dressed frame, Eq.~\eqref{eq:htilde_sum}. However, for simplicity, and because it does not contribute significantly to the resonance $\Delta = \Omega/\np$, we will drop the last term which is proportional to $\Omega_{2,z}$. (We can treat it separately, however, to obtain an equivalent `low' $\omega_2$ resonance under the different condition $\omega_2 = \Omega/\np$.)
Thus, from Eq.~\eqref{eq:htilde_sum} we have 
\bea
\hat{\tilde H} &=& \Omega \hat F_z\label{eq:htilde_sum_local}\\
&+& \Omega_{2,+}\left[\cos\theta\cos\Delta t\hat F_x + \sin\theta\cos\Delta t\hat F_z + \sin\Delta t\hat F_y\right] \,. \nonumber
\eea
This Hamiltonian is dominated by the diagonal $\Omega \hat F_z $ term under the condition that $\Omega_{2,+}$ is rather weak. To proceed, we follow the argument of \cite{Pegg1970} (see also \cite{Allegrini1971,Pegg1972,Garraway1997})
and we further transform the Hamiltonian to an interaction representation by performing a rotation about $z$ that completely removes the diagonal term.That is, 
\begin{eqnarray}
  \hat{\tilde H} \rightarrow \hat{\tilde H}'
&=&
   e^{i \frac{\phi(t)}{\hbar}\hat{F}_z }
   \hat{\tilde H}
   e^{-i \frac{\phi(t)}{\hbar}\hat{F}_z }
 - \hbar \phi'(t) \hat F_z \nonumber\\
 &=& 
 \Omega_{2,+}
\cos(\Delta t )\cos\theta
    \left[ \hat F_x \cos\phi(t)  - \hat F_y \sin\phi(t)   \right]
 \nonumber\\ 
 &+&
 \Omega_{2,+}
\sin(\Delta t )
    \left[ \hat F_y \cos\phi(t)  + \hat F_x \sin\phi(t)   \right] 
\label{eq:hprime_off-diagonal}
\end{eqnarray}
where $\phi(t)$ comes from integrating the coefficient of $\hat F_z$ in Eq.~\eqref{eq:htilde_sum_local}, i.e.\
\bea
  \phi(t) &=& \int_0^t \, dt' \left( \Omega + \Omega_{2,+} \sin\theta\cos\Delta t' \right)\nonumber\\
 &=& \Omega t + \frac{  \Omega_{2,+} \sin\theta }{\Delta}  \sin(\Delta t) .\label{eq:phi-defn}
\eea

We now inspect the trigonometric functions in $\hat{\tilde H}'$, Eq.~\eqref{eq:hprime_off-diagonal}. To deal with the time-dependence, involving a trigonometric function of a trigonometric function, we follow \cite{Pegg1970} and use a Bessel function expansion to rewrite the exponential of $\phi(t)$, Eq.~\eqref{eq:phi-defn}: i.e.\ we observe that we have the Fourier expansion
\begin{equation}
  e^{i\phi(t)} = e^{i \Omega  t} \cdot e^{i b \sin(\Delta t)} =
   e^{i \Omega t} \sum_{n=-\infty}^\infty    e^{i n \Delta t}
   J_n(b)
\,,
  \label{eq:besseldef}
\end{equation}
where we let $b= \sin\theta \,\Omega_{2,+} /\Delta $, which we expect to have a small value. The real and imaginary parts of Eq.~\eqref{eq:besseldef} provide the needed trigonometric functions in Eq.~\eqref{eq:hprime_off-diagonal} as infinite sums.

Under the $\np$-photon resonance condition $\np \Delta = \Omega$ we find that the Hamiltonian of Eq.~\eqref{eq:hprime_off-diagonal} contains purely harmonic terms and we can now make a second rotating wave approximation \citep{Pegg1973,Garraway1997} where we keep only the static terms and drop all the rotating terms.
After doing this we are left with a new Hamiltonian in the 2nd RWA where the infinite sums are now lost, i.e.\
\bea
\hat{\tilde H}'_{\np,\text{RWA}} &=&
\left[
  (\cos\theta-1)  J_{\np+1}(b) + (\cos\theta+1)  J_{\np-1}(b)
\right]\nonumber\\
&\times& (-1)^{\np+1}\frac{  \Omega_{2,+} }{2} \hat F_x.\label{eq:h_multi_approx1}
\eea
This coupling is proportional to the weak probe strength $\Omega_{2,+} $, and couples the dressed states when $\np \Delta = \Omega$. The strength of this coupling can also be shown to be a measure of the width of the resonance. 

From Eq.~\eqref{eq:h_multi_approx1} we can read off an effective Rabi frequency $\Omega_{\np,\text{eff}}$ for the coupling (i.e.\ the coefficient of $\hat F_x$). Given that the parameter $b= \sin\theta \,\Omega_{2,+} /\Delta $, is so small when $\np\Delta = \Omega$, we simplify this for practical purposes by using the Bessel function expansion
\begin{equation}
  \label{eq:bessel_expanded}
J_n(x) \simeq \frac{1}{n!} \left(\frac{x}{2}\right)^n  \,, \qquad\qquad n \in\mathbb{N}
\end{equation}
and keeping only the most significant terms (which means neglecting $J_{\np+1}(b) $ compared to $ J_{\np-1}(b) $), so that from Eq.~\eqref{eq:h_multi_approx1}, and for $\np\ge 1$,
\begin{eqnarray}
&& (\cos\theta+1)  J_{\np-1}(b) + (\cos\theta-1)  J_{\np+1}(b) 
 \nonumber\\
&&\quad\simeq (\cos\theta+1) \frac{1}{(\np-1)!} \left(\frac{b}{2}\right)^{\np-1}, \nonumber
 \end{eqnarray}
such that
\begin{eqnarray}
\Omega_{\np,\text{eff}} &\simeq&(-1)^{\np-1} \frac{  \Omega_{2,+} }{2}\nonumber\\
&\times& (\cos\theta+1) \frac{1}{(\np-1)!} \left(
\frac{\np\Omega_{2,+}\sin\theta }{2\Omega }
\right)^{\np-1}
  \label{eq:mp_RAbi_effective}
\end{eqnarray}
where in the last line we have used the expression for $b$ and the value of
$\Delta$ on resonance. 

We note that when $\np=1$ we obtain from Eq.~\eqref{eq:mp_RAbi_effective} the previous result of Eq.~\eqref{eq:h2p}, i.e.\
\begin{equation}
  \label{eq:N1}
\Omega_{\np=1,\text{eff}} =  \frac{  \Omega_{2,+} }{2}  (\cos\theta+1)\,.
\end{equation}
For $\np=-1$, we can use the relation
\begin{equation}
  \label{eq:bessel-negative-order}
  J_{-n}(x) = (-1)^n J_n(x)
\end{equation}
in Eq.~\eqref{eq:h_multi_approx1} and then we find that the other Bessel function dominates so that $(\cos\theta+1)$ is replaced by $(\cos\theta-1)$ in Eq.~\eqref{eq:N1}, in agreement with Eq.~\eqref{eq:h2m}.

The case discussed at the start of this section occurs when $\np=2$. Then
\begin{equation}
  \label{eq:N2}
\Omega_{\np=2,\text{eff}} = - \frac{ \Omega_{2,+}^2 }{2\Omega } \sin\theta (\cos\theta+1).
\end{equation}
For $\np=-2$ a similar result emerges, but again with  $(\cos\theta+1)$ replaced by $(\cos\theta-1)$.

\begin{figure}[t]
\centering
\includegraphics[width=\linewidth]{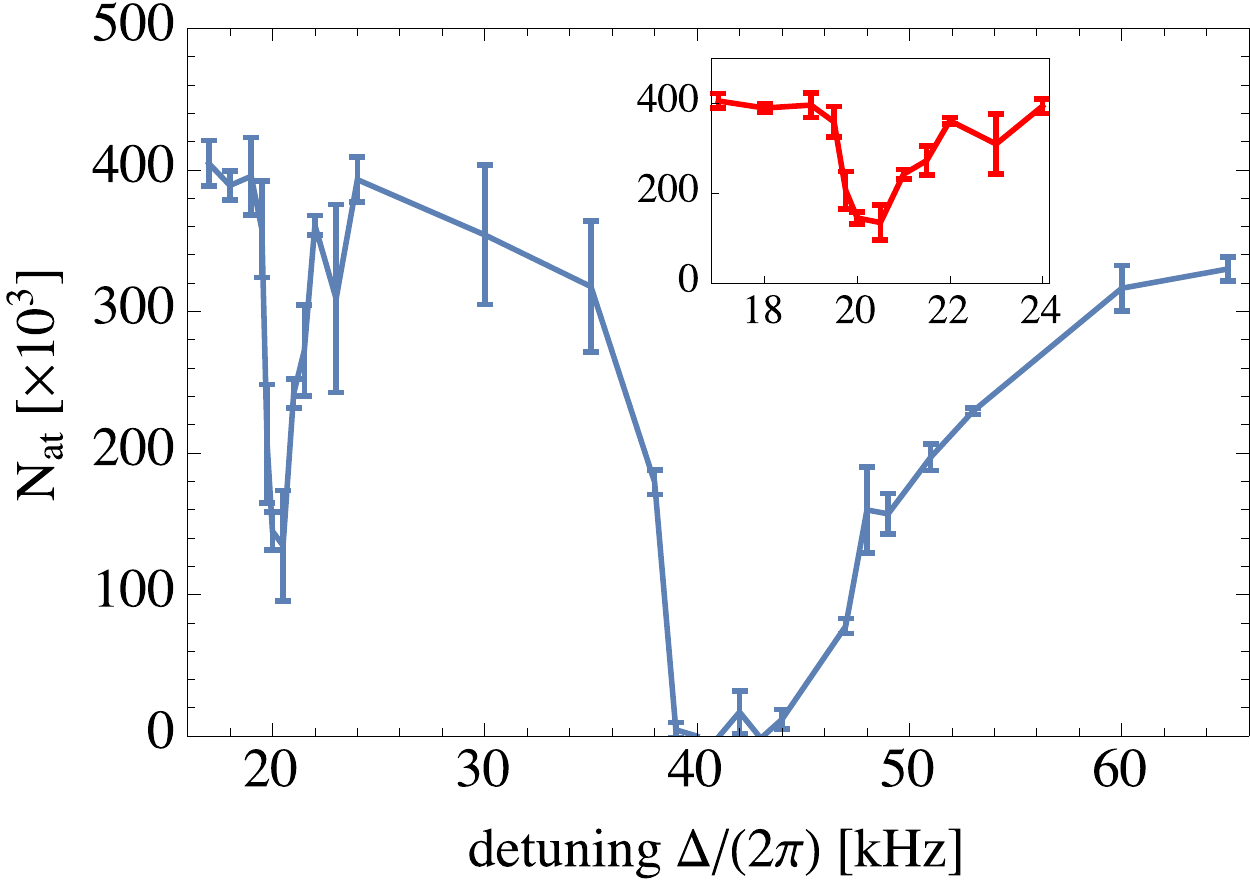}
 \caption{\label{fig:N2}
Recent data taken at LPL for the $\np=\pm2$ multi-photon process in a rf-dressed quadruple trap, with a rf dressing frequency $\omega_1/(2\pi)=1200$~kHz and a Rabi frequency $\Omega_{1,+}/(2\pi)=41$~kHz. A probe at frequency $\omega_2$ is applied for 1~s to the trapped atoms, and the number of remaining atoms is measured. The probe Rabi frequency is of order 2~kHz, much larger than the amplitude normally used for the observation of the single-photon resonance. The probe is scanned between 1216 and 1266 kHz, corresponding to a detuning $\Delta/(2\pi)$ scanned between 16 and 66~kHz. Two resonances occur: a broad resonance (because of the large probe amplitude) at 1241~kHz, i.e., $\Delta/(2\pi)=41$~kHz, corresponding to the single-photon resonance at $\omega_2=\omega_1+\Omega_{1,+}$; and a narrow resonance at 1220.5~kHz, i.e., $\Delta/(2\pi)=20.5$~kHz, corresponding to the two-photon resonance at $\omega_2=\omega_1+\Omega_{1,+}/2$. Inset: zoom around the two-photon resonance.}
\end{figure}

Equation~\eqref{eq:h_multi_approx1} shows a dependence on $\Omega_{2,+}^\np$ which means that the resonances where $\Delta=\Omega/\np$ will become extremely weak for high order $\np$. 
In addition the resonances are correspondingly narrow, and the $\sin\theta$ dependence in Eq.~\eqref{eq:mp_RAbi_effective} may mean that the gravitational sag, which reduces $\sin\theta$, will reduce the coupling strength yet further.
As discussed above, the case $\np=\pm1$ is used in evaporative cooling of rf dressed atom traps and for rf trap spectroscopy. The case $\np=\pm2$ requires a higher probe strength to become visible and can be regarded as a consequence of a probe that is too strong. This means that stray rf fields must be avoided, not only close to the usual probe resonance at $|\Delta|=\Omega_{1,+}$, but also close to other frequencies and especially $\Omega_{1,+}/2$. This has recently been observed at LPL (see \figref{fig:N2}). 

Finally, we note that if we keep the  $\Omega_{2,z}$ terms in Eq.~\eqref{eq:htilde_sum}, but as part of an approximation drop the $\Omega_{2,+}$ terms; then the analysis above produces resonances under the  condition $\omega_2=\Omega/\np$ at low frequency. In this case,  Eq.~\eqref{eq:mp_RAbi_effective} becomes (for $\np\ge 1$)
\begin{eqnarray}
\Omega_{\np,\text{eff}} &=& 
(-1)^{\np} \Omega_{2,z} \sin\theta
\frac{1}{(\np-1)!} \left(
\frac{ \cos\theta \,\Omega_{2,z} }{\omega_2 }
\right)^{\np-1}.
  \label{eq:mp_RAbi_effective_z}
\end{eqnarray}
This result also agrees with section~\secref{sec:quantum_spectro} when $\np=1$.

\section{Practical issues with rf dressed atom traps}
\label{sec:practical}
\subsection{Estimates for the decay of rf traps}
\label{sec:LZlosses}

In this section, we discuss in more detail the validity of the adiabatic following of the adiabatic states. With the notations of \secref{sec:adiabatic_potentials}, the unitary operator $\hat{U}(\GG{r},t)$ applied to $\hat{H}\ind{spin}(\GG{r},t)$ is a position dependent rotation operator, of the form
\beq
\hat{U}(\GG{r},t) = \hat{R}_{\GG{u}}\left[s\omega t + s\phi(\GG{r})\right]\hat{R}_{\GG{u}_\perp}\left[\theta(\GG{r})\right].
\label{eq:unitary_transform}
\eeq
The \GG{r} dependence has been omitted in $\GG{u}(\GG{r})$ and in $\GG{u}_\perp(\GG{r})$, the vector orthogonal to $\GG{u}(\GG{r})$ around which the last rotation is performed.

More precisely, the application of $\hat{U}(\GG{\hat{R}},t)$ on $\hat{H}\ind{spin}(\GG{\hat{R}},t)$ writes:
\beq
\overline{\hat{U}(\GG{\hat{R}},t)^\dagger \hat{H}\ind{spin}(\GG{\hat{R}},t)\hat{U}(\GG{\hat{R}},t)} - i\hbar\overline{\hat{U}^\dagger\left[\partial_t \hat{U}\right]} = \Omega(\GG{\hat{R}})\hat{F}_z
\eeq
where the bar represents the time-averaging applied within the rotating wave approximation.

As $\hat{U}(\GG{\hat{R}},t)$ now depends on the position operator, the diagonalisation procedure, when applied to the total Hamiltonian of Eq.~\eqref{eq:hamil_adiab_pot_recast}, gives rise to extra terms due to the non commutation of \GG{\hat{R}} and \GG{\hat{P}}, and the full transformed Hamiltonian reads
\beq
\hat{H}' = \hat{T} +  \Omega(\GG{\hat{R}})\hat{F}_z + \Delta \hat{T} = \hat{H}\ind{adia} + \Delta \hat{T}.
\eeq
￼In this expression $\hat T = \hat{\mathbf{P}}^2/2M$ and $\hat{H}\ind{adia}=\hat{T} +  \Omega(\GG{\hat{R}})\hat{F}_z$ is called the adiabatic Hamiltonian. The internal and external degrees of freedom are decoupled, such that for each spin state, it describes the motion of a particle in a different \emph{adiabatic potential}, see \secref{sec:adiabatic_potentials}, without any coupling between spin states.

The correction to this hamiltonian is given by
\beq
\Delta \hat{T} = \overline{\hat{U}(\GG{\hat{R}},t)^\dagger \hat{T}\hat{U}(\GG{\hat{R}},t)} - \hat{T}.
\label{eq:LZoutcoupling}
\eeq
Its effect is to couple the spin eigenstates of $\hat{H}\ind{adia}$, in a way which depends on the atomic velocity. The adiabatic approximation consists in neglecting this correction $\Delta\hat{T}$ in front of the energy splitting $\hbar\Omega$ induced by the rf coupling. The purpose of this section is to discuss the effect of $\Delta\hat{T}$ and give a condition for the application of the adiabatic approximation.

Let us first write the transform of the momentum operator $\hat{\GG{P}}$ under $\hat{U}(\GG{\hat{R}},t)$:
\bea
\hat{U}(\GG{\hat{R}},t)^\dagger\hat{\GG{P}}\hat{U}(\GG{\hat{R}},t) &=& \hat{\GG{P}} -i\hbar\hat{U}(\GG{\hat{R}},t)^\dagger\nablagras\hat{U}(\GG{\hat{R}},t) \nonumber\\
&=& \hat{\GG{P}} + \hat{\GG{A}}(\GG{\hat{R}},t).
\eea
The transformed kinetic operator thus reads
\bea
\hat{T}' &=& \frac{1}{2M}\left[\hat{\GG{P}} + \hat{\GG{A}}(\GG{\hat{R}},t)\right]^2 \nonumber\\
&=& \hat{T} + \frac{1}{2M}\left[\hat{\GG{P}}\cdot\hat{\GG{A}} + \hat{\GG{A}}\cdot\hat{\GG{P}}\right] + \frac{\hat{\GG{A}}^2}{2M}\nonumber\\
&=&\hat{T} + \hat{\GG{A}}\cdot\frac{\hat{\GG{P}}}{M} - i\frac{\hbar}{2M}\nablagras\cdot\hat{\GG{A}} + \frac{\hat{\GG{A}}^2}{2M}.\nonumber
\eea
After time-averaging we get:
\beq
\Delta\hat{T} = \overline{\hat{\GG{A}}(\GG{\hat{R}},t)}\cdot\frac{\hat{\GG{P}}}{M} - i\frac{\hbar}{2M}\nablagras\cdot\overline{\hat{\GG{A}}(\GG{\hat{R}},t)} + \frac{\overline{\hat{\GG{A}}(\GG{\hat{R}},t)^2}}{2M}.
\eeq

As $\hat{U}$ is a rotation operator, $\hat{\GG{A}}$ is a combination of the angular momentum operators $\hat{F}_z$, $\hat{F}_\pm$, with position dependent vectorial weights oscillating at $\omega$:
\beq
\hat{\GG{A}}(\GG{\hat{R}},t) = \GG{A}_z(\GG{\hat{R}},t)\hat{F}_z + \GG{A}_+(\GG{\hat{R}},t)\hat{F}_+ + \GG{A}_-(\GG{\hat{R}},t)\hat{F}_-.
\eeq
The effect of $\Delta\hat{T}$ is thus twofold: (i) it produces an energy shift due to the terms in $\hat{F}_z$ or $\hat{F}_z^2$, and (ii) it couples the adiabatic state $\ket{m}_{\theta(\mathbf{r})}$ to $\ket{m\pm 1}_\theta$ (terms in $\hat{F}_\pm$) or to $\ket{m\pm 2}_\theta$ (terms in $\hat{F}_\pm^2$). This second effect (ii) is responsible for non adiabatic Landau-Zener losses.

Computing the coefficients in front of $\hat F_\pm$ and $\hat F_\pm^2$ in the general case is beyond the scope of this review. A recent derivation of loss rates in some particular cases can be found in \cite{Burrows2017}. Instead, we will consider a simple toy model where the static magnetic field has a fixed direction $z$ and a spatial gradient along the $x$ direction:
\beq
\GG{B}_0(\GG{r}) = b'x\, \GG{e}_z = \frac{\hbar\alpha}{\mu}x\,\GG{e}_z.
\label{eq:linear_mag_field}
\eeq
It is a `dummy field' in the sense that it does verify Maxwell's equation $\nablagras\times\GG{B}_0=\GG{0}$, but it is enough to give an insight in the loss processes. We also assume that the rf phase $\phi$ and Rabi frequency $\Omega_1$ are homogeneous, and we chose $\phi=0$. $\GG{u}_\perp$ is then equal to $\GG{e}_y$. In this case, $\hat U$ is the product of a position-independent rotation and a rotation which depends on position through $\theta(\GG{r})$:
\beq
\hat U(\GG{\hat{R}},t) = \hat R_z(s\omega t)\hat R_y[\theta(\GG{\hat{R}})].
\eeq

Using the analogue of Eq.~\eqref{eq:diff_rot_op_pm} with gradients, we find that
\bea
\GG{\hat A}&=&-i\hbar\hat{U}(\GG{\hat{R}},t)^\dagger\nablagras\hat{U}(\GG{\hat{R}},t) = -i\hbar\hat{R}^\dagger_y(\theta)\nablagras \hat{R}_y(\theta)\nonumber\\
\GG{\hat A}& =& - \nablagras{\theta}\hat{F}_y = i\frac{\nablagras\theta}{2}\left(\hat F_+-\hat F_-\right).
\label{eq:diff_rot_op_pm_space}
\eea
The coupling term outside the adiabatic states is thus proportional to $\nablagras\theta$. The situation is analogue to the case where the rf frequency or the Rabi frequency varies in time, which produces a coupling outside the dressed states of the form $\dot\theta\hat F_+$, see Eq.~\eqref{eq:outcoupling} and the adiabatic condition Eq.~\eqref{eq:adia_thetadot}.

Let's consider atoms initially in the extreme $\ket{m=F}_{\theta(\mathbf{r})}$ adiabatic state. The matrix element from $\ket{F}_{\theta(\mathbf{r})}$ to $\ket{F- 1}_{\theta(\mathbf{r})}$ is thus of order
\beq
V_{-} = i\hbar\sqrt{2F}\,\hat{\GG{v}}\cdot\nablagras\theta + \sqrt{2F}\frac{\hbar^2}{2M}\triangle\theta
\eeq
where $\hat{\GG{v}}=\hat{\GG{P}}/M$, while the matrix element from $\ket{F}_{\theta(\mathbf{r})}$ to $\ket{F- 2}_{\theta(\mathbf{r})}$ is of order
\beq
V_{2-}\simeq F\frac{\hbar^2}{2M}|\nablagras\theta|^2.
\eeq
In the example of a linear magnetic field, we have
\beq
\tan\theta=\frac{-\delta}{\Omega_1}=\frac{\alpha x-\omega}{\Omega_1}\Rightarrow\nablagras\theta=\frac{\alpha\Omega_1}{\Omega_1^2+\left(\alpha x-\omega\right)^2}\GG{e}_x.
\eeq
$\nablagras\theta$ is largest at the resonance surface $x=\omega/\alpha$, where it reaches $\alpha/\Omega_1$. On the other hand, $\triangle\theta$ cancels on the resonance surface $x=\omega/\alpha$ and is largest at positions $x=(\omega\pm\Omega_1)/\alpha$ where it reaches $\alpha^2/(2\Omega_1)^2$.

The adiabatic potential confines particles along the $x$ direction, with an oscillation frequency at the bottom of the extreme $m_\theta=F$ adiabatic potential given as in Eq.~\eqref{eq:general_transverse_frequency} by
\beq
\omega_x=\alpha\sqrt{\frac{\hbar F}{M\Omega_1}}.
\label{eq:osc_freq_linear_x}
\eeq
The matrix elements can then be recast under the form
\bea
V_- &\lesssim& \hbar\sqrt{2F} \frac{\alpha |v_x|}{\Omega_1} + \frac{1}{\sqrt{2F}}\left(\frac{\omega_x}{\Omega_1}\right)^2\frac{\hbar\Omega_1}{2} , \nonumber\\
V_{2-}&\lesssim& \left(\frac{\omega_x}{\Omega_1}\right)^2\frac{\hbar\Omega_1}{2}.
\eea
Let us discuss first the terms in $(\omega_x/\Omega_1)^2\Omega_1$. They have to remain very small as compared to the oscillation at the Rabi frequency $\Omega_1$ responsible for the rf dressing and the existence of the adiabatic states. This requires $\omega_x\ll\Omega_1$, which states exactly that the motion in the adiabatic potential should be adiabatic with respect to the level splitting $\Omega_1$.

On the other hand, the first term of $V_-$ is reminiscent of the Landau-Zener paradigm, see \secref{sec:land-zener-paradigm}, and its extension to a motion in space described in \secref{sec:an-adiabatic-trap}, with a rate in energy change $\hbar\alpha v_x$. The Landau-Zener adiabatic parameter for the linear field situation reads
\beq
\Lambda = \frac{V^2}{\hbar^2\alpha |v_x|} = \frac{\Omega_1^2}{\alpha |v_x|}
\eeq
which is equivalent to the ratio of the rf coupling $\Omega_1$ to the first term of $V_-$.

The order of magnitude of this first term $V_-$ for a particle with total energy $(n+1/2)\hbar\omega_x$ in the trap is
\bea
\hbar\sqrt{2F} \frac{\alpha |v_x|}{\Omega_1} &\simeq& \hbar\alpha\sqrt{\frac{(2n+1)F\hbar\omega_x}{M\Omega_1^2}}\nonumber\\
&\simeq&\hbar\sqrt{2n+1}\left(\frac{\omega_x}{\Omega_1}\right)^{3/2}\Omega_1.
\eea
This shows that this coupling term is the largest, even in the ground state $n=0$. The condition for low Landau-Zener losses is again $\omega_x\ll\Omega_1$, or equivalently
\beq
\frac{\hbar\alpha^2}{M}\ll\Omega_1^3,
\eeq
and the adiabaticity parameter for the $n^{\rm th}$ level of the adiabatic trap is of order
\beq
\Lambda \simeq \frac{1}{\sqrt{2n+1}}\left(\frac{\Omega_1}{\omega_x}\right)^{3/2} = \frac{1}{\sqrt{2n+1}}\left(\frac{M\Omega_1^3}{\hbar\alpha^2F}\right)^{3/4}.
\eeq
For a given trap geometry, adiabaticity is better ensured for the lowest harmonic states. This can lead to selective losses of excited atoms, which provides a cooling mechanism, as observed experimentally at LPL \citep{MerlotiThese}.

 \subsection{rf stability and other experimental considerations}
 \label{sec:noise}
 
\begin{figure}[t]
\centering
\subfigure[Bad choice for the initial $\omega$.]{\includegraphics[width=0.49\linewidth]{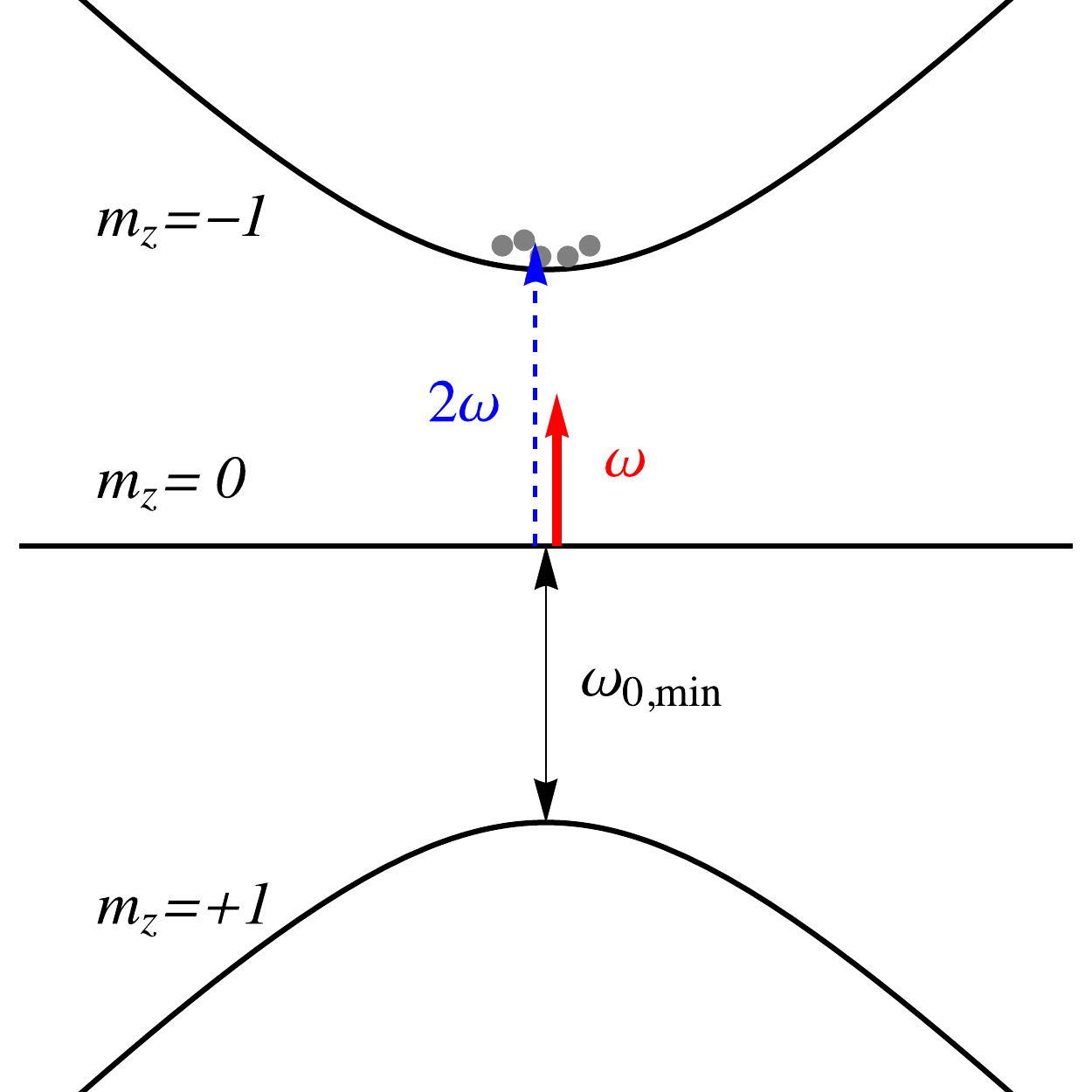}}
\subfigure[Good choice for the initial $\omega$.]{\includegraphics[width=0.49\linewidth]{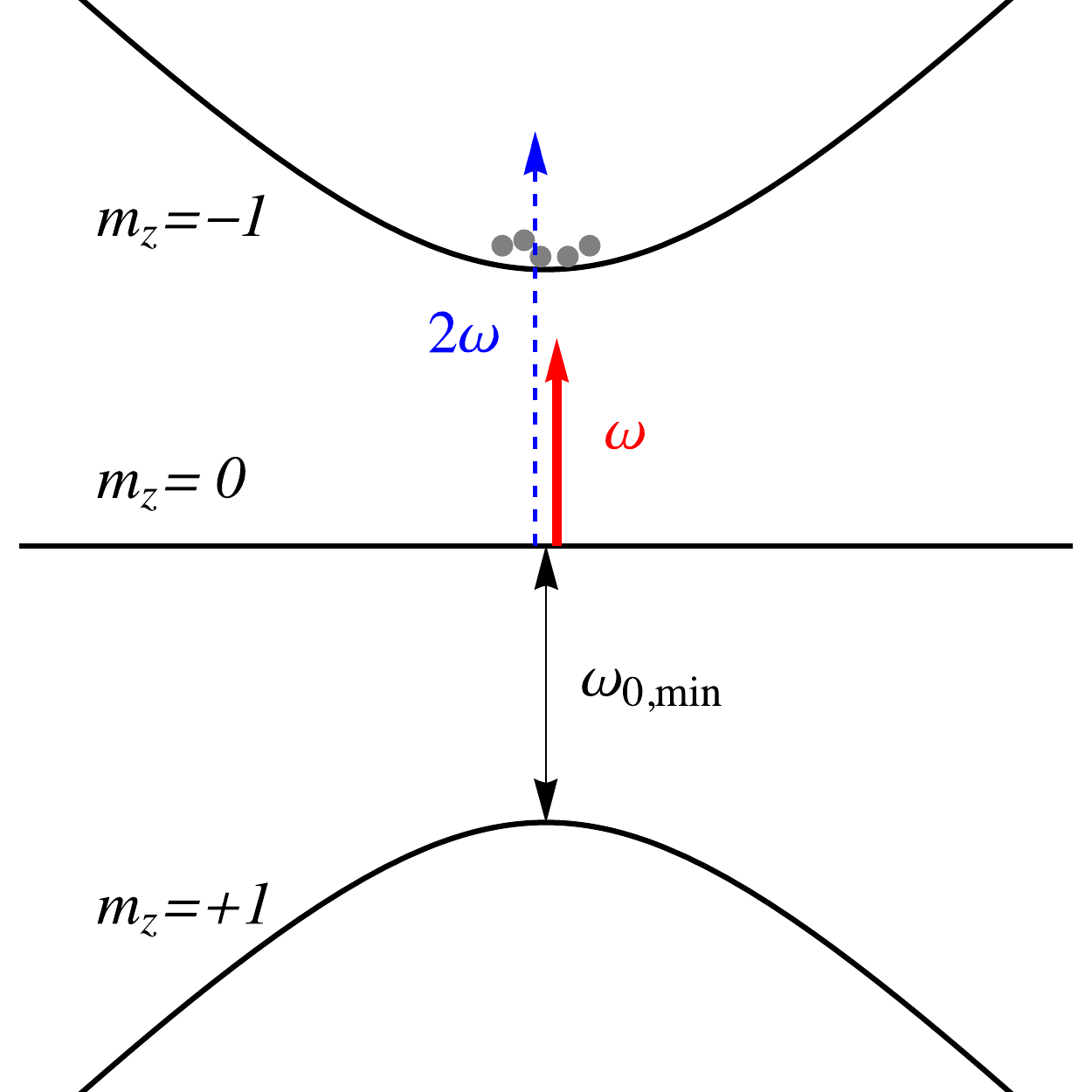}}
 \caption{\label{fig:second_harmonic}
Effect of a second harmonic at $2\omega$ (blue dashed arrow) during the loading phase, where the rf source (red arrow) is switched on from below the minimum Larmor frequency $\omega\ind{$0$,min}$ (black arrow) in the initial magnetic trap. The $F=1$ levels are those of the unperturbed bare states $\ket{m}_z$. Discs symbolise atoms. (a) The frequency $2\omega$ is resonant with the Zeeman splitting at a position where atoms are present. This leads to losses. (b) While the rf frequency $\omega$ is below $\omega\ind{$0$,min}$ as it should be, the second harmonic $2\omega$ is far above $\omega\ind{$0$,min}$, at an energy that no atom can reach. This avoids unwanted losses.
}
\end{figure}

The first experimental issue that may arise when using rf fields to dress atoms is the harmonic distortion in the rf signal, which could come from the amplification process of the rf source. The presence of rf fields at frequencies $2\omega$, $3\omega$\dots\ could lead to a deformation of the adiabatic potential or to rf evaporation, see \secref{sec:multipleRF}. In particular, a field at a frequency $2\omega$, even with a small amplitude, could lead to atom losses if some atoms reach an energy $\hbar(\omega-\Omega_1)$ above the bottom of the adiabatic potential. This may be an issue during the loading phase where the dressing field is switched on from below the minimum Larmor frequency $\omega\ind{$0$,min}$, see \secref{sec:loading} and \figref{fig:second_harmonic}a. As non-linearity is hard to avoid totally in an rf amplifier, the initial rf frequency must be chosen in such a way that the energy distribution of the atoms in the initial magnetic field lie between $\hbar\omega\ind{$0$,min}$ and $2\hbar\omega$, as in \figref{fig:second_harmonic}b.

Another possible experimental issue for adiabatic trapping with rf fields concerns noise sources in rf frequency, amplitude, polarization, or in the amplitude of the static magnetic field. In particular, phase or frequency jitter should be avoided to prevent unwanted spin flips, during the loading ramp (see \secref{sec:loading}) or while operating the trap. For this reason, rf sources based on Direct Digital Synthesis (DDS) are generally used to realise adiabatic traps. This problem has been addressed in detail by \cite{Morizot2008}, together with an analysis on the possible sources of heating, related to frequency, amplitude or phase fluctuations. In this section, we will briefly review the effect of noise in the different experimental parameters, in the simple case of a linear magnetic field as in Eq.~\eqref{eq:linear_mag_field}.

The fluctuations in the experimental parameters can have two main effects: displacing the position of the resonance surface and thus the trap centre, see \secref{sec:bubble}, or modifying the trapping frequency Eq.~\eqref{eq:general_transverse_frequency}. The first effect gives rise to linear dipolar heating, the second to exponential parametric heating.

We first start with dipolar heating. We recall the heating rate $\dot E$ \citep{Gehm1998}
\beq
\dot{E}=\frac{1}{4} M \omega_x^4 \, S_x(\nu_x) \label{eq:dipolar_heating}
\eeq
where $\nu_x=\omega_x/(2\pi)$, $\omega_x$ is given by Eq.~\eqref{eq:osc_freq_linear_x} and $S_x$ is the one-sided Power Spectral Density (PSD) of the position fluctuations $\delta x$, defined as the Fourier transform of the time correlation function~\citep{Gehm1998}
\beq
S_x(\nu) = 4 \int_0^{\infty} \! \! d\tau \, \cos(2 \pi \nu \tau) \langle \delta x(t) \, \delta x(t+ \tau) \rangle.
\label{eq:Sx}
\eeq
The average trap position $x_0$ is fixed by the resonance condition $\omega=\alpha x_0$. Position fluctuations can thus be due either\footnote{In the case of an external force like gravity in the direction $x$, the trap position is shifted and rf amplitude or phase fluctuations can also lead to dipolar heating.} to rf frequency or static field amplitude fluctuations. If these noise sources are not correlated to each other, $S_x(\nu)$ is related to the sum of the relative frequency ($\delta\omega/\omega$) PSD $S_y(\nu)$ and the relative static field ($\delta\alpha/\alpha$) PSD $S_b(\nu)$ taken at the trap frequency $\nu_x$ through
$$
S_x(\nu_x) = x_0^2 \, [S_y(\nu_x) + S_b(\nu_x)] = \frac{\omega^2}{\alpha^2}\,[S_y(\nu_x) + S_b(\nu_x)].
$$
Replacing $\omega_x$ by its expression Eq.~\eqref{eq:osc_freq_linear_x}, we find that the linear heating rate is
\bea
\dot E &=& \frac{F}{4}\hbar\omega_x\frac{\omega_x}{\Omega_+}\omega^2 \,[S_y(\nu_x) + S_b(\nu_x)] \nonumber\\
&=& \frac{\hbar^2F^2\alpha^2\omega^2}{4M\Omega_+^2}\,[S_y(\nu_x) + S_b(\nu_x)].\nonumber
\eea

The contribution of the rf frequency noise can also be written in terms of the phase noise PSD $S_\varphi$ of the source:
$$
\dot E\ind{phase noise} = \frac{F}{4}\frac{\hbar\omega_x^4}{\Omega_+}S_\varphi(\nu_x).
$$

Relative frequency noise of DDS sources can be very low. A typical standard figure for the phase noise is $S_\varphi=10^{-10}$~Hz$^{-1}$ in the kHz range. If we take $\Omega_+/\omega_x=10$ to ensure adiabaticity (see \secref{sec:LZlosses}) and $\nu_x=1$~kHz, we find a very small heating rate of order 5~pK$\cdot$s$^{-1}$, which is not a problem for trapping degenerate gases for seconds in an adiabatic trap.

On the other hand, the contribution of the magnetic field fluctuations reads:
$$
\dot E\ind{static field} = \frac{F}{4}\frac{\hbar\omega_x^2\omega^2}{\Omega_+}S_b(\nu_x).
$$
This contribution to the linear heating is expected to be larger than the previous one, because of the large $\omega^2/\omega_x^2$ factor between $\dot E\ind{static field}$ and $\dot E\ind{phase noise}$. For the low noise power supply used in our experiment\footnote{SM15-100 from Delta Elektronika.} we have measured a PSD of the relative amplitude of $S_b= 10^{-14}$~Hz$^{-1}$ around 1~kHz, which is quite low. This corresponds to a heating rate due to the magnetic fluctuation of order 5~nK$\cdot$s$^{-1}$ for a dressing frequency of $\omega/(2\pi)=1$~MHz and a trapping frequency of $\nu_x = 1$~kHz. In the experiment \onlinecite{Merloti2013a} measure indeed a heating rate of few nK$\cdot$s$^{-1}$. Batteries can also provide low noise current sources.

We now turn to parametric heating, which is due to fluctuations in the trapping frequency. The temperature is expected to increase exponentially, with a rate $\Gamma\ind{param}$ given by \citep{Gehm1998}:
\beq
\Gamma\ind{param} = \pi^2 \nu_x^2 S_k(2 \nu_x)
\label{eq:parametric_heating}
\eeq
where $S_k$ is the one-sided power spectrum of the fractional fluctuation in the spring constant $k=M\omega_x^2=\hbar F\alpha^2/\Omega_+$, taken at twice the trap frequency. $S_k$ can be due to static magnetic field amplitude fluctuations via the $\alpha^2$ term, or to fluctuations in $\Omega_+$, in turn due to rf amplitude noise or noise in the relative orientation between the rf and static field.

Let us consider the contribution of rf amplitude noise to parametric heating. We have
$$
\Gamma\ind{rf amplitude} = \pi^2 \nu_x^2 S_a(2 \nu_x)
$$
where $S_a$ is the PSD of relative rf amplitude noise ($\delta\Omega_+/\Omega_+$). For $\Gamma\ind{rf amplitude}$ to be lower than $0.001$~s$^{-1}$ in the $\nu_x=1$~kHz trap considered here, the PSD of the relative rf amplitude should be less than $S_a=10^{-10}$~Hz$^{-1}$ around 2~kHz. Such low noise rf sources are easily accessible, and $S_a=10^{-11}$~Hz$^{-1}$ is typical. Care must, however, be taken with the amplification chain that may be installed at the output of the rf source. In particular, \onlinecite{Merloti2013a} found that rf attenuators can bring additional noise, and a full digital control of the rf amplitude with the DDS provides better performance.

The noise in the relative static magnetic field contributes twice more, because of the square dependence on $\alpha$ mentioned above:
$$
\Gamma\ind{static field} = 2 \pi^2 \nu_x^2 S_b(2 \nu_x).
$$
For the same performance, the requirement is now $S_b(2\nu_x)<5\times 10^{-11}$~Hz$^{-1}$. We see that this condition is less stringent than the one on linear heating.

Finally, the strongest experimental requirements for quiet adiabatic potentials are on the noise in the static magnetic field, mostly because of linear dipolar heating. While a good DDS source, together with low noise amplifiers, will be good enough to prevent heating due to the rf source, the static field power supply is required to have an excellent noise performance.

\subsection{Beyond the rotating wave approximation}
\label{sec:beyondRWA}
In the calculation of the adiabatic potentials, we have ignored the contribution of terms rotating at a frequency $2\omega$ after the transformation to the frame rotating with the rf frequency, see e.g. Eqs.~\eqref{eq:Hrot} and \eqref{eq:Heff}. This holds either when the local polarisation is purely circular, which in general is not true especially because the direction of the static field is not homogeneous, or when both the local detuning to the rf frequency and the local Rabi coupling are small as compared to the Rabi frequency: $|\delta|,\Omega_1\ll\omega$.

In real experiments, this is not always true, in particular when moderate rf frequencies (below 1~MHz) are used. Counter-rotating terms have then to be included in the calculation of the adiabatic potential. The effect of beyond RWA terms has been studied in the late 60s in the context of the development of the dressed state approach by \cite{Cohen1969b}, see also \citep{CohenPhotonsAtomesAnglais}. It is also relevant to estimate the trapping potential due to far-off resonant laser beams \citep{Grimm2000}. More recently, it has been studied in the context of rf-dressed adiabatic potentials from atom chips by \cite{Hofferberth2007}, in a situation where the rf frequency is chosen below the minimum Larmor frequency ($\omega<\omega\ind{0,min}$). 

In particular, the Bloch-Siegert shift \citep{Bloch1940,Ramsey1955} has been observed by \cite{Hofferberth2007} using rf spectroscopy with a weak probe field, see \secref{sec:quantum_spectro}. If $\Omega_-$ becomes of order $\omega$, an exact treatment with a numerical approach is required, as done by \cite{Hofferberth2007}. When however $\Omega_-/\omega$ is not too large, this shift can be estimated by second order perturbation theory of the bare states due to the presence of the counter-rotating term. A simple calculation of the shift for the bare state $\ket{m,N}_z$ similar to Eq.~\eqref{eq:VF-large-delta-limit} leads to the new energy
\beq
\tilde{E}_{m,N} = N\hbar\omega - m\hbar\delta + m\frac{\hbar|\Omega_-|^2}{2(\omega+\omega_0)}.
\label{eq:Bloch_Siegert}
\eeq
The detuning $\delta$ should thus be replaced in the equations by a shifted detuning
$$
\tilde{\delta} = \delta - \frac{|\Omega_-|^2}{2(\omega+\omega_0)} = \delta - \frac{|\Omega_-|^2}{2(2\omega-\delta)}.
$$
As a result, the rf resonance for the co-rotating field is shifted and occurs now for a detuning $\delta\ind{res}=\omega-\omega_0$ such that
$$
\delta\ind{res} = \frac{|\Omega_-|^2}{2(2\omega-\delta\ind{res})},
$$
which corresponds for an inhomogeneous magnetic field to a resonance surface defined by
$$
\omega_0(\GG{r}) = \sqrt{\omega^2-|\Omega_-|^2}.
$$
The adiabatic potential should then be corrected and now reads
\beq
\tilde{V}_m(\GG{r}) = m\hbar\sqrt{\left[\omega-\omega_0(\GG{r}) - \frac{|\Omega_-(\GG{r})|^2}{2(\omega+\omega_0(\GG{r}))}\right]^2 + |\Omega_+(\GG{r})|^2}.
\label{eq:potential_beyond_RWA}
\eeq

\begin{figure}[t]
\centering
\includegraphics[width=0.7\linewidth]{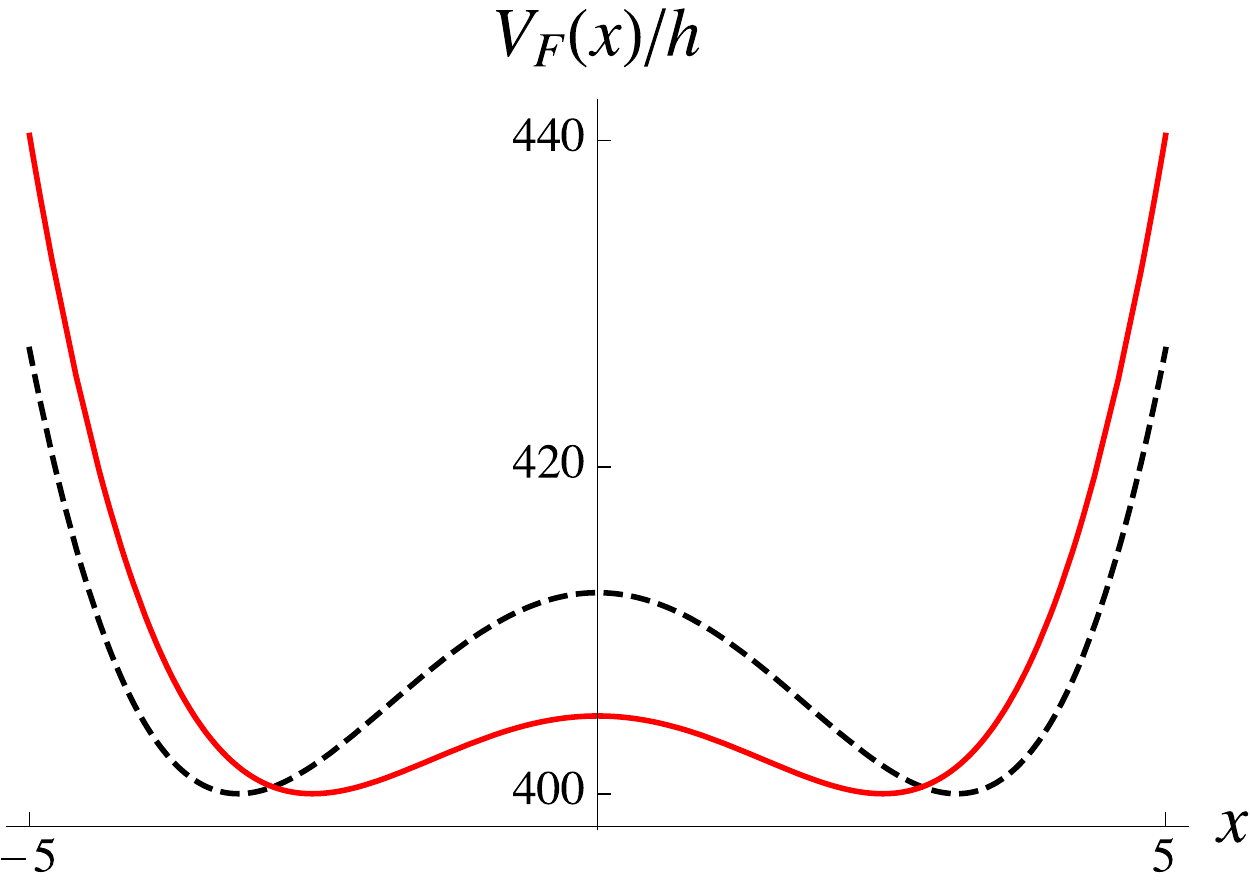}
\caption{Effect of beyond RWA terms in the shape of a double-well potential arising from dressing an initially harmonic magnetic potential $V_0(x)=\hbar\omega\ind{0,min}+\hbar\kappa x^2/2$ with a rf frequency $\omega>\omega\ind{0,min}$ and a particularly large value of $\Omega_\pm/\omega$. Dashed black line: potential calculated within RWA. Red full line: potential $\tilde V_F$ calculated with Eq.~\eqref{eq:potential_beyond_RWA}, including non RWA terms to second order. The shape of the double-well is significantly different. The resonance occurs at a smaller value of the magnetic field, resulting in closer wells. Parameters: $\omega\ind{0,min}/(2\pi)=1$~MHz, $\omega/(2\pi)=1.1$~MHz\label{fig:beyondRWA}, $\Omega_+/(2\pi)=\Omega_-/(2\pi)=400$~kHz, bare trap curvature $\kappa/(2\pi)=20$~kHz$\cdot\mu$m$^{-2}$. Units: $x$ in $\mu$m, $\tilde V_F/h$ in kHz.}
\end{figure}

The effect of this correction can be important when adiabatic potentials are used to produce double-wells, as the tunnel coupling between the two wells depends exponentially on the barrier height. \Figref{fig:beyondRWA} shows the effect of beyond RWA terms on the potential, in the case of dressed magnetic trap with a non-zero minimum, for a very strong Rabi frequency.

\subsection{Misalignment effects of the rf polarization}
\label{sec:misalignment}
In the beginning of \secref{sec:polarization} we stated that the component of the rf field aligned with the static magnetic field doesn't couple different states of angular momentum, such that the rf polarisation should always be chosen with a non-zero projection in the plane orthogonal to the static field. In fact, this is not strictly true. When the rf field has some component $\Omega_z$ along the axis of the magnetic field, taken as the $z$ axis, its effect is to modify the Land\'e factor, by a factor $J_0(\Omega_z/\omega)$ \citep{Haroche1970a}. It can also lead to transitions at submultiples of the Larmor frequency.

The hamiltonian for a static field along $z$ and an rf field with a non-zero projection along $z$ writes:
\beq
\hat{H} = \left(\omega_0+\Omega_z\sin\omega t\right)\,\hat{F}_z +\left\{ \frac{\Omega_+}{2}\, e^{-i\omega t}\hat{F}_+ + h.c.\right\}.
\eeq
The rf field is circularly polarised, but also has a linear component of its polarization along $z$.

To understand the effect of this $\Omega_z$ term, we will look for the solution of a spin rotated through $\hat R=\hat R_z[(n+1)\omega t-\frac{\Omega_z}{\omega}\cos\omega t]$ where $n\in\mathbb{N}$ is a fixed integer. This rotation is chosen to cancel the $\Omega_z$ term of the initial hamiltonian. We get for the rotated hamiltonian Eq.~\eqref{eq:hrot}:
\bea
\hat{H}\ind{rot} &=& \hat R^\dagger \hat{H} \hat R - [(n+1)\omega+\Omega_z\sin\omega t]\,\hat{F}_z \\
&=& -\delta_n\hat{F}_z + \left\{ \frac{\Omega_+}{2}\, e^{in\omega t}e^{-i\frac{\Omega_z}{\omega}\cos\omega t}\hat{F}_+ + h.c.\right\} \nonumber
\eea
where $\delta_n=(n+1)\omega-\omega_0$. The exponential of the cosine may be expanded in terms of Bessel functions of the first kind. This gives:
\bea
&&\hat{H}' = -\delta_n\hat{F}_z \\
&&+ \left\{ \frac{\Omega_+}{2}\, \sum_{n'=-\infty}^{+\infty} (-i)^{n'}J_{n'}\left(\frac{\Omega_z}{\omega}\right)\,e^{i(n-n')\omega t}\hat{F}_+ + h.c.\right\}.\nonumber
\eea

It is now clear that the term with $n'=n$ is the sum is stationary. This means that resonances appear at frequencies $\omega$ such that $\delta_n=0$, with a coupling amplitude given by the Bessel function:
\beq
\mbox{coupling } \Omega_+ J_n\left(\frac{\Omega_z}{\omega}\right) \mbox{ at frequency } \omega = \frac{\omega_0}{n+1}.
\eeq
The $n=0$ case is the usual, expected transition. However, the rf coupling is modified from $\Omega_+$ to $ \Omega_+ J_0\left(\frac{\Omega_z}{\omega}\right)$. We recover a coupling $\Omega_+$ when $\Omega_z$ vanishes. Everything happens as if the Land\'e factor had been modified by a factor $J_0\left(\frac{\Omega_z}{\omega}\right)$, smaller than one, which can even change sign if $\Omega_z$ is comparable with $\omega$ \citep{Cohen1969b,Haroche1970a}. As a consequence, if an accurate description of the adiabatic potential is required, the corrected value of the Land\'e factor must be used instead of its bare value.

The cases $n>0$ correspond to resonances at submultiples of the Larmor frequencies~\citep{Cohen1969b,Pegg1973}, with smaller amplitudes $ \Omega_+ J_n\left(\frac{\Omega_z}{\omega}\right)$. For $\Omega_z\ll\omega$, the coupling amplitude scales as $\left(\frac{\Omega_z}{\omega}\right)^n$ and is very small. For practical purposes in rf-dressed adiabatic potentials, the rf source often has harmonics of the frequency $\omega$ due to non linear amplification. This misalignment effect is another reason, apart from possible anharmonicity of the rf source, for avoiding to have atoms at a position where the Larmor frequency is close to $2\omega$ (see \secref{sec:noise}).

\section{Conclusion}
\label{sec:conclusion}

Magnetic resonance has a long and rich history (see for example \cite{Abragam1961}).
However, in combination with spatially varying fields a new set of phenomena have emerged such as the resonant and off-resonant trapping of ultra-cold atomic gases. 
By using the adiabatic potentials from magnetic resonance to manipulate atoms, radio-frequency dressing has now also become a standard tool for preparation of cold atomic systems.
The approach is also applicable to dressed microwave potentials (\onlinecite{Agosta1989}, \onlinecite{Spreeuw1994}, \onlinecite{Treutlein2006}, \onlinecite{Boehi2009}, \onlinecite{Boehi2010}, \onlinecite{Ammar2015}) where there may also be applications to atomic clocks \citep{Sarkany2014,Kazakov2015}.

In this article we focused on generic experimental set-up configurations, not discussing much the way the fields are produced, by macroscopic or microscopic arrangement of wires. However, it must be stressed that atom chips provide strong fields and gradients that make new field configurations which are strongly influenced by the geometric design of the wires on the chips. These cases include radio-frequency chips as well as microwave potentials from chips such as those mentioned above. The compactness of atom chip systems makes this an area of interest for research into Quantum Technologies. 
Emerging possibilities for cold atom experiments in space \citep{Lundblad2015} further broaden the possibilities of unusual topologies such as the bubble states of matter \citep{Zobay2004}, when atoms are released from the gravitational potential.

\section*{Acknowledgments}
HP thanks the {\'E}cole de physique des Houches, where part of the text has been written for a lecture series, and the whole BEC group at LPL for their contribution to the experiment over the last 15 years. The work presented here is connected to the PhD thesis of Yves \onlinecite{ColombeThese}, Olivier \onlinecite{MorizotThese}, Raghavan \onlinecite{KollengodeEaswaranThese}, Thomas \onlinecite{LiennardThese}, Karina \onlinecite{MerlotiThese}, Camilla \onlinecite{DeRossiThese}, Kathryn \onlinecite{BurrowsThese} and Mathieu de Go\"er de Herve. BMG would like to thank the Leverhulme Trust, the CNRS, the UK EPSRC (grants EP/I010394/1 and EP/M005453/1) and the University of Sussex for supporting the research that contributed to this review.

Laboratoire de physique des lasers is UMR 7538 of CNRS and Paris 13 University. LPL is member of the Institut Francilien de Recherche sur les Atomes Froids (IFRAF).

\setlength{\bibhang}{1em}

\end{document}